\documentclass[twocolumn]{aa} 
\usepackage{graphicx}
\usepackage{amsmath}
\usepackage{amssymb}
\usepackage{pifont}
\usepackage{lscape}
\usepackage{rotating}
\usepackage{natbib}
\usepackage{subfig}
  \def\cT2{c_T^2}

\begin{document}

   \title{Uncertainties in water chemistry in disks: An application to TW Hya}
   \thanks{{\it Herschel} is an ESA space observatory with science instruments provided by European-led Principal Investigator consortia and with important participation from NASA.}

\titlerunning{Uncertainties in water chemistry in disks}

   \author{I. Kamp\inst{1}
         \and
          W.-F. Thi\inst{2}
          \and
          G. Meeus\inst{3}
          \and
          P. Woitke\inst{4}
          \and
          C. Pinte\inst{2}
          \and
          R. Meijerink\inst{1}
          \and
          M. Spaans\inst{1}
          \and
          I. Pascucci\inst{5}
          \and
          G. Aresu\inst{1}
          \and
          W.R.F. Dent\inst{6}
            }

         \institute{
             Kapteyn Astronomical Institute, Postbus 800,
             9700 AV Groningen, The Netherlands
         \and
             Universit\'{e} Joseph Fourier Grenoble-1, CNRS-INSU, Institut de Plan\'{e}tologie et d'Astrophysique (IPAG) UMR 5274, 
             Grenoble, F-38041, France
          \and
             Universidad Autonoma de Madrid, Dpt. Fisica Teorica, Campus Cantoblanco, Spain
             \and
             SUPA, School of Physics \& Astronomy, University of St. Andrews, North Haugh, ST. Andrews, KY16 9SS, UK
             \and
             Lunar and Planetary Laboratory, 1629 E. University Blvd., Tucson AZ 85271-0092, USA
             \and
             ALMA SCO, Alonso de Cordova 3107, Vitacura, Santiago, Chile
            }

   \date{Received 24.10.2012; accepted 2.8.2013}

 
\abstract
   {This paper discusses the sensitivity of water lines to chemical processes and radiative transfer for the protoplanetary disk around TW~Hya. The study focuses on the Herschel spectral range in the context of new line detections with the PACS instrument from the Gas in Protoplanetary Systems project (GASPS).}
    {The paper presents an overview of the chemistry in the main water reservoirs in the disk around TW~Hya. It discusses the limitations in the interpretation of observed water line fluxes.}
    {We use a previously published thermo-chemical \underline{Pro}toplanetary \underline{Di}sk \underline{Mo}del (ProDiMo) of the disk around TW~Hya and study a range of chemical modeling uncertainties: metallicity, C/O ratio, and reaction pathways and rates leading to the formation of water. We provide results for the simplified assumption of $T_{\rm gas}\!=\!T_{\rm dust}$ to quantify uncertainties arising for the complex heating/cooling processes of the gas and elaborate on limitations due to water line radiative transfer.}
    {We report new line detections of p-H$_2$O ($3_{22}-2_{11}$) at $89.99~\mu$m and CO\,J=18--17 at $144.78~\mu$m for the disk around TW~Hya. Disk modeling shows that the far-IR fine structure lines ([O\,{\sc i}], [C\,{\sc ii}]) and molecular submm lines are very robust to uncertainties in the chemistry, while the water line fluxes can change by factors of a few. The water lines are optically thick, sub-thermally excited and can couple to the background continuum radiation field. The low-excitation water lines are also sensitive to uncertainties in the collision rates, e.g.\ with neutral hydrogen. The gas temperature plays an important role for the [O\,{\sc i}] fine structure line fluxes, the water line fluxes originating from the inner disk as well as the high excitation CO, CH$^+$ and OH lines. 
    }
    {Due to their sensitivity on chemical input data and radiative transfer, water lines have to be used cautiously for understanding details of the disk structure. 
    Water lines covering a wide range of excitation energies provide access to the various gas phase water reservoirs (inside and outside the snow line) in protoplanetary disks and thus provide important information on where gas-phase water is potentially located. Experimental and/or theoretical collision rates for H$_2$O with atomic hydrogen are needed to diminish uncertainties from water line radiative transfer.}



   \maketitle


\section{Introduction}

Since the first detection of water in the disk around the low mass star AA~Tau with the Spitzer Space Telescope by \citet{Carr2008}, protoplanetary disks have seen an increasing boom in water detections \citep{Salyk2008,Najita2010,Pontoppidan2010,Riviere-Marichalar2012}. While Spitzer observations are spectrally unresolved, ground based observations have confirmed in several cases that the water is indeed circumstellar in origin by spatially and spectrally resolving the emission \citep{Pontoppidan2010a}. The excitation temperatures of the far-IR and submm water lines range from tens of K to 1000~K. Hence, they originate in very different radial zones of the disk. The strong mid-IR water lines discovered by Spitzer/IRS \citep[e.g.][]{Carr2008, Salyk2008} and from the ground with VLT/CRIRES \citep{Pontoppidan2010} are inferred to originate inside a few AU and hence to trace the warm water reservoir inside the snow line. \citet{Glassgold2009} proposed that the warm water in the innermost disk is formed through warm neutral gas-phase chemistry and a more extensive discussion on the relevance of heating processes for the column densities of warm water can be found in \citet{Najita2011}. \citet{Woitke2009b} identified three distinct water reservoirs in the disk around a typical Herbig Ae star: a warm high abundance inner water reservoir, a lower abundance warm surface layer driven by neutral-neutral chemistry and an outer cold water belt from photodesorption of icy grains.

While the presence of water in these disks is thus clearly established, the question remains how sensitive the interpretation of the water lines is to details in the physics and chemistry of the underlying disk models. Detailed multi-wavelengths disk observations and modeling indicates that even relatively young disks can be very different from a `prototypical disk' with a canonical gas-to-dust mass ratio of 100 \citep[e.g.][]{Tilling2012,Thi2011b,Woitke2011,Meeus2010,Chapillon2010}. Hence, we start answering these questions by choosing the well known  nearby and isolated disk around TW Hya \citep[$51 \pm 4$~pc,][]{Mamajek2005} that has an age ranging from $3$~Myr \citep{Vacca2011} to $10^{+10}_{-7}$~Myr \citep{Barrado2006}. TW Hya has ---  despite its possibly unusual gas-to-dust mass ratio ---, a wide wavelength coverage in observations of water.

TW Hya is a well studied disk with a rich set of observed atomic/molecular lines from the near-IR to the submm. \citet{Kastner1997} were the first to observe molecules such as $^{12}$CO, $^{13}$CO, HCO$^+$, HCN, and CN in the disk around TW Hya. Subsequently, \citet{Thi2004} observed a series of CO, HCN, and HCO$^+$ lines, whose profiles indicate an origin in the outer disk ($R_{\rm out}\!\sim\!200$~AU). The molecular emission arises from gas in Keplerian rotation \citep{Qi2006} and the line ratios require gas temperature gradients different from that of the dust.  \citet{Salyk2007} detected also a series of CO v=1--0 ro-vibrational lines from inside 1~AU. \citet{Najita2010} report the non-detection of water in the Spitzer/IRS data of TW Hya at wavelengths of \mbox{$5\!-\!14$}~$\mu$m. [Ne\,{\sc ii}] and optical forbidden lines originating at $1\!-\!10$~AU seem to trace a photoevaporative wind \citep{Pascucci2011}. A strong atomic oxygen line at $63~\mu$m has been detected with Herschel/PACS \citep{Thi2010a}. Recently, \citet{Hogerheijde2011} have detected the ortho and para ground state water line in TW\,Hya with the HIFI instrument onboard the Herschel Space Observatory. The width of the lines point towards an origin in the outer disk. 

Based on Herschel Science Demonstration Phase data, \citet{Thi2010a} found that a model with a gas-to-dust mass ratio of 1 can fit the current SED and PACS line data for TW Hya. The recent publication by \citet{Fogel2011} describes an alternative disk model for TW\,Hya with special focus also on the water abundances in the outer disk, suggesting that photodesorption by stellar UV and Ly\,$\alpha$ photons generates gas phase water abundances\footnote{The water abundance is defined as $\epsilon_{\rm H_2O}=n({\rm H_2O})/n_{\langle {\rm H} \rangle}$, with $n_{\langle {\rm H} \rangle}$ being the total hydrogen number density $n({\rm H})+2n({\rm H_2})$} at the level of $10^{-6}$ in regions below $100$~K. \citet{Hogerheijde2011} present another model for the disk around TW\,Hya especially focussing on the explanation of the observed HIFI water line fluxes. In both papers, dust settling of larger icy grains is invoked to yield lower water abundances of the order of $10^{-8}$; such a low water abundance can then explain the observed HIFI water line fluxes from the outer disk. In addition to that, \citet{Gorti2011} present yet another TW Hya model consistent with a large set of atomic/molecular line diagnostics; their model contains $6\!\times\!10^{-2}$~M$_\odot$ of gas in the outer disk ($4\!-\!200$~AU).

In this paper, we report the new detection of the $89.99~\mu$m para water line ($3_{22}-2_{11}$) and the \mbox{CO J=18--17} line in a deep PACS integration of TW Hya within the GASPS Open Time Key Program on the Herschel Space Observatory \citep{Dent2013}. However, instead of finding another `best fitting' model, we rather explore with the existing TW Hya model the range of water vapor abundances in the disk and  the impact of chemical processes and input parameters on our disk model: elemental abundances, X-rays, photodesorption yields, water formation on dust surfaces, and water desorption channels. We also investigate the radiative transfer of water in more detail and assess the diagnostic value of this species for the overall disk structure.

\section{Observations}
\label{Sect:obs}

We describe in the following the new Herschel/PACS observations of TW Hya, selected sub-mm lines from the literature and also Spitzer archival data used to support the analysis of water in this particular disk.

\subsection{{\it Herschel} PACS observations}
\label{s_pacs}

\begin{figure*}[!htbp]
\includegraphics[width=6.5cm, angle=90]{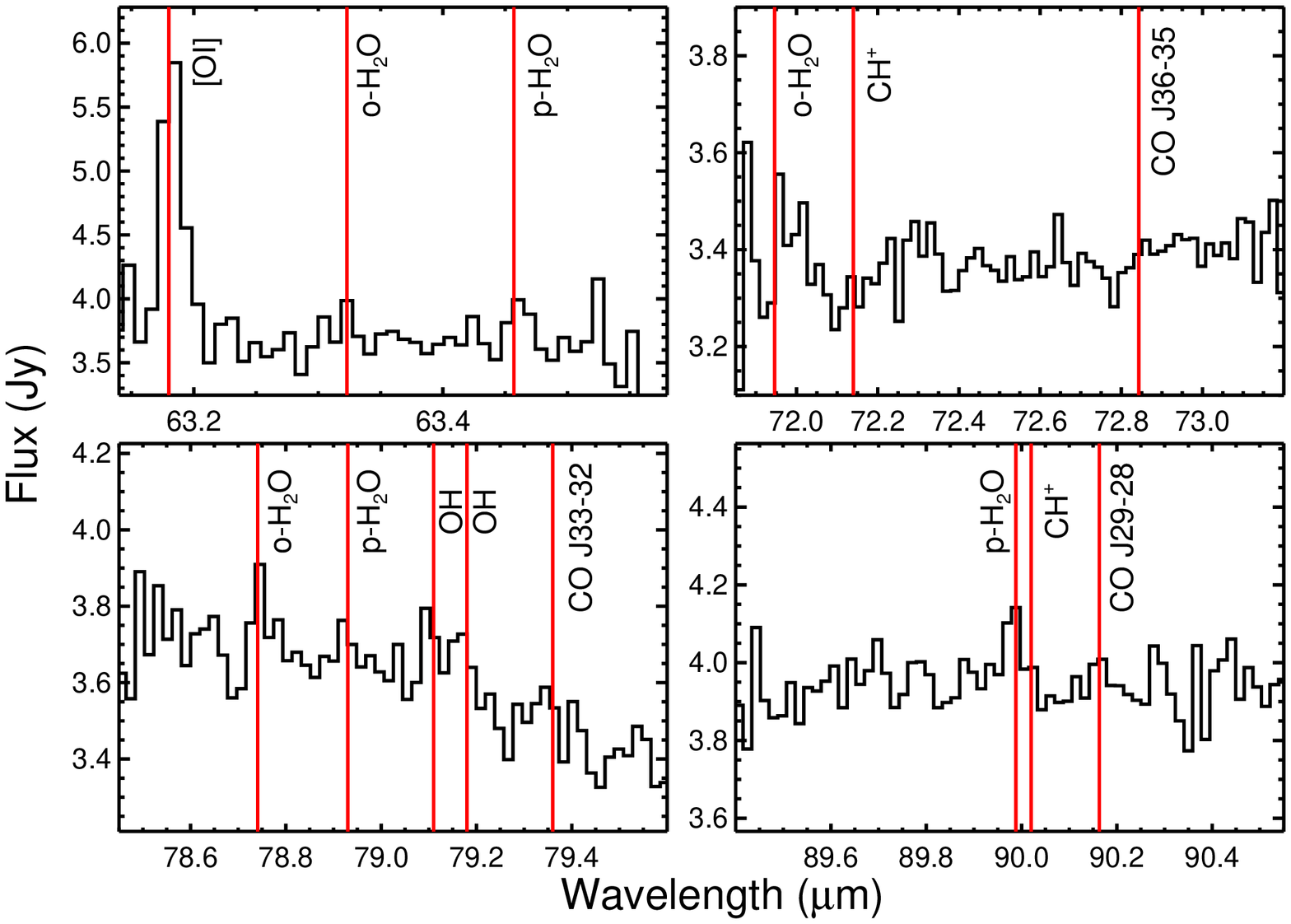}
\includegraphics[width=6.5cm, angle=90]{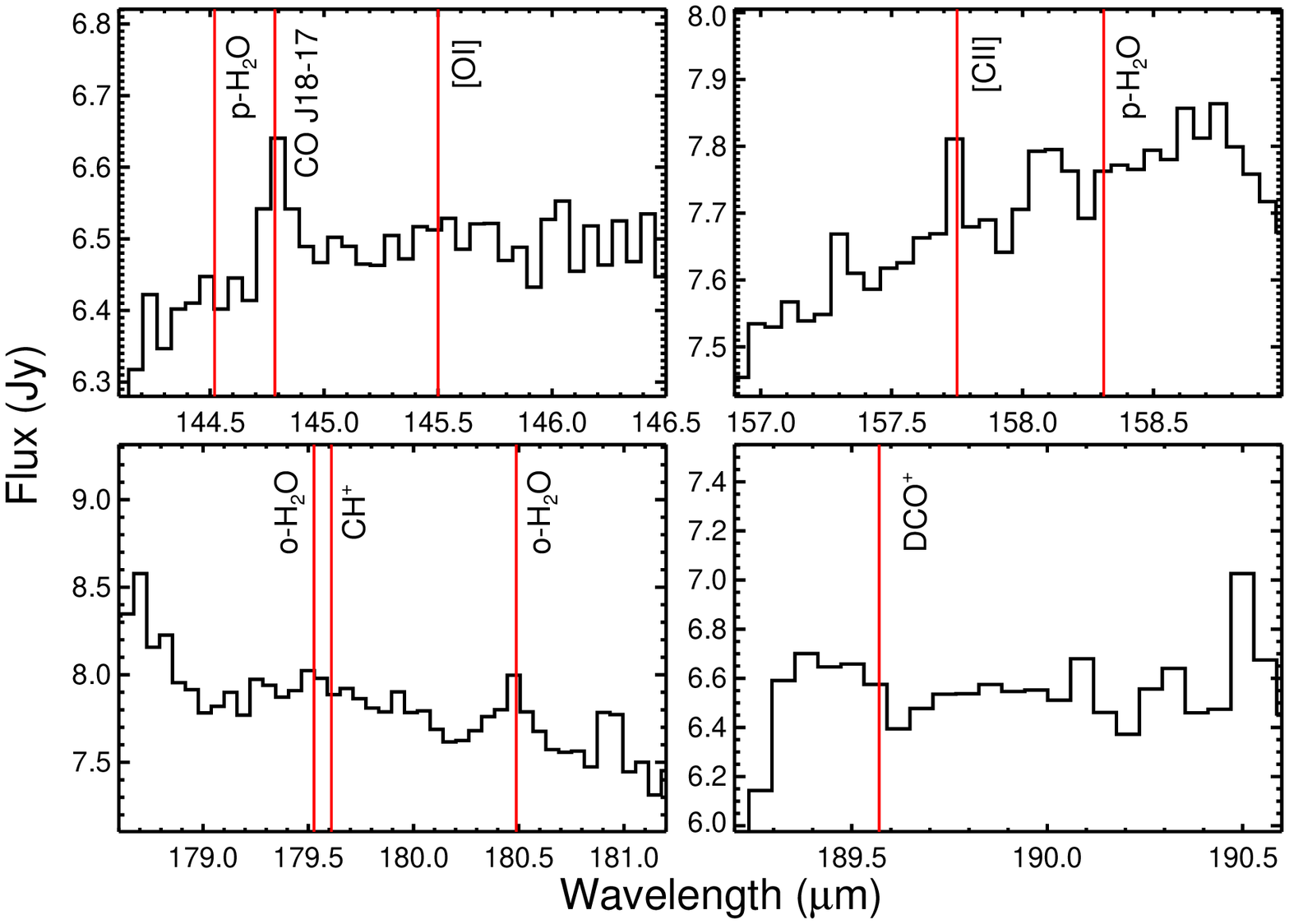}
\caption{The full set of TWHya PACS spectra obtained from the GASPS program including the deeper range scan. }
\label{fig:observations}
\end{figure*}

Our TW Hya observations are part of the {\it Herschel} Open Time Key Project GASPS (P.I.\ Dent),  see \citet{Dent2013}. We initially obtained PACS \citep{Poglitsch2010} spectroscopy (obsid 1342187238 PacsRangeSpec, 5150\,s and obsid 1342187127, PacsLineSpec, 1669\,s) centered on 8 selected lines. In a later stage, we obtained a deeper range scan to confirm tentative detections (obsid 1342211845, 41107\,s). The observations were carried out in ChopNod mode, in order to be able to remove the emission of the telescope.

\begin{table}[h]
\caption{Herschel/PACS line fluxes are in $10^{-18}$~W/m$^2$. For non-detections, $3\sigma$ upper limits are listed.}
\begin{tabular}{llll}
\hline
\hline
Species           & observed wavelength & line flux & error \\
                  & [$\mu$m]   &           & ($1\sigma$) \\
\hline
O 	          & 63.18      & 37.0     & 3.3 \\	
	          & 145.52     & $<3.5$    & \\
CII               & 157.75     & $<3.6$    & \\
o-H$_2$O            & 63.32       & $< 6.1$ & \\
p-H$_2$O              & 63.46     & $<7.2$ & \\
o-H$_2$O               & 71.946     & $<6.4$    & \\
o-H$_2$O                & 78.74      & $<6.0$  & \\
p-H$_2$O                & 78.93      & $<6.4$  & \\
p-H$_2$O 	          & 89.988 \tablefootmark{1}     & 3.1      & 0.9 \\
p-H$_2$O                 & 144.52 & $<6.6$ & \\
p-H$_2$O                 & 158.31   & $<5.2$ & \\	
o-H$_2$O 	          & 179.52     & $<7.9$   & \\
o-H$_2$O 	          & 180.45     & $<8.32$   & \\
OH 	          & 79.11/79.18 & $<7.8$    & \\
CO J=36-35       & 72.86  & $<4.2$ & \\	
CO J=33-32        & 79.36      & $<8.3$    & \\
CO J=29-28        & 90.16      & $<3.4$    & \\
CO J=18-17        & 144.78     & 3.5      & 1.2 \\	
CH$^+$ J=5-4         & 72.14      & $<7.1$   & \\	
CH$^+$ J=4-3         & 90.02      & water blend & \\	
CH$^+$ J=2-1         & 179.61     & $<7.8$ & \\	
\hline
\end{tabular}
\\ \tablefoottext{1}{blend with CH$^+$ 90.02~$\mu$m}
\label{tab:PACSobsfluxes}
\end{table}

The spectroscopic data were reduced with the stable developers build version 8.0.1559 of the Herschel Interactive Processing Environment \citep[HIPE; ][]{Ott2010}, using standard tasks provided in HIPE. These include bad pixel flagging; Chop On/Off subtraction; spectral response function division; rebinning with oversample$=\!2$ and upsample$=\!1$ corresponding to the native resolution of the instrument; spectral flatfielding and finally averaging of the 2 Nod positions. The spectral resolution varies between 1100 and 3400. In order to conserve the best signal and not to introduce additional noise, we only extracted the central spaxel, and corrected for the flux loss with an aperture correction provided by the PACS instrument team ('pointSourceLossCorrection.py'). This was possible as the source was well-centered.

We extracted the line fluxes using a gaussian fit to the emission lines with a first-order  polynomial to the continuum, using the RMS on the continuum (excluding the line) to derive a 1$\sigma$ error on the line by integrating a gaussian with height equal to the continuum RMS and width of the instrumental FWHM. This approach is necessary as HIPE currently does not deliver errors. The absolute flux calibration error given by the PACS instrument team is currently 30\%. 

The spectra obtained in line and range scan show the presence of several lines of water, CO, and [O\,{\sc i}]. Fig.~\ref{fig:observations} shows in fact the tentative presence of many more molecular lines such as the hyperfine doublet of OH ($79.11, 79.18~\mu$m). However, formally these are not $3\sigma$ detections. Table~\ref{tab:PACSobsfluxes} summarizes the measured integrated fluxes and upper limits.

\subsection{Data from the literature}

In addition to Herschel PACS data, we use the two water line detections reported by \citet{Hogerheijde2011} from the WISH program \citep{vanDishoeck2011} using the HIFI instrument onboard the Herschel space observatory. Water has not been detected with Spitzer IRS at wavelength below 20~$\mu$m \citep{Najita2010}. The Spitzer IRS spectra that we use stems from \citet{Szulagyi2012}  (SL module). In addition we use the upper limit for the [Fe\,{\sc ii}] fine structure line at 25.99~$\mu$m (Carr \& Najita private communication)  to inform our choice of metal abundances.

We also include single dish sub-mm data on four lines, \mbox{$^{12}$CO~3--2}, \mbox{$^{13}$CO~3--2}, \mbox{HCO$^+$~4--3}, and \mbox{HCN~4--3} from \citet{Thi2004} (see Table~\ref{tab:PACSmodfluxes1} and \ref{tab:PACSmodfluxes2}). These lines are included because they provide stringent constraints on the outer disk structure and gas mass. The errors for the submm lines are calculated from the thermal noise provided in Table~3 of \citet{Thi2004} and a drift+systematic noise estimated to be 20\% of the line $T_{\rm mb}$. Since these observations originate from very deep integrations, an additional error could come from a wrong value of the beam efficiency correction which can vary with time. The fluxes agree with the previous single dish observations of \citet{Kastner1997} to within 10\%. In the final comparison between observed and modeled line fluxes, we also take into account their  \mbox{$^{12}$CO~2--1} and \mbox{$^{13}$CO~2--1} fluxes with their respective error bars: $0.136\pm 0.0067$ and $0.0163\pm 0.0058\!\times\!10^{-18}$~W/m$^2$.






\section{The TW\,Hya disk model}
\label{sect:standardmodel}

The `standard' disk model used in this work is the \citet{Thi2010a} MCFOST/ProDiMo model with a gas mass of $3\!\times\!10^{-3}$~M$_\odot$ (surface density power law $\Sigma \propto r^{-1}$). The grain size distribution is a continuous power law between $3\!\times\!10^{-2}~\mu$m and 10~cm with a power law index of 3.4 and the opacities are those for amorphous olivine. The large $a_{\rm max}$ in our model is driven by the VLA detection of TW Hya at cm wavelength \citep{Wilner2000}\footnote{Recent modeling of the photo evaporative wind shows that the VLA data could partially originate from free-free emission \citep{Pascucci2012}.}. It is a fully parametrized disk model without any vertical structure iteration, but the chemistry and gas temperature are calculated self-consistently. The model has an optically thin inner dust disk from 0.25 to 4~AU containing $1.2\!\times\!10^{-9}$~M$_\odot$ in grains up to 1~mm ($2 \!\times\!10^{-8}$~M$_\odot$ in grains up to 10~cm). The outer dust disk (4-196~AU) has a mass of $1.9\!\times\!10^{-4}$~M$_\odot$ in grains up to 1~mm ($3\!\times\!10^{-3}$~M$_\odot$ in grains up to 10~cm) and extends to 196~AU. The total dust mass in grains up to 10~cm implies a gas-to-dust mass ratio of 1. The mean surface area of the dust is $\langle a^2\rangle\!=\!5.4\,10^{-11}$~cm$^{2}$. The PAH abundance is 0.01 wrt the standard ISM abundance ($10^{-6.52}$~PAH molecules/H-nucleus). The `standard' disk model does not include X-rays to enable a direct comparison with the previous results of \citet{Thi2010a}. However, we explore the role of X-rays on the water chemistry in the general sensitivity analysis. The resolution of the models presented here --- unless noted otherwise --- is 200x140 grid points; a logarithmic radial grid enables us to equally well sample the inner and outer disk. More details and a full table of modeling input parameters can be found in \citet{Thi2010a}. Details about the 3D Monte Carlo radiative transfer code MCFOST can be found in \citet{Pinte2006,Pinte2009}. Details on the thermo-chemical disk modeling code \underline{Pro}toplanetary \underline{Di}sk \underline{Mo}dels (ProDiMo) are discussed in a series of papers \citep{Woitke2009a,Kamp2010,Thi2011a,Woitke2011,Aresu2011,Meijerink2012}. Our chemistry is based on the UMIST2007 database \citep{UMIST2007}. Details on the atomic and molecular data used to calculate the statistical equilibrium and line radiative transfer are provided in Appendix~\ref{app:moldata}.

The elemental abundances are changed with respect to the original paper to reflect the low metallicity of the original ISM material that formed the disk in the first place (Table~\ref{Tab:elementabus}). The data are taken from \citet{Jenkins2009}, column~7 of his Table~4 and are referred to as `low metal' case. 

TW Hya has an observed $L_{\rm X}$ of $1.3\!\times\!10^{30}$~erg/s \citep{Kastner2002}. This is an average X-ray luminosity for low mass young stellar objects \citep{Guedel2007}. However, based on our recent modeling of the impact of $L_{\rm X}$ versus $L_{\rm FUV}$ \citep{Meijerink2012,Aresu2012}, we expect the disk to be dominated by FUV. We ran two extra models including X-rays to show the impact on the gas emission lines studied here. 

\subsection{Physical and chemical disk structure}

\begin{figure*}[!htbp]
\includegraphics[width=6cm]{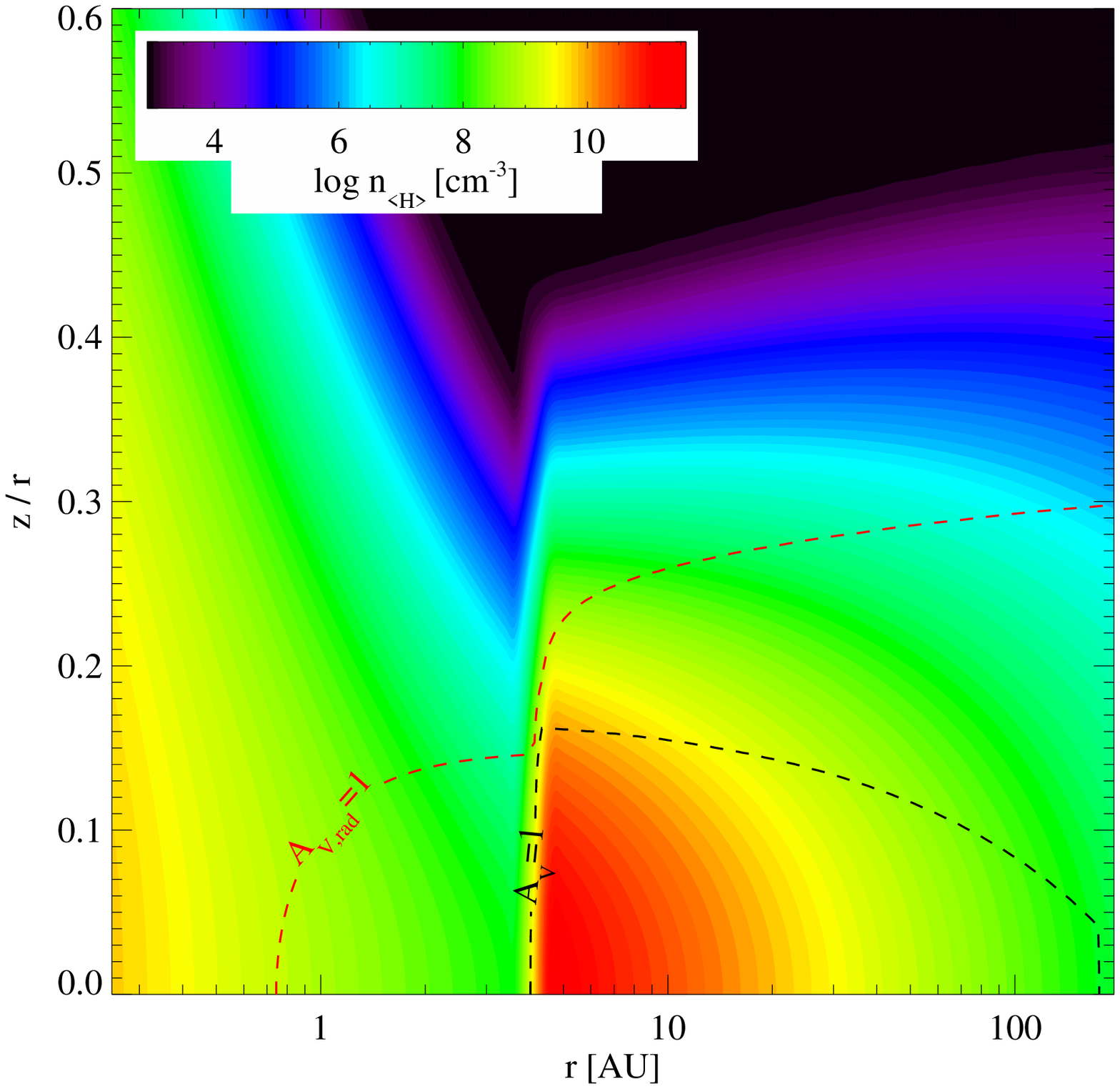} 
\includegraphics[width=6cm]{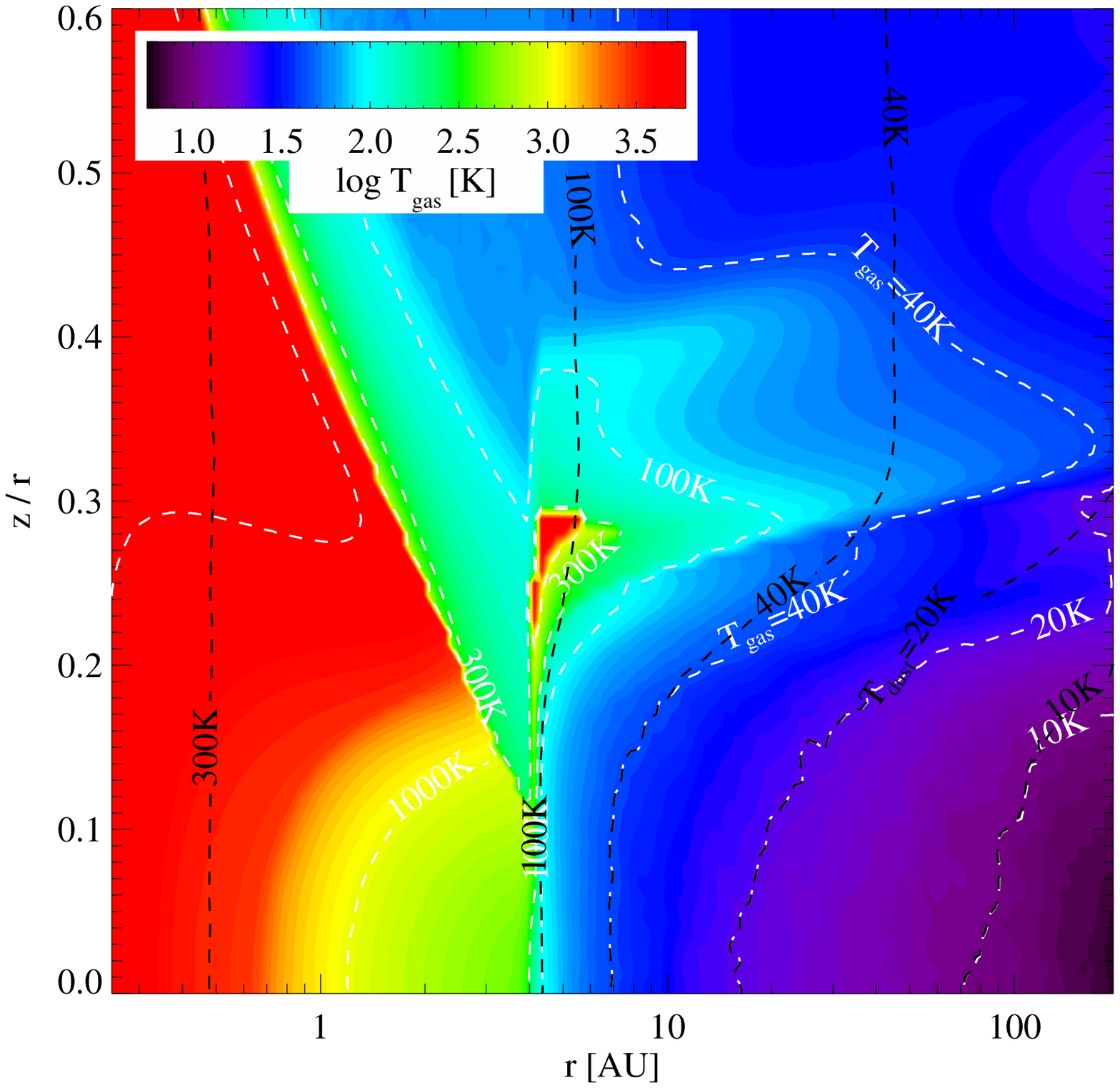} 
\includegraphics[width=6cm]{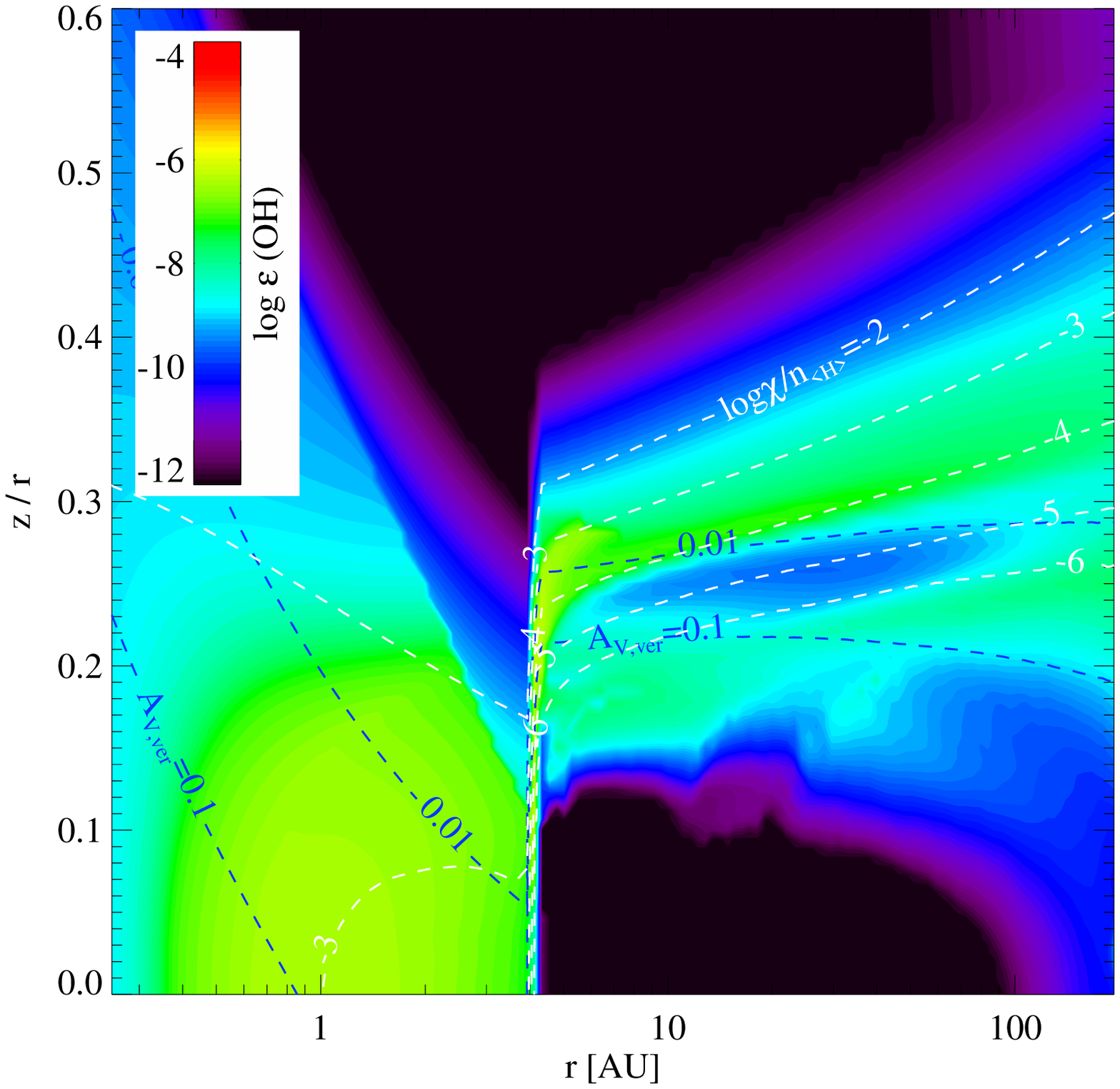} 
\includegraphics[width=6cm]{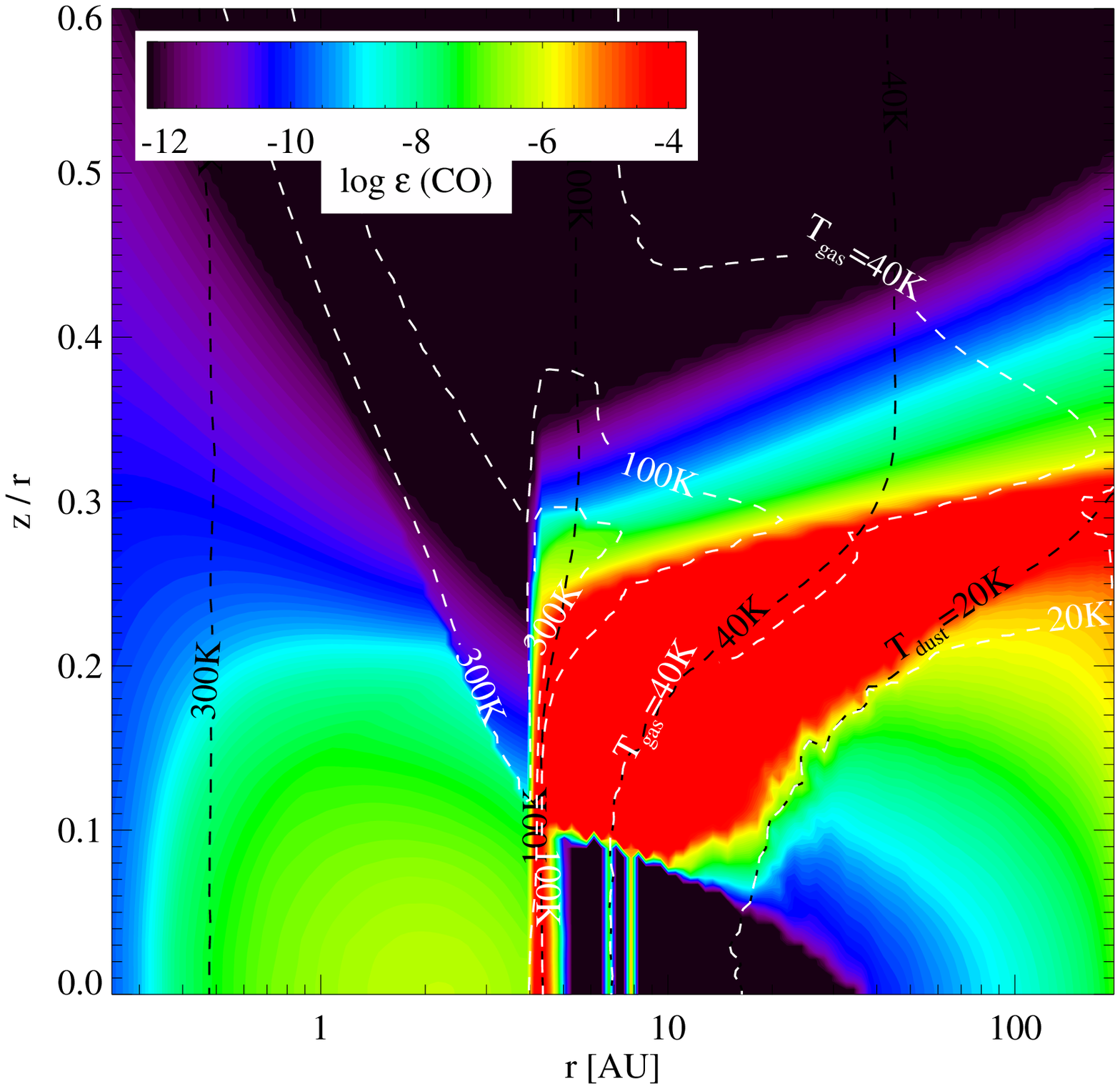} 
\includegraphics[width=6cm]{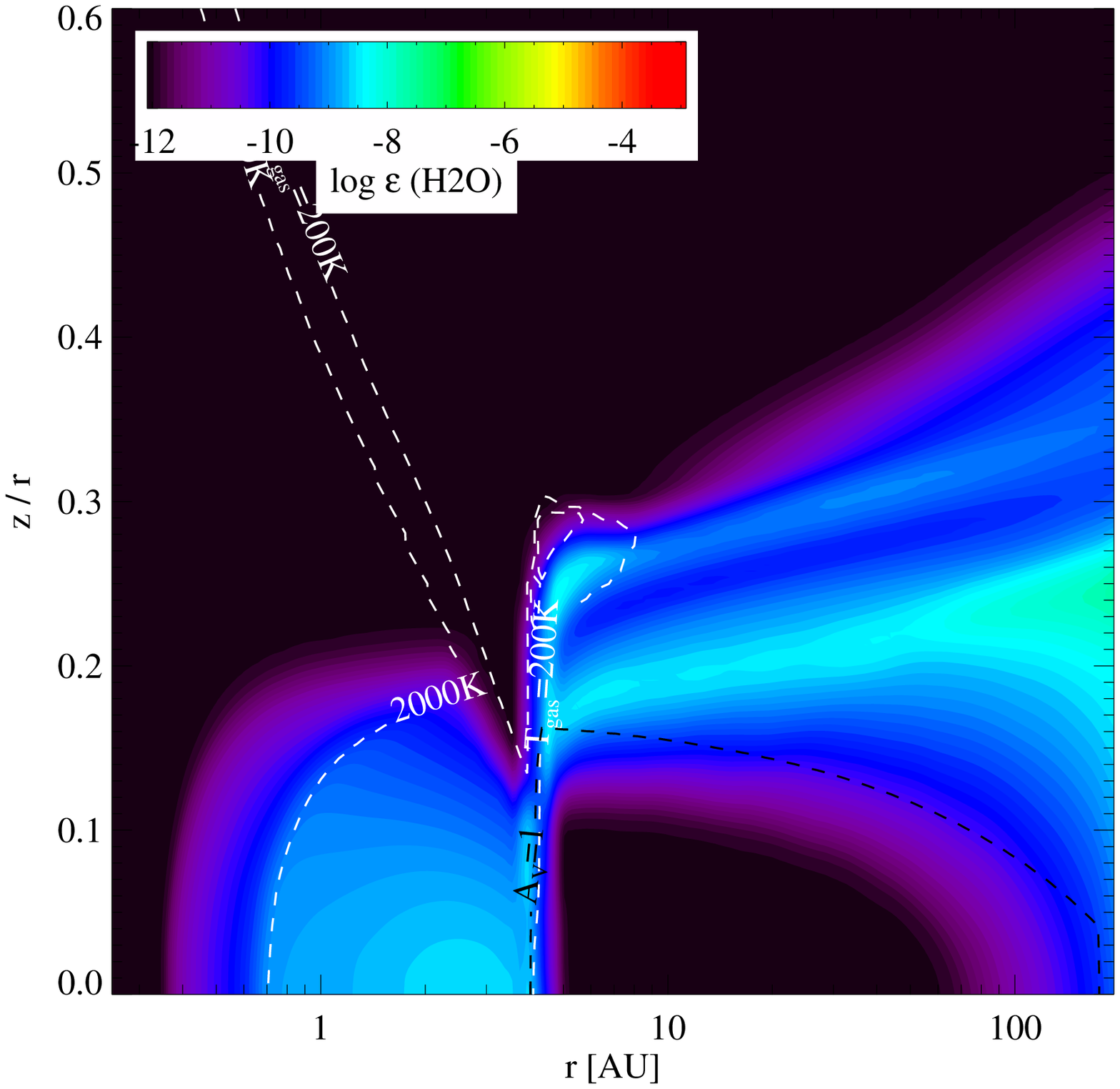} 
\includegraphics[width=6cm]{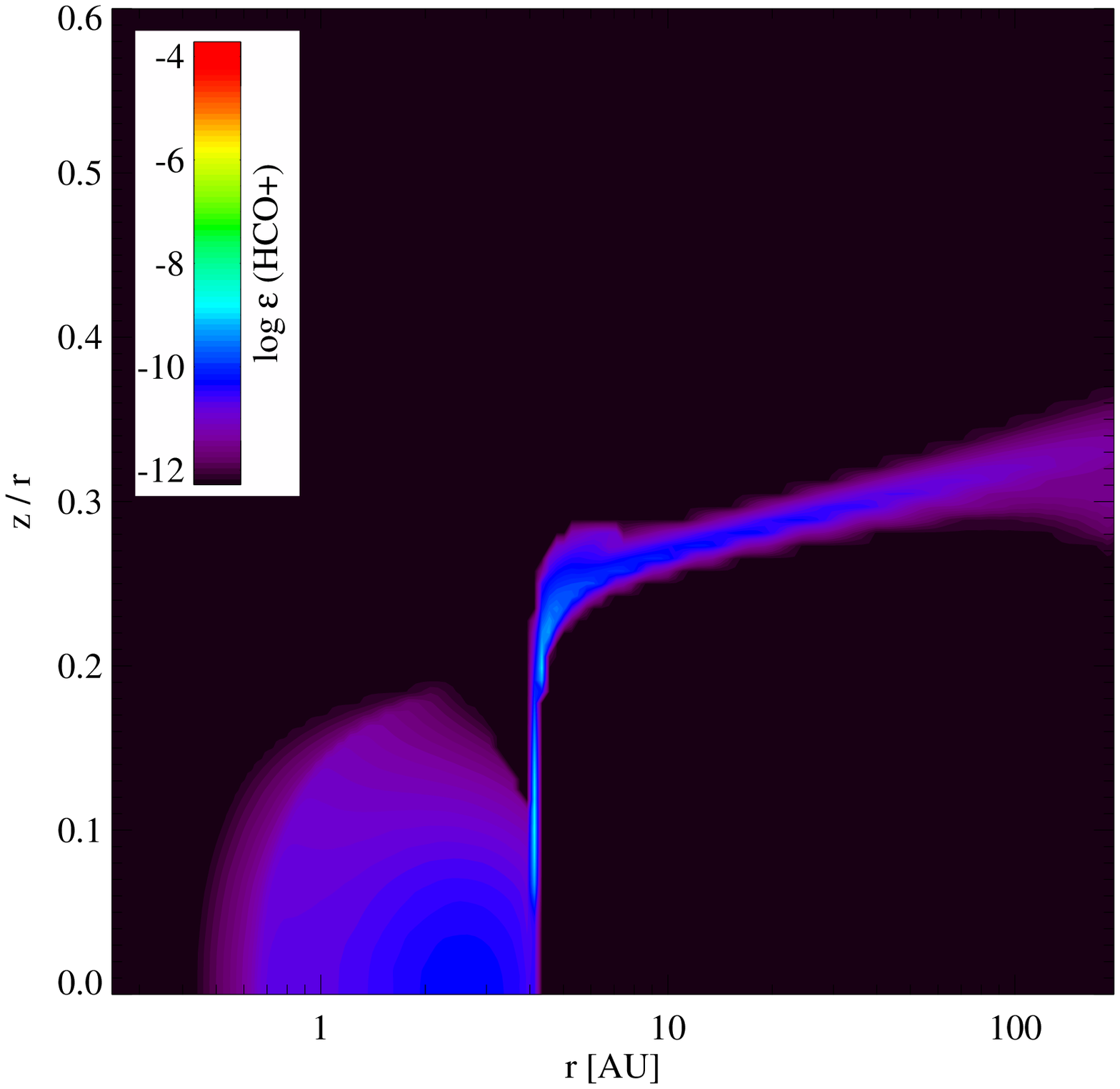} 
\caption{Basic plots for the standard disk model: (1) The total hydrogen number density with red/black contours overlaid showing the radial/total extinction of 1, (2) the gas temperature with contours overlaid for the gas temperature (white) and dust temperature (black), (3) the OH abundance with the vertical extinction of 0.01 and 0.1 overlaid by blue contours and the PDR parameter $\log \chi/n_{\langle \rm H\rangle}\!=\!-2, ...-6$ overlaid by white contours, (4) the CO abundances with gas temperature contours (white) and dust temperature contours (black), (5) the water abundance with the total extinction $A_{\rm V}\!=\!1$ overlaid in black and the gas temperature contours (white) and (6) the HCO$^+$ abundance.}
\label{fig:standard-basic}
\end{figure*}

Fig.~\ref{fig:standard-basic} illustrates the total hydrogen number density, radial and vertical extinction, gas and dust temperatures and the OH, CO, H$_2$O and HCO$^+$ abundances of the `standard' model. 

The inner disk is optically thin to interstellar UV photons and thus contains very low molecular abundances (CO and water). In the inner tenuous disk, OH mainly forms through the endothermic reaction (activation energy $E_a\!=\!3150$~K) of molecular hydrogen with atomic oxygen. Gas temperatures in the inner disk are of the order of 1000~K. 

In the outer disk, the top OH layer builds up through reactions of atomic oxygen with photo-excited H$_2$ and NH;  NH itself also being the product of N collisions with excited molecular hydrogen. The reaction of atomic oxygen with NH can either lead to NO or OH with a branching ratio of 7\% \citep{Adamson1994}. The rate leading to OH though varies in the chemistry databases ranging between $1.16\!\times\!10^{-11}$~cm$^3$~s$^{-1}$ (UMIST, OSU) and zero (KIDA).
The HCO$^+$ chemistry and its dependence on electron density (and thus metallicity of the gas) is discussed in some more detail in Appendix~\ref{app:metalabus}. CO in the outer disk forms at the disk surface via ions and ion-molecules such as C$^+$, CO$^+$ and HCO$^+$, HOC$^+$. At intermediate layers neutral-neutral chemistry dominates (CH\,$+$\,O, C\,$+$\,NO and O\,$+$\,C$_2$). In all layers, the main destruction mechanisms are photodissociation and/or freeze-out of CO. Deeper into the disk, an equilibrium is reached between thermal-/photo-desorption and freeze-out, except in very shielded regions where H$_3^+$ (generated by cosmic rays) adds to the destruction channel. The water chemistry will be discussed in detail in Sect.~\ref{sect:waterreservoirs}.

\begin{table}[h]
\caption{Elemental abundances for the standard model taken from \citet{Jenkins2009}.}
\begin{tabular}{cc|cc}
\hline
\hline
Element & $12\!+\!\log \epsilon$ & Element & $12\!+\!\log \epsilon$ \\
\hline
H  & 12.00 & Na & 5.08 \\
He & 10.99 & Mg & 6.35 \\
C  & 8.25 & Si & 6.25 \\
N  & 7.79 & S  & 5.96 \\
O  & 8.52 & Fe & 5.30 \\
\hline
\end{tabular}
\label{Tab:elementabus}
\end{table}

\subsection{Comparison with other disk models}

We do not include Ly\,$\alpha$ emission in our input spectrum. Ly\,$\alpha$ line photons can be more efficiently resonantly scattered deep into the disk compared to the UV continuum \citep{Bergin2003}. The presence of Ly\,$\alpha$ photons deep inside a non-settled disk model of TW Hya does change the abundance pattern of HCN, but has very little impact on e.g.\ CO \citep{Fogel2011}; for some species such as water and OH, Ly\,$\alpha$ dissociation is roughly compensated by increased Ly\,$\alpha$ photodesorption.

\citet{Hogerheijde2011} used the same underlying 2D density structure and grain size distribution as \citet{Thi2010a} and hence as our `standard model'. However, the dust temperatures and gas chemical composition are recalculated using different codes and assuming $T_{\rm gas}\!=\!T_{\rm dust}$. According to our disk model, the latter assumption is fairly good in the regions where the fundamental water lines originate; differences in those regions between gas and dust temperatures are lower than a factor 2 in our `standard model'. Another important difference in our modeling strategy is that \citet{Hogerheijde2011} assume that the ice reservoir remains the primordial one even if they decrease the disk gas mass. We assume in this paper steady state chemistry which involves an overall lower mass ice reservoir on small grains in an evolved (lower gas-to-dust mass ratio) disk. In models that take into account the time-dependent chemistry in the cosmic-ray dominated midplane, water ice can even form at late times (one to a few Myr) processing oxygen from CO into water  \citep{Chaparro2012a}.

\begin{table}
\caption{Radial zones in which the water lines form. Listed are the radii within which 15 and 85\% of the line flux have built up (note that these values are calculated from the vertical escape probability approach).}
\begin{tabular}{llllllll}
\hline
\hline
  wavelength & desig.                        & $E_{\rm up}$  & $A_{ij}$    & $R_{\rm 15\%}$ & $R_{\rm 85\%}$ \\
  \, [$\mu$m]     &                                               &        [K]              & [s$^{-1}$] &  [AU]                    & [AU] \\
\hline
\multicolumn{6}{c}{o-H$_2$O}\\
\hline
 538.28           & $1_{\rm 10} - 1_{\rm 01}$ &     60.96           & 3.458(-3) &  35 & 160  \\
 180.49           & $2_{\rm 21} - 2_{\rm 12}$ &    194.1            & 3.058(-2) & 4 & 11 \\
 179.53           & $2_{\rm 12} - 1_{\rm 01}$ &    114.4            & 5.593(-2) & 5 & 110 \\
   78.74           & $4_{\rm 23} - 3_{\rm 12}$ &    432.2            & 4.838(-1) & 4 & 4.5 \\
   71.95           & $7_{\rm 07} - 6_{\rm 16}$ &    843.5            & 1.157 & 4 & 4.5 \\
   63.32           & $8_{\rm 18} - 7_{\rm 07}$ &   1071.             & 1.751 & 4 & 4.5 \\
\hline
\multicolumn{6}{c}{p-H$_2$O}\\
\hline
 269.27           & $1_{\rm 11} - 0_{\rm 00}$ &       53.43         & 1.842(-2) & 22 & 150 \\
 158.31           & $3_{\rm 31} - 4_{\rm 04}$ &   410.4            & 1.630(-4) & 4 & 4.5\\
 144.52           & $4_{\rm 13} - 3_{\rm 22}$ &   396.4            & 3.316(-2) & 4 & 4.5 \\
    89.99          & $3_{\rm 22} - 2_{\rm 11}$ &     296.8          & 3.524(-1) & 4 & 5.5 \\
    78.93          & $6_{\rm 15} - 5_{\rm 24}$ &    781.1           & 4.526(-1) & 4 & 4.5 \\	
    63.46          & $8_{\rm 08} - 7_{\rm 17}$ &  1070.6           & 1.742 &  4  & 4.5 \\
\hline
\end{tabular}
\label{tab:originH2O}
\end{table}


\section{The water reservoirs in the standard model}
\label{sect:waterreservoirs}

The PACS lines studied within the GASPS program have upper level energies between 100 and 1000~K. Water lines in the $4\!-\!30~\mu$m range have excitation energies of $1000\!-\!3000$~K. The two HIFI lines at $538.28$ and $269.27~\mu$m have upper level energies well below 100~K. Such low excitation lines are therefore thought to originate in the outer disk, where the temperatures are low enough for water to freeze out on dust grains.

Table~\ref{tab:originH2O} shows where the Herschel lines form according to our standard disk model. The radii within which 15\% and 85\% of the line emission arises are calculated using a simplified vertical escape probability. Since the inclination of the disk around TW\,Hya is only $7^{\rm o}$, these values should be a good approximation. The lines fall into three categories: (1) lines from the warm gas of the inner rim of the outer disk ($4\!-\!4.5$~AU), (2) lines from the cold gas of the outer disk (beyond 20~AU), and (3) lines that originate in a range spanning both parts of the disk ($4\!-\!100$~AU). There is a single line in the latter category, the $179.53~\mu$m line with a low upper level energy of $\sim\!114.4$~K. Interestingly, the higher excitation lines are preferentially detected in the disk around TW\,Hya with Herschel (e.g.\ $89.99~\mu$m, $E_{\rm up}\!=\!296.8$~K); according to our model, these lines originate from the inner rim of the outer disk. As a side note, the one CO line clearly detected with PACS ($144.78~\mu$m) originates from a similar radial region: $4\!-\!5$~AU.

\begin{table*}[!bhtp]
\caption{Overview of the model series and respective parameters touched. 100xGG denotes the model in which the grain surface area has been scaled by a factor 100 in the chemical rates. [M/H] denotes the metallicity, where `low' means the low ISM metal abundances from \citet{Jenkins2009}. Model~14 is the `standard' model with a 100 times higher gas mass in the disk inside 4~AU.}
\begin{tabular}{l|c|c|c|c|c|c|c|c|c}
\hline
\hline
model name                 & model   & O\#, & surface   & self-      & $T_g\!=\!T_d$ & $N_{\rm Lay}$ & X-rays & C/O & [M/H]\\
                           & number  & OH\#  & reactions & shielding  &        &     &   &   &     \\
\hline
standard                                  & 1  &           &           &   \ding{51}    &         &   2  &   & 0.54  &    low    \\
standard-Xrays                       & 2  &           &           &   \ding{51}    &        &   2  &  \ding{51}  &  0.54  &    low     \\
standard-lowmetal1000      & 3  &           &           &   \ding{51}    &         &   2  &   &  0.54  &    low/1000    \\
standard-CoverOhigh          & 4  &           &           &   \ding{51}    &         &   2  &   &  1.86 &    low   \\
standard-withice-Tg=Td       & 5  & \ding{51} &           &   \ding{51}    & \ding{51}   &   2  &   & 0.54   & low   \\
standard-withice-withoutSS & 6  & \ding{51} &           &                &          & 2    &   &  0.54  &   low  \\[0mm]
\hline
wateronsurface                       & 7  & \ding{51} & \ding{51} &   \ding{51}    &          & 2    &   &  0.54  &    low  \\
wateronsurface-Xrays           & 8  & \ding{51} & \ding{51} &   \ding{51}    &        &  2   &  \ding{51}  & 0.54   &    low   \\
wateronsurface-withoutadsorption \tablefootmark{a}      & 9  & \ding{51} & \ding{51}  &   \ding{51}    &     & 2    &   & 0.54   &  low  \\
wateronsurface-100xGG      & 10  & \ding{51} & \ding{51} $\times 100$ &   \ding{51}     &     &  2 &   & 0.54    &      low    \\
wateronsurface-yields          & 11  & \ding{51} & \ding{51} &   \ding{51}    &       &  2   &   &  0.54 &     low  \\
wateronsurface-nlayer10    & 12  & \ding{51} & \ding{51} &   \ding{51}    &           & 10   &   & 0.54  &  low \\
wateronsurface-nowaterphotodes. & 13  & \ding{51} & \ding{51} &   \ding{51}    &          & 2    &   &  0.54  &    low  \\[0mm]
\hline
standard-twozone-delta0.01              & 14  &           &           &   \ding{51}    &         &   2  &   & 0.54  &    low    \\
\hline
\end{tabular}
\label{Tab:modelseries}
\end{table*}

The mass of water in the inner disk ($<\!4$~AU) is $1.4\!\times\!10^{-14}$~M$_\odot$ and its average gas temperature\footnote{The average gas temperature for water is defined as $\langle T_{\rm g}({\rm H_2O}) \rangle \!=\! \frac{m({\rm H_2O})}{M({\rm H_2O})} \int n({\rm H_2O})(r,z) T_g(r,z) dV$, with $n$ being the density and $m$ the mass of water. $M({\rm H_2O})$ is the total mass of water in the disk model and $dV, r, z$ are the volume element and the radial and vertical coordinate.} is 670~K. The outer gas-phase water reservoir ($4\!-\!196$~AU) contains $2.2\!\times\!10^{-11}$~M$_\odot$, which is at temperatures below 20~K. Water ice in the outer disk ($4\!-\!196$~AU) adds up to a mass of $7.7\!\times\!10^{-6}$~M$_\odot$. 


\begin{figure*}[!htbp]
\hspace*{-7mm}\includegraphics[width=9.8cm]{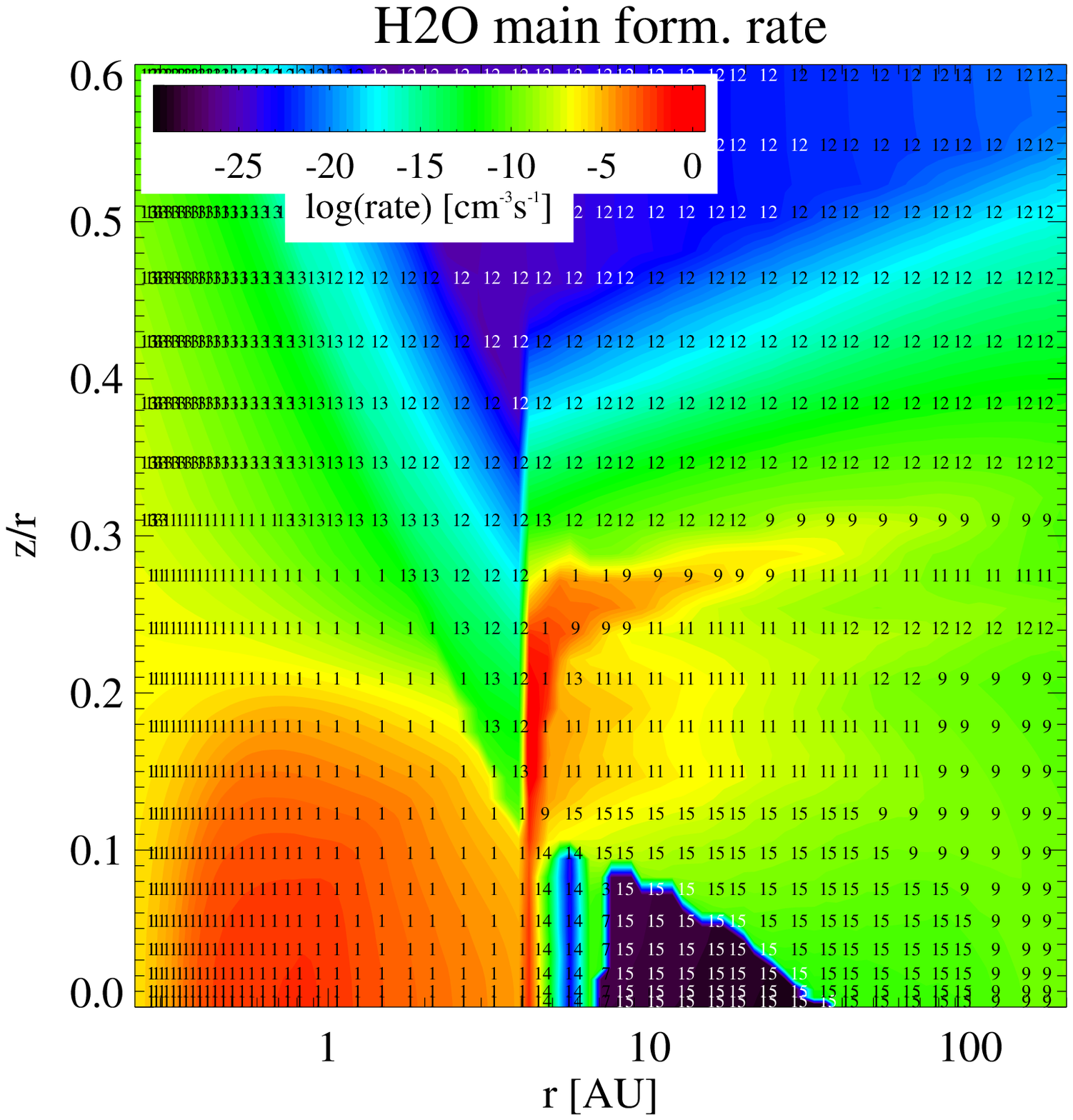}
\hspace*{-7mm}\includegraphics[width=9.8cm]{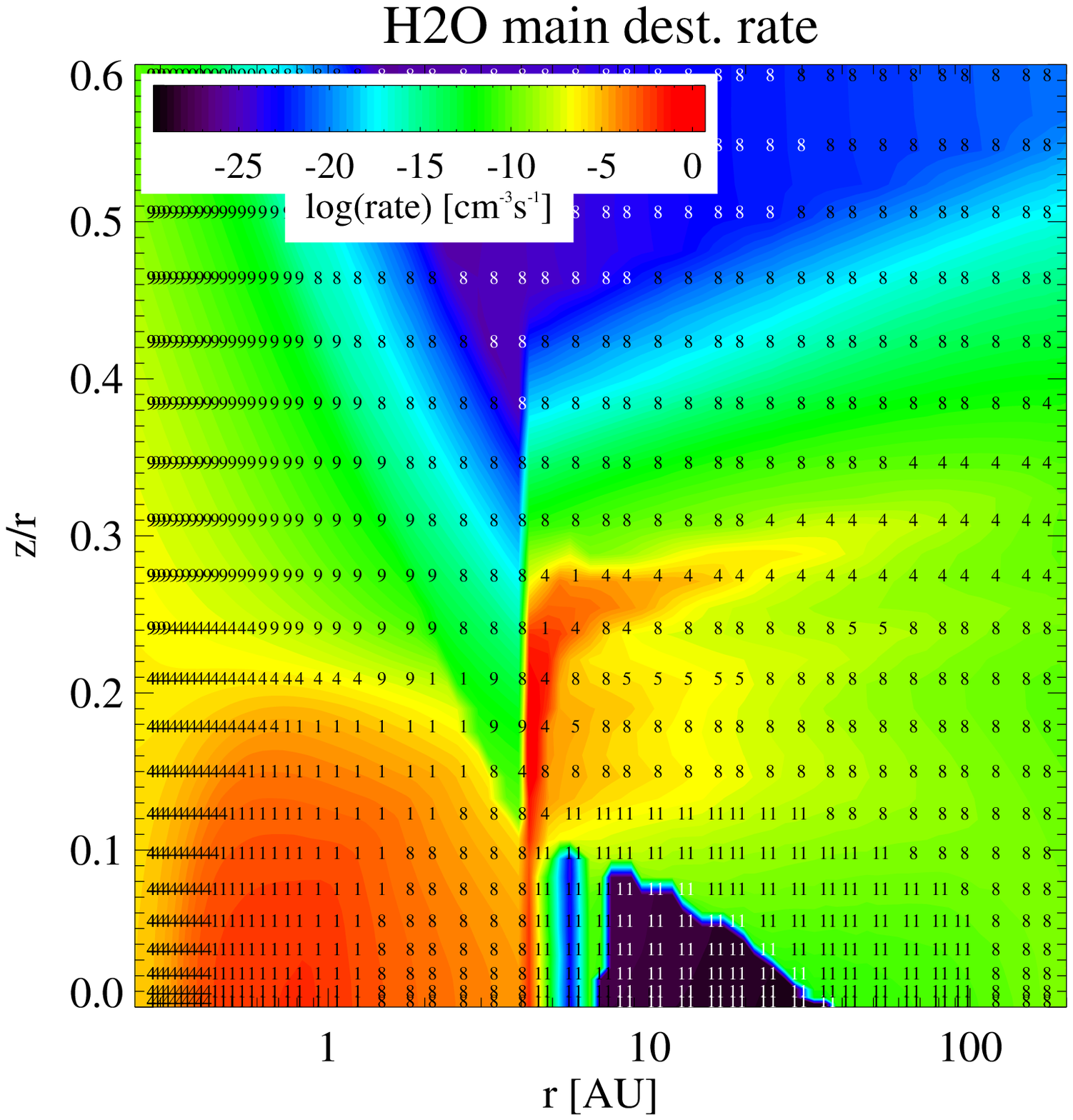}
\hspace*{-7mm}\includegraphics[width=9.8cm]{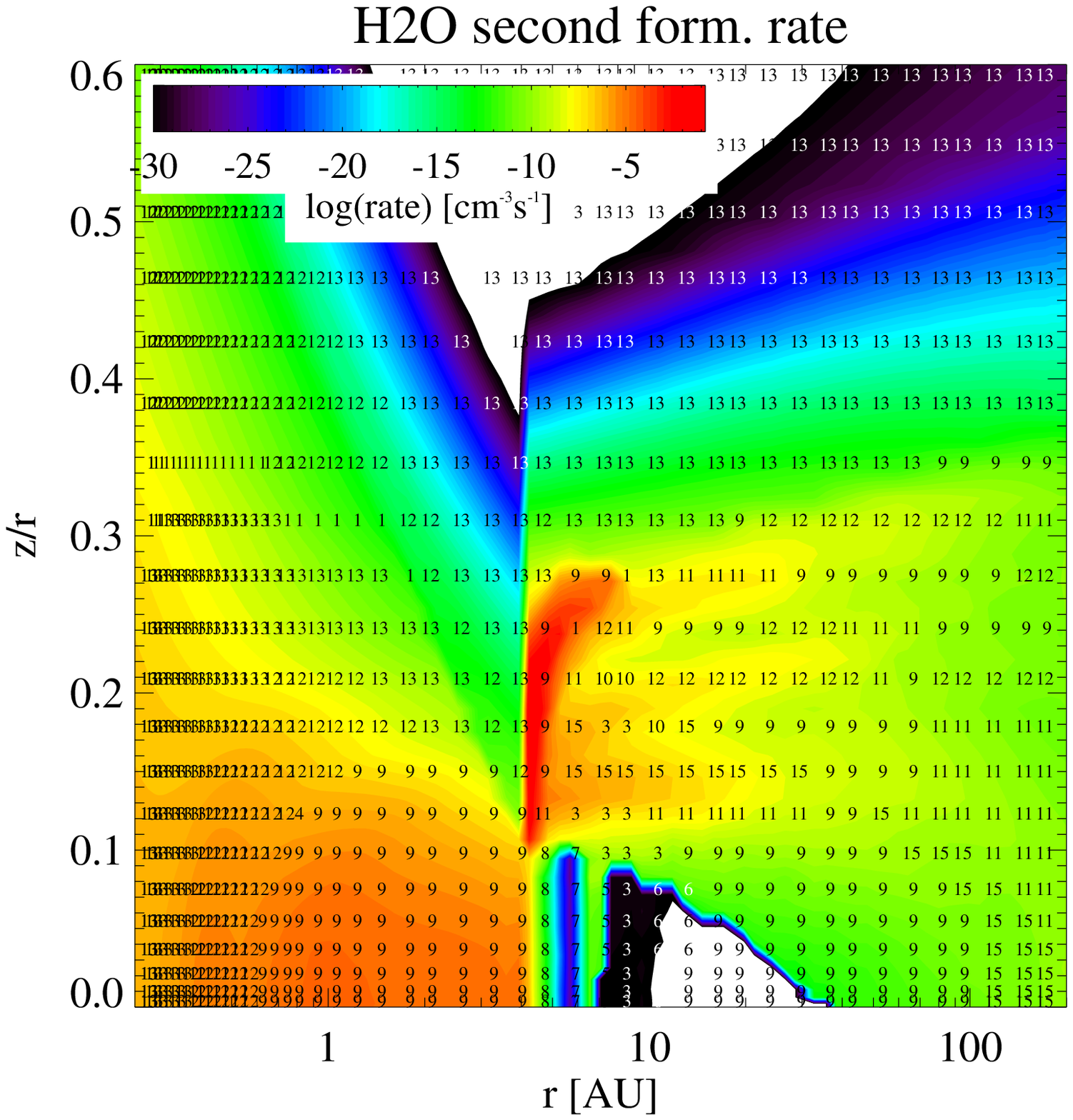}
\hspace*{-7mm}\includegraphics[width=9.8cm]{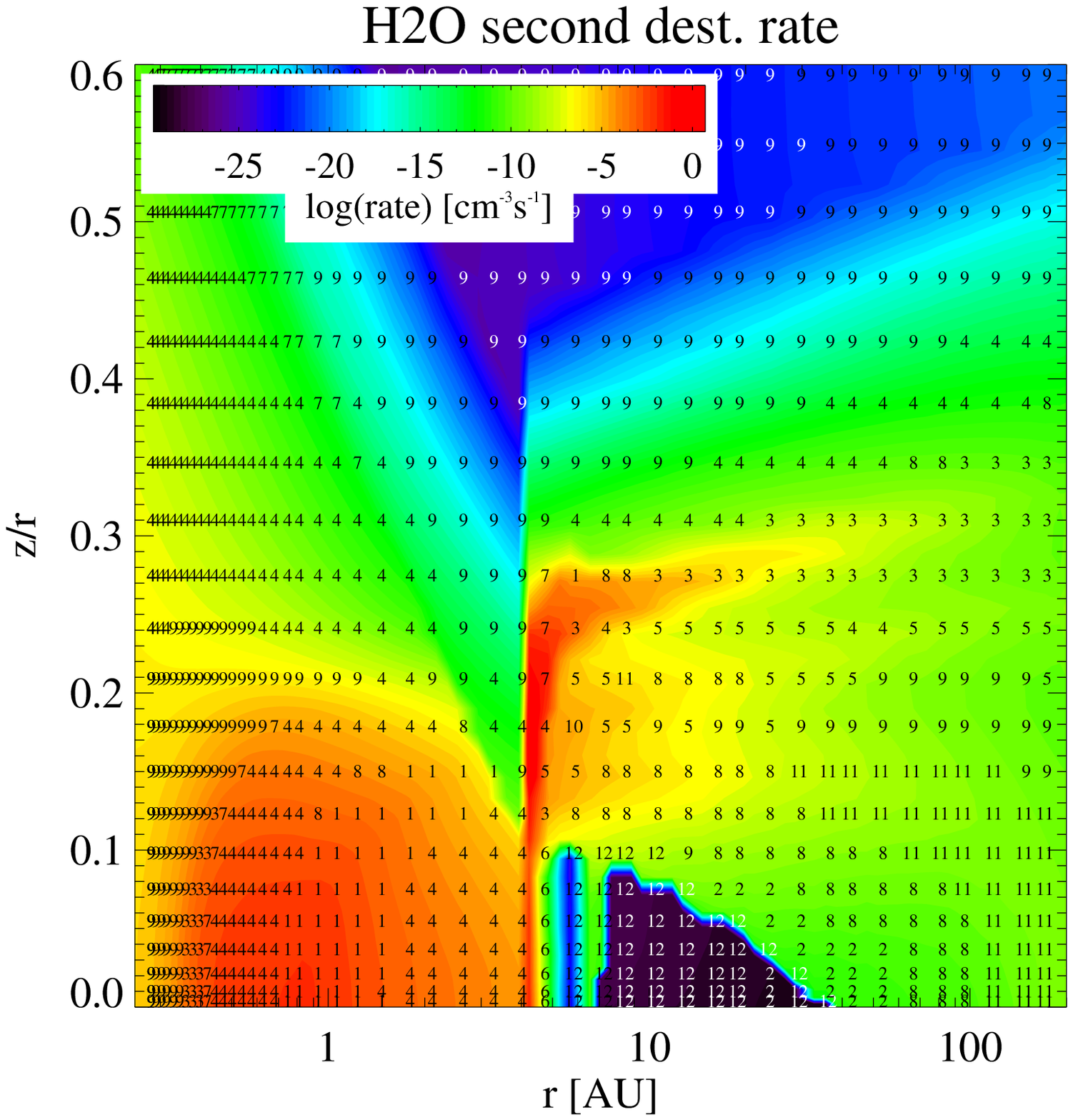}
\vspace*{-4mm}
\caption{The primary and secondary formation and destruction rates for gas phase water in the `standard' disk model (grid resolution 100x70). The numbers listed for every second grid point refer to the reactions listed in Table~\ref{tab:chemical reactions-water}.}
\label{fig:H2Ochem-at32AU}
\end{figure*}

\subsection{Water chemistry in the disk}

We analyse in some more detail the water chemistry in the surface layers of the outer disk at $0.01\!<\!A_{\rm V}\!<\!3$ to understand the role of gas phase and surface chemistry and water photodesorption in particular.

The main water desorption channel in the outer disk is photodesorption. The reason for this are the cool temperatures, below 100~K for gas and dust, in the outer disk of TW\,Hya. At such low temperatures, thermal desorption is negligible and water abundances reach levels of $\sim\!10^{-8}$. Fig.~\ref{fig:H2Ochem-at32AU} summarizes the two main formation/destruction channels for water vapour in the disk. Interestingly, gas phase formation of water dominates the chemical balance at $A_{\rm V}\!\lesssim\!1$. Only at higher extinction ($A_{\rm V}\sim\!3$), the detailed balance is between water photodesorption from and adsorption on grains (formation reaction 15, destruction reaction 11). Water destruction in the upper atmosphere is not only through UV photons, but also through collisions with C$^+$, Si$^+$ and H$^+$. One of the most important gas phase formation routes is radiative association through OH\,$+$\,H. This channel is efficient due to the relative high abundance of H (H/H$_2$ of the order of $10^{-3 \ldots 4}$) in layers where the OH abundance peaks. Gas temperatures here are $\sim\!40$~K, high enough for some atoms/molecules to overcome the activation barrier of this reaction (90~K).

Inside 2~AU, efficient warm gas phase chemistry ($T_{\rm gas}\!>\!250$~K) can increase the water gas phase abundances up to levels of $10^{-6}$ for higher gas disk masses (gas-to-dust ratio of 100) \citep[see][for a more detailed discussion]{Thi2010b}. The main channel is here H$_2$\,$+$\,OH. A similar conclusion on the warm neutral chemistry leading to water formation was reached by \citet{Glassgold2009} and \citet{Najita2011}.

\begin{table}[htb]
\caption{Formation and destruction reactions for water. The numbers in the first column correspond to the main formation/destruction channels indicated in Fig.~\ref{fig:H2Ochem-at32AU}.}
\begin{tabular}{llll}
\hline
No. & \multicolumn{3}{c}{formation reaction} \\
\hline
1   & H$_2$ + OH & $\rightarrow$ & H$_2$O + H \\
3   & NH$_2$ + NO & $\rightarrow$ & N$_2$ + H$_2$O\\
6   & H$_3^+$ + CH$_3$OH & $\rightarrow$ & CH$_3^+$ + H$_2$O \\
7   &  H$_3$O$^+$ + HCN & $\rightarrow$ & HCNH$^+$ + H$_2$O \\
9   &  H$_3$O$^+$ + e$^-$ & $\rightarrow$ & H$_2$O + H \\
11 & H$_2$ + O$^-$ & $\rightarrow$ & H$_2$O + e$^-$\\
12 &  H + OH & $\rightarrow$ & H$_2$O + h$\nu$\\
13 &  H$_2^*$ + OH & $\rightarrow$ & H$_2$O + H\\
14 &  H$_2$O\# & $\rightarrow$ & H$_2$O (thermal desorption)\\
15 &  H$_2$O\# & $\rightarrow$ & H$_2$O (photodesorption)\\
\hline
No. & \multicolumn{3}{c}{destruction reaction} \\
\hline
1    & H + H$_2$O & $\rightarrow$ & OH + H$_2$\\
2    &  H$_3^+$ + H$_2$O & $\rightarrow$ & H$_3$O$^+$ H$_2$\\
3    &  C$^+$ + H$_2$O & $\rightarrow$ & HCO$^+$ + H\\
4    &  C$^+$ + H$_2$O & $\rightarrow$ &HOC$^+$ + H \\ 
5    &  Si$^+$ + H$_2$O & $\rightarrow$ & SiOH$^+$ + H\\
7    & H$^+$ + H$_2$O & $\rightarrow$ & H$_2$O$^+$ + H\\
8    &  H$_2$O + h$\nu$ & $\rightarrow$ & OH + H\\
9    &  H$_2$O + h$\nu$ & $\rightarrow$ & H$_2$O$^+$ + e$^-$\\
11  &  H$_2$O + dust & $\rightarrow$ & H$_2$O\# (adsorption)\\
12  &  CH + H$_2$O & $\rightarrow$ & CH$_2$OH \\
\hline
\end{tabular}
\label{tab:chemical reactions-water}
\end{table}

Fig.~\ref{fig:H2Ochem-at32AU} shows a small region close to the inner rim where water is formed through NH$_2$ and/or OH. Formation of OH involves the reaction NH\,$+$\,O with disputed rate coefficients (see Sect.~\ref{sect:standardmodel}). Removing all nitrogen chemistry from the `standard' model has a minor impact on the water abundance structure, confirming that other channels such as water photodissociation contribute at roughly equal efficiency to the formation of OH. 

A comparison of the OH (Fig.~\ref{fig:standard-basic}) and the water abundance distribution in the standard model, shows that the gas-phase formation of water in the upper disk layers is a process that scales strongly with density. A first step is the formation of OH and the first peak of the OH abundance is above $A_{\rm V}\!=\!0.01$. Water only starts to become abundant below the OH layer. The OH abundance shows two distinct layers. In the first one around $A_{\rm V}\!=\!0.01$, dissociative recombination of ion molecules such as H$_2$O$^+$ plays a large role in OH formation. The second layer of higher OH abundance originates besides nitrogen chemistry from water photodissociation (see also Fig~\ref{fig:H2Ochem-at32AU}). The low abundance layer of OH, NO, NH around the layer $r\!=\!10\!-\!100$~AU, $z\!=\!2.4\!-\!28$~AU coincides with $\chi/n_{\langle H \rangle}\!\sim\!10^{-5}$ (The quantity $\chi/n_{\langle H \rangle}$ is the classical PDR parameter, the ratio between the strength of the FUV radiation field, $912\!-\!2050$~\AA, and the total hydrogen number density).

\begin{figure}[!tbhp]
\begin{center}
\includegraphics[width=8cm]{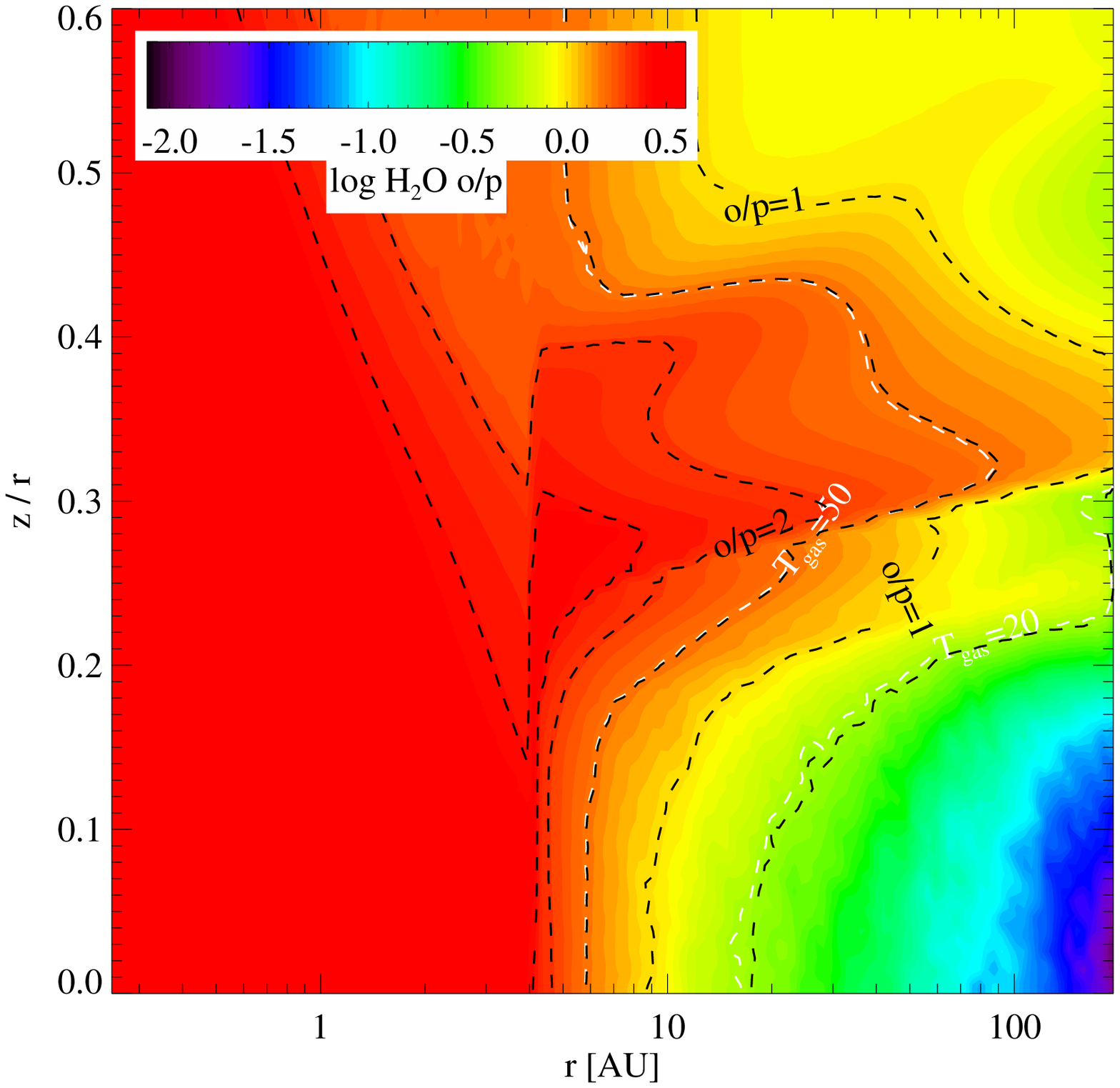} 
\vspace*{-5mm}
\end{center}
\caption{Water ortho/para ratio in the standard model. Black contours show o/p ratios of 0.5, 1, 1.5, 2, 2.5 and 3. White contours denote gas temperatures of 20 and 50~K.}
\label{fig:ortho-para-water}
\end{figure}

\subsection{The water ortho/para ratio}

ProDiMo calculates the LTE ortho/para ratio (o/p ratio) using the gas temperature found from the gas thermal equilibrium. Fig.~\ref{fig:ortho-para-water} shows the o/p ratio of water in the `standard' disk model. 

There is a strong vertical $T_{\rm gas}$ gradient above the radial extinction $A_{\rm V,rad}\!=\!1$ layer (Fig.~\ref{fig:standard-basic}); here the o/p water ratio changes quickly reaching the maximum value of 3 at the inner rim of the outer disk ($4\!-\!8$~AU). In general, the radial gradient of the o/p ratio reflects the radial gas temperature gradient. Where the gas temperature reaches values below 20~K in the outer disk, the o/p ratio drops below 1.

\section{A study of underlying modeling uncertainties}

\begin{figure*}[!bhtp]
\includegraphics[width=18cm]{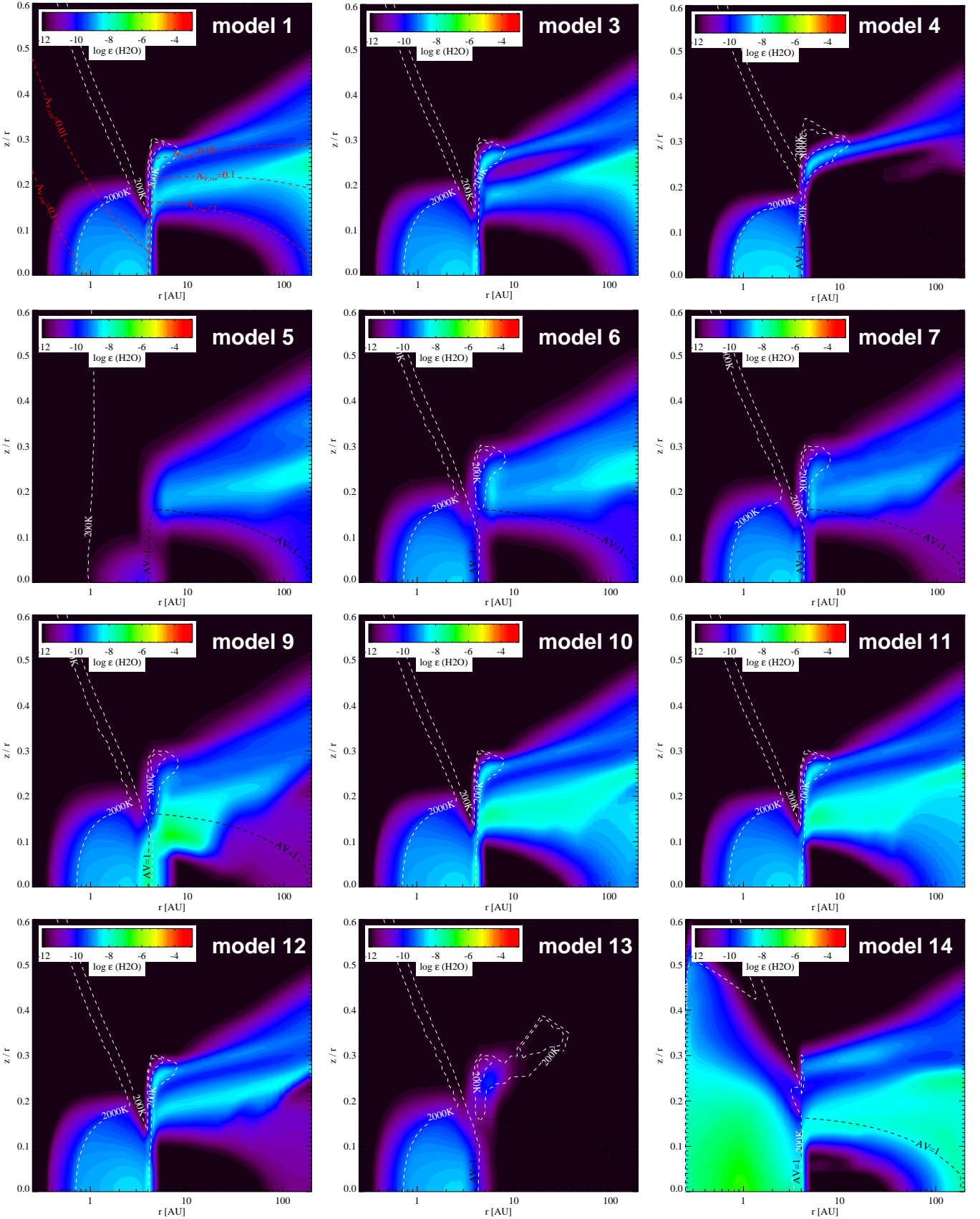}
\vspace*{-2mm}
\caption{Water gas phase abundance in the TW Hya disk model (from top left to bottom right): Standard model (model 1), with 1/1000 of the low ISM metallicity (model 3), C/O ratio of 1.86 (model 4), 
$T_{\rm gas}\!=\!T_{\rm dust}$ (model 5), without water self shielding (model~6), water on surfaces (model~7),  without water adsorption (model~9), with 100 times higher surface reaction rates (model~10), with two different desorption channels (model~11), with 10 active grain surface layers instead of 2 (model~12), without water photodesorption (the outer water reservoir is gone, model~13), and the standard model with a 100 times higher gas mass in the disk inside 4~AU (model~14). White dashed contours: gas temperatures of 200 and 2000~K. Red dashed contours: vertical $A_{\rm V}$. Black dashed contours: minimum of radial and vertical $A_{\rm V}$.}
\label{fig:waterabus}
\end{figure*}

\begin{figure*}[!htbp]
\ContinuedFloat
\centering
\includegraphics[width=12cm]{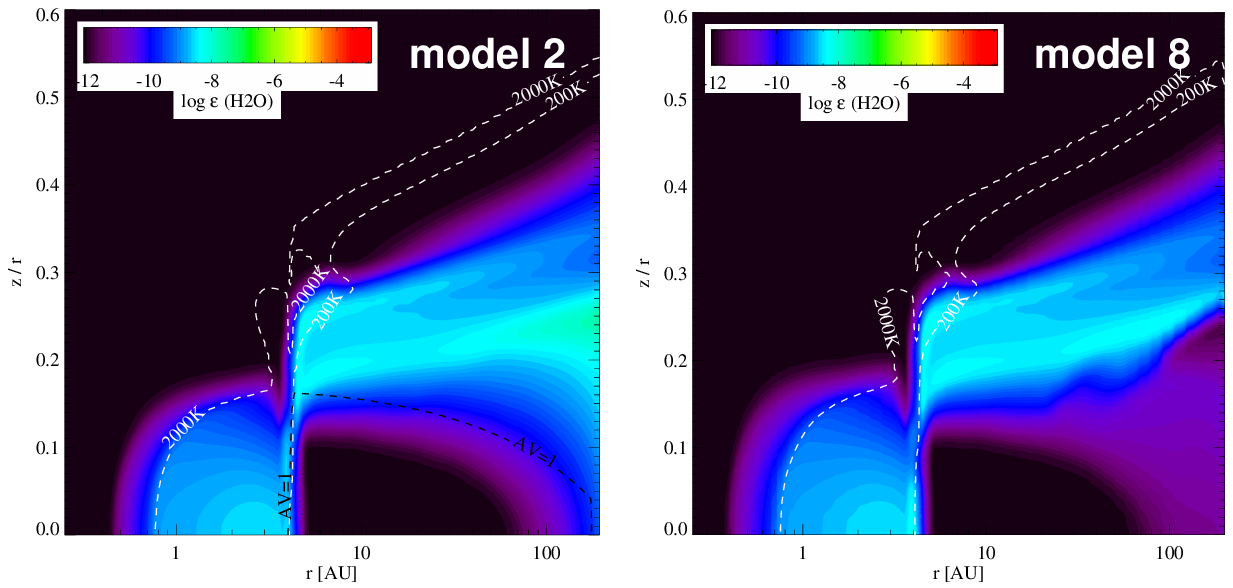}
\caption{Continued. Water gas phase abundance in the TW Hya disk model. Left: standard model with X-rays (model 2). Right: water on surfaces with X-rays (model~8). White dashed contours: gas temperatures of 200 and 2000~K. Black dashed contours: minimum of radial and vertical $A_{\rm V}$.}
\end{figure*}

A disk model is defined by a set of input parameters, but also by a set of processes (chemical and physical) such as photodesorption and surface chemistry and a set of numerical recipes such as the radiative transfer method. The following sections contain `numerical experiments' to understand underlying modeling uncertainties due to the implementation of processes and radiative transfer methods. The goal is to assess the diagnostic power of individual water line observations and to set the stage for physics beyond the homogeneously mixed stationary disk models, such as invoking inhomogeneous dust properties (e.g. dust settling, radial gradient in dust properties), radial and vertical mixing of gas, and non-equilibrium chemistry. During these `experiments', the main parameters leading to the density structure, dust temperatures/opacities and stellar/interstellar radiation are kept unchanged. However, chemical abundances, gas temperatures, species level populations ($1\!+\!1$D escape probability) and line radiative transfer are intrinsically coupled, thereby changing the emergent fluxes in a complex way.
The `experiments' carried out include the C/O ratio, water self-shielding, X-rays, photodesorption and surface chemistry. The basic model series is summarized in Table~\ref{Tab:modelseries} and the resulting water abundances are shown in Fig.~\ref{fig:waterabus}. The impact of each `experiment' on the resulting water abundance structure is discussed in Sect.~\ref{sect:chemistry}.

For all these models, we ran the line radiative transfer of the series of atomic and molecular lines described in Sect.~\ref{Sect:obs} based on the level populations obtained from a $1\!+\!1$D escape probability \citep{Woitke2009a}. The inclination for the TW Hya disk model is fixed to $7^{\rm o}$. Tables~\ref{tab:PACSmodfluxes1} and \ref{tab:PACSmodfluxes2} provide an overview of the resulting fluxes in comparison with the observed values and a detailed discussion is presented in Sect.~\ref{sect:emergentfluxes}. 


\subsection{Chemistry}
\label{sect:chemistry}

The following paragraphs focus on a sensitivity analysis for the water abundance structure within the model series described in Table~\ref{Tab:modelseries}.

\subsubsection{X-rays (model~2 and 8)}
\label{subsect:X-rays}

Including X-rays increases the gas temperature in the disk surface at heights above $A_{\rm V,rad}\!>\!1$ due to Coulomb heating. X-rays slightly enhance the abundance of H$_3^+$ in the disk surface, resulting in a vertically more extended HCO$^+$ layer. That same layer now also hosts H$_2$O$^+$ and H$_3$O$^+$, albeit at low abundance levels ($10^{-9}\!-\!10^{-10}$). The abundances of the other species, atomic oxygen, ionized carbon, CH$^+$, CO and HCN are not significantly affected. The same behaviour of these species with respect to X-rays has been noted by \citet{Najita2011}. Due to the higher abundances of H$_2$O$^+$ and H$_3$O$^+$, water formation in the gas phase becomes more efficient and the water vapor layer becomes thicker (i.e.\ vertically more extended, Fig.~\ref{fig:waterabus} model~2). However, the total amount of water vapor and also the hot water reservoir changes by less than 10\%.

\subsubsection{C/O ratio (model~4)}
\label{subsect:CoverO}

The C/O ratio is crucial in determining whether carbon or oxygen-rich chemistry develops. If the limiting abundance is that of oxygen, the formation of CO binds the entire oxygen reservoir hence leaving the remaining C atoms to form CH chains and nitrogen bearing molecules molecules such as HCN. If a surplus of oxygen exists, CO formation will consume most of the carbon and water can still form at high abundances.

There is limited evidence for how the C/O ratio evolves during the star and planet formation phase. The C/O ratio in the Sun is 0.4 \citep{Anders1989}, so very oxygen rich. We have also evidence for the C/O ratio in the precursor dense ISM material to be smaller than one, ${\rm C/O}\!=\!0.54$ \citep[e.g.][]{Jenkins2009}. In a mature planetary system, our calibration point are cometary compositions in the Solar System \citep[e.g.][]{Bockelee-Morvan2010}, ${\rm C/O}\!=\!0.45$, which are representing the ice phase. To which extent the gas-phase C/O ratio is constant throughout the protoplanetary disks and how it may change as the disk evolves is unclear. Large scale mixing processes, dust settling, the formation of ices all can affect locally as well as globally the gas-phase C/O ratio in a disk \citep[e.g.][]{Ciesla2006,Hogerheijde2011}.

In model~4, we assume an elemental C/O ratio larger than one, i.e.\ ${\rm C/O}\!=\!1.86$ ($12\!+\!\log \epsilon({\rm O})\!=\!7.98$, $12\!+\!\log \epsilon({\rm C})\!=\!8.25$). As a result, the water abundance in the disk is suppressed (compare model 1 and 4 in Fig.~\ref{fig:waterabus}), especially in deeper surface layers. This is largely a chemical effect because the temperature structure of the disk does not change significantly. In a carbon rich environment, CO takes up most of the oxygen until freeze-out temperatures of CO are reached around 20~K. There is a substantially smaller and lower mass water ice reservoir, $M({\rm H_2O\#})\!=\!3.7\!\times\!10^{-7}$~M$_\odot$, a factor 20 less than in the `standard' model.


\subsubsection{$T_{\rm gas}\!=\!T_{\rm dust}$ (model~5)}

Model~5 contains O and OH ice as species, but now with the assumption that the gas temperature equals the dust temperature throughout the disk model. The disk inside a few AU becomes then too cold for an efficient gas phase formation of water (model~5 in Fig.~\ref{fig:waterabus}, where the inner water reservoir is absent). Even the outer disk gas phase water formation via ion-molecule chemistry is less efficient.

\subsubsection{No water self-shielding (model~6)}

We study the impact of the self-shielding on the water abundance by switching the process off (model~6). This should maximize the ionization/dissociation of atoms/molecules for all species except CO, and H$_2$, where the self-shielding is taken into account via extensive tables of shielding coefficients. Also for C, the self-shielding is calculated directly from the photoionization cross section provided in the code.

For the `standard' TW Hya disk model, the impact of water self-shielding is rather small (Fig.~\ref{fig:waterabus}, model~6) compared to what is seen by \citet{Bethell2009} for a disk model with strong dust settling. In the absence of grain settling, continuum absorption by dust dominates the shielding of the photodissociation process.


\subsubsection{Water formation on grain surfaces (model~7-10)}

\begin{figure*}[!htp]
\includegraphics[width=6cm]{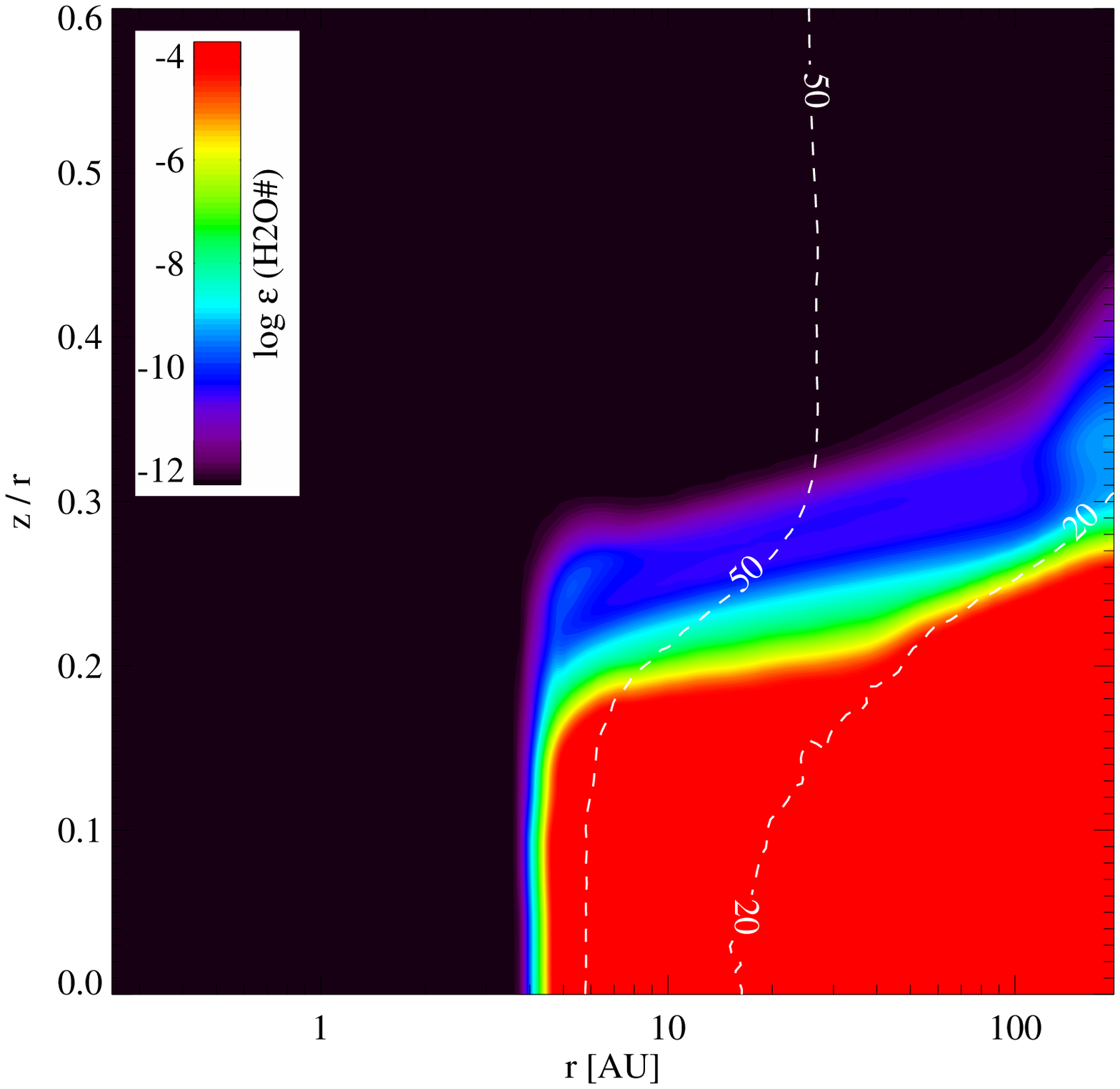}
\includegraphics[width=6cm]{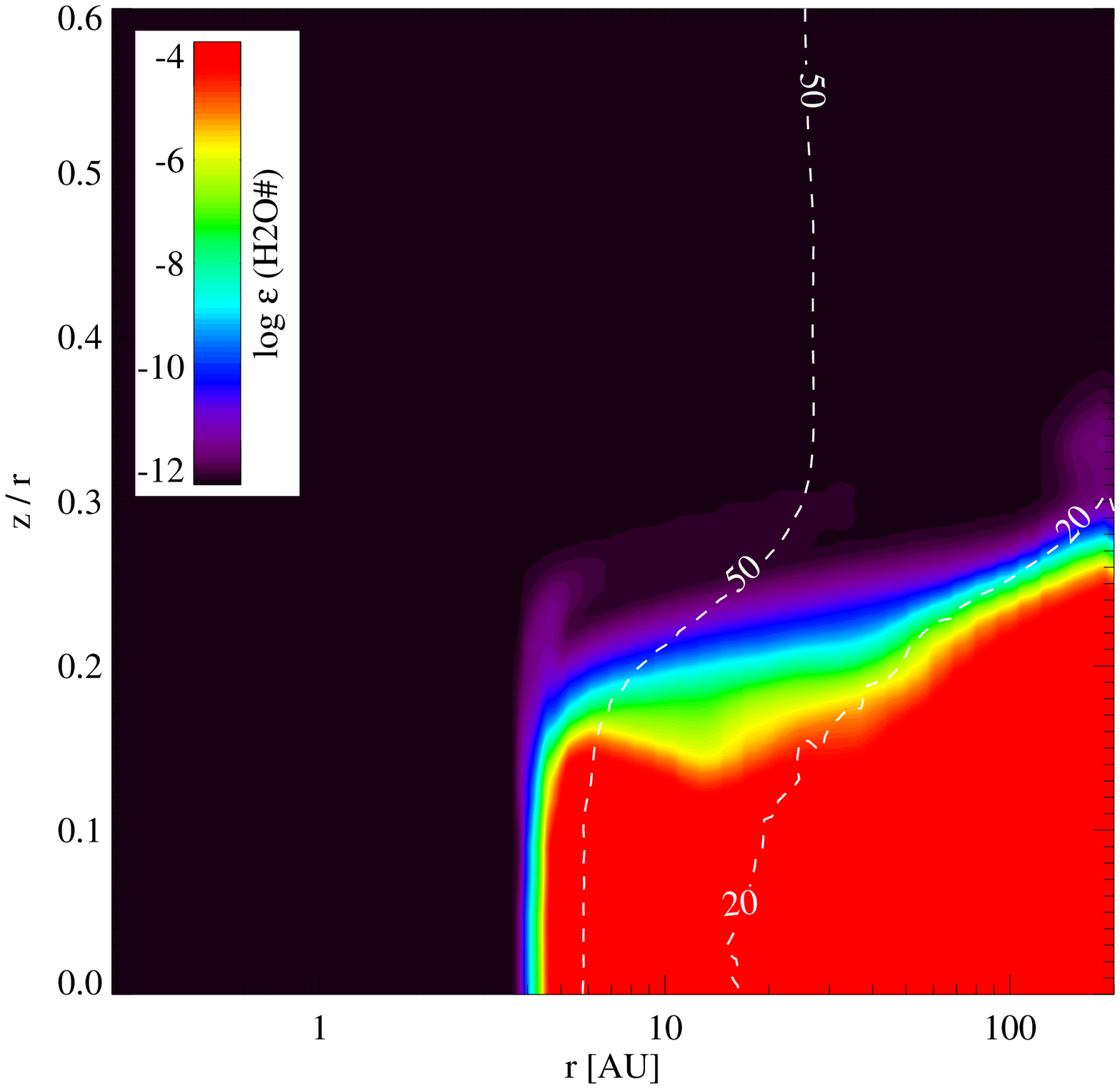}
\includegraphics[width=6cm]{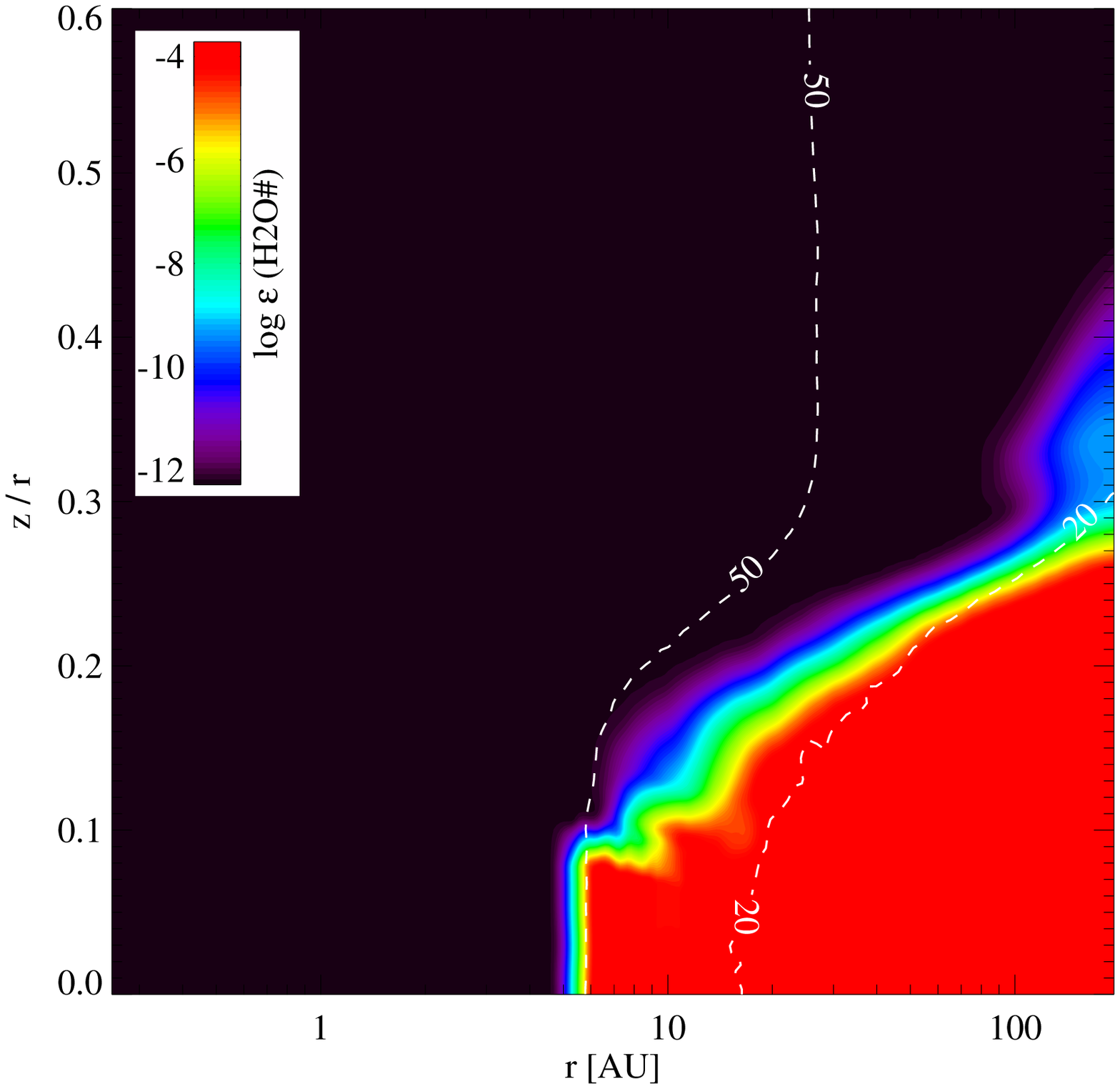}
\caption{Water ice abundance with surface reactions (left, model~7), with changed water desorption processes (middle, model~11) and without water adsorption from the gas phase (model~9) --- all water ice being formed by surface reactions (right). The white contours denote dust temperatures of 20 and 50~K.}
\label{fig:H2Oice}
\end{figure*}


If we include O and OH ice as species and also surface reactions, these ices are further processed on the grain into water ice (see Appendix~\ref{app:watersurf}). The water ice abundance for model~7 is shown in Fig.~\ref{fig:H2Oice} (left panel). The efficiency of water formation on the grain surface itself has been tested by switching off the water adsorption reaction (model~9); then water ice can only form through reactions of O and OH ice on the grain surface (see right panel of Fig.~\ref{fig:H2Oice}).

In the presence of water formation on surfaces, the water vapour abundance at 150~AU close to the midplane ($z/r\!<\!0.2$) shows a much steeper decline from $10^{-8}$ to $10^{-11}$ compared to the model without water formation on grain surfaces. In addition, the model with grain surface chemistry shows much lower levels of atomic oxygen below $z/r < 0.2$. Oxygen freezes out at these temperatures and --- besides water ice being the main reservoir --- ends up preferentially at the same abundance level in OH ice and in gas phase CO.

\citet{Fogel2011} assume a continuous power law grain size distribution with an exponent of $-3.5$ with $a_{\rm min}\!=\!0.005$ and $a_{\rm max,small}\!=\!0.25~\mu$m and a single population of large 1~mm grains. In contrast to that, our disk model contains a continuous distribution of grains between $0.03$ and 10~cm with a power law exponent of $-3.4$. Hence our model has a largely reduced grain surface area available for surface reactions compared to the \citet{Fogel2011} model. To assess the importance of the available grain surface, we artificially scale the rate coefficient for grain surface reactions by a factor 100 (model~10). The abundance distribution of water vapor in the outer disk increases (Fig.~\ref{fig:waterabus}). The larger grain surface area increases the photodesorption rates of ices back into the gas phase and thus provides a source of oxygen for the gas phase water formation.

\subsubsection{Water desorption (model~11-13)}

\citet{Ceccarelli2005} noted the possibility that water can be desorbed from the cold icy grains in the outer disk by UV and/or X-ray photons. This process has then been studied more quantitatively by \citet{Dominik2005}, using experimental water photodesorption yields from \citet{Westley1995} and the binding energy measured by \citet{Fraser2001} ($E_{\rm ads}\!=\!5600$~K). They find typical water vapor abundances of $\sim\!10^{-7}$ in the disk surface.

\citet{Oeberg2009} have recently measured again the photodesorption yields in the lab resulting in a total yield (OH and H$_2$O) of $Y\!=\!1.94\!\times\!10^{-3}$~photon$^{-1}$ at 20~K. This is close to the previous experiments by \citet{Westley1995} and also within a factor four of the theoretically calculated ones \citep{Arasa2010}. We assume here the photodesorption yields and adsorption energy used by \citet{Hollenbach2009} for comparison reasons: $Y({\rm H_2O})\!=\!10^{-3}$ and $Y({\rm OH})\!=\!2\!\times\!10^{-3}$, $E_{\rm ads}\!=\!4800$~K \citep{Aikawa1996}. The exact value of the photodesorption yield in the range $10^{-3}\!-\!10^{-5}$ was found to be less important \citep{Semenov2011}.

\citet{Andersson2008} showed that roughly two third of the water ice is actually dissociated by the incoming UV photon and desorbs as OH and H. Using this additional OH photodesorption channel, the outer cold water reservoir does not change significantly (Fig.~\ref{fig:waterabus}, model~11). Changing the adsorption energy to the higher value of $5600$~K has a minor effect on the inner water vapor reservoir, but enhances the mass of the outer cold water vapor reservoir from $8.7\!\times\!10^{-7}$ to $9.0\!\times\!10^{-7}$~M$_{\rm Earth}$. 

If the number of active layers for photoevaporation from ices is increased from $N_{\rm lay}\!=\!2$ to 10, the change in the gas phase water abundances is negligible (Fig.~\ref{fig:waterabus}, model~12).

The entire outer water reservoir is due to photodesorption of water ice from the grain surfaces. Removing that  process from the chemical network leaves only the inner water reservoir remaining (Fig.~\ref{fig:waterabus}, model~13); in the outer disk, oxygen is not returned back into the gas phase. The `standard' model has a total water reservoir of $7.3\!\times\!10^{-6}$~~M$_{\rm Earth}$ of which less than 1\% resides inside the gap. Removing photodesorption as the main source of cold water in the outer disk reduces the mass of water there by four orders of magnitude. However, reducing the photodesorption yields by only a factor 10 does not affect the water abundance in the outer disk.

%

\subsection{Emergent line fluxes}
\label{sect:emergentfluxes}

For all models described above, we calculate the atomic/molecular line fluxes that are in the PACS range in Table~\ref{tab:PACSmodfluxes1} and \ref{tab:PACSmodfluxes2}. In addition to those lines, we also list the water lines accessible with HIFI and a selection of submm lines of CO, HCO$^+$ and HCN. These lines provide additional constraints on the irradiation environment, the electron density and the disk gas mass. As an example, X-rays are found to  enhance the HCN abundance \citep[e.g.][]{Aikawa2001} and high electron abundances can keep the HCO$^+$ abundance low \citep[e.g.][]{Qi2003}. Also water self-shielding can affect OH abundances, because one of the processes forming OH --- among many others --- is photodissociation of water. The following sections focus on the water line fluxes and some high excitation lines originating at the inner rim. A brief discussion on the fine structure lines and submmm line fluxes can be found in Appendix~\ref{app:linefluxes}.


\subsubsection{Water lines}
\label{subsect:waterlines}

\begin{figure}[bth]
\hspace*{-4mm}{\includegraphics[width=9.5cm]{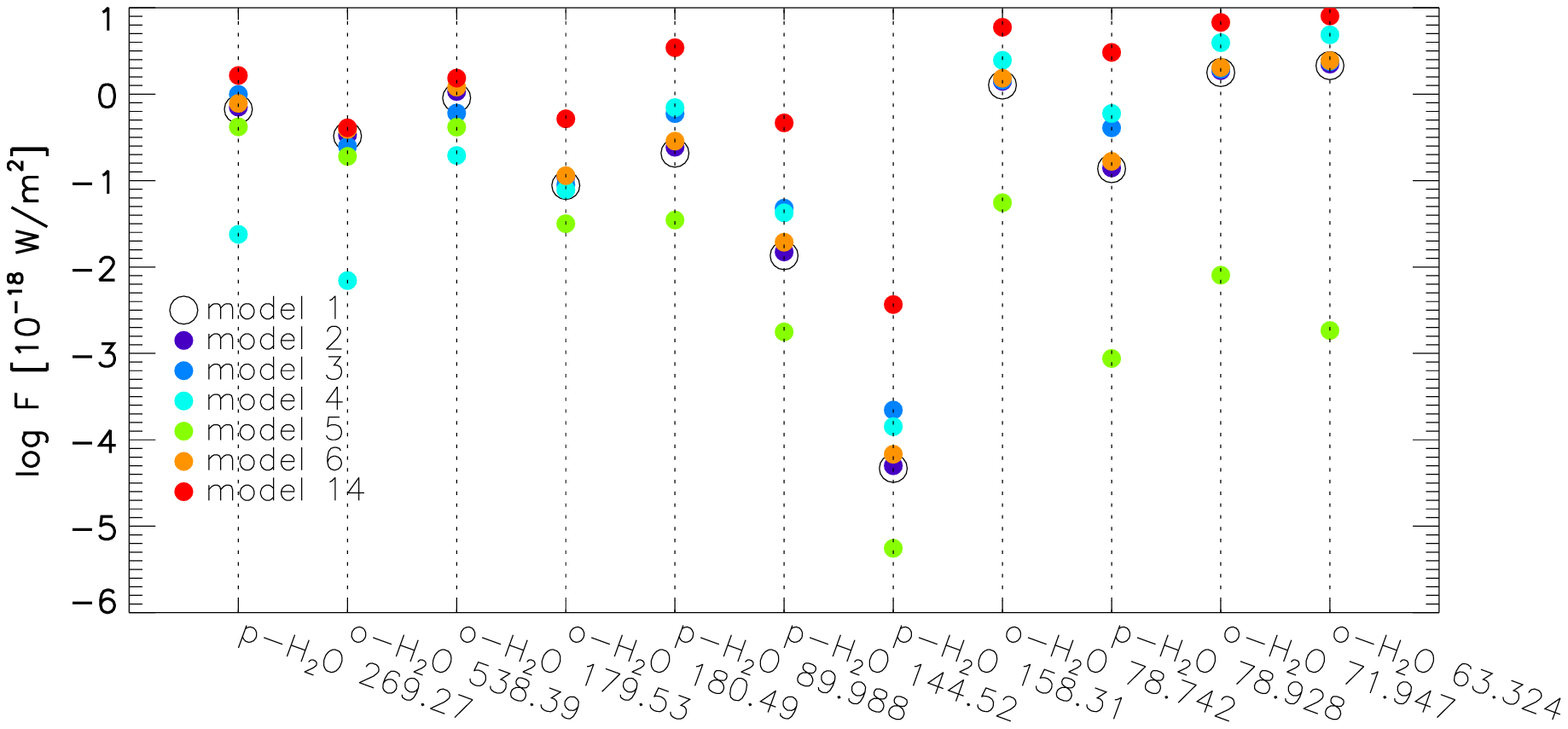}}
\hspace*{-4mm}{\includegraphics[width=9.5cm]{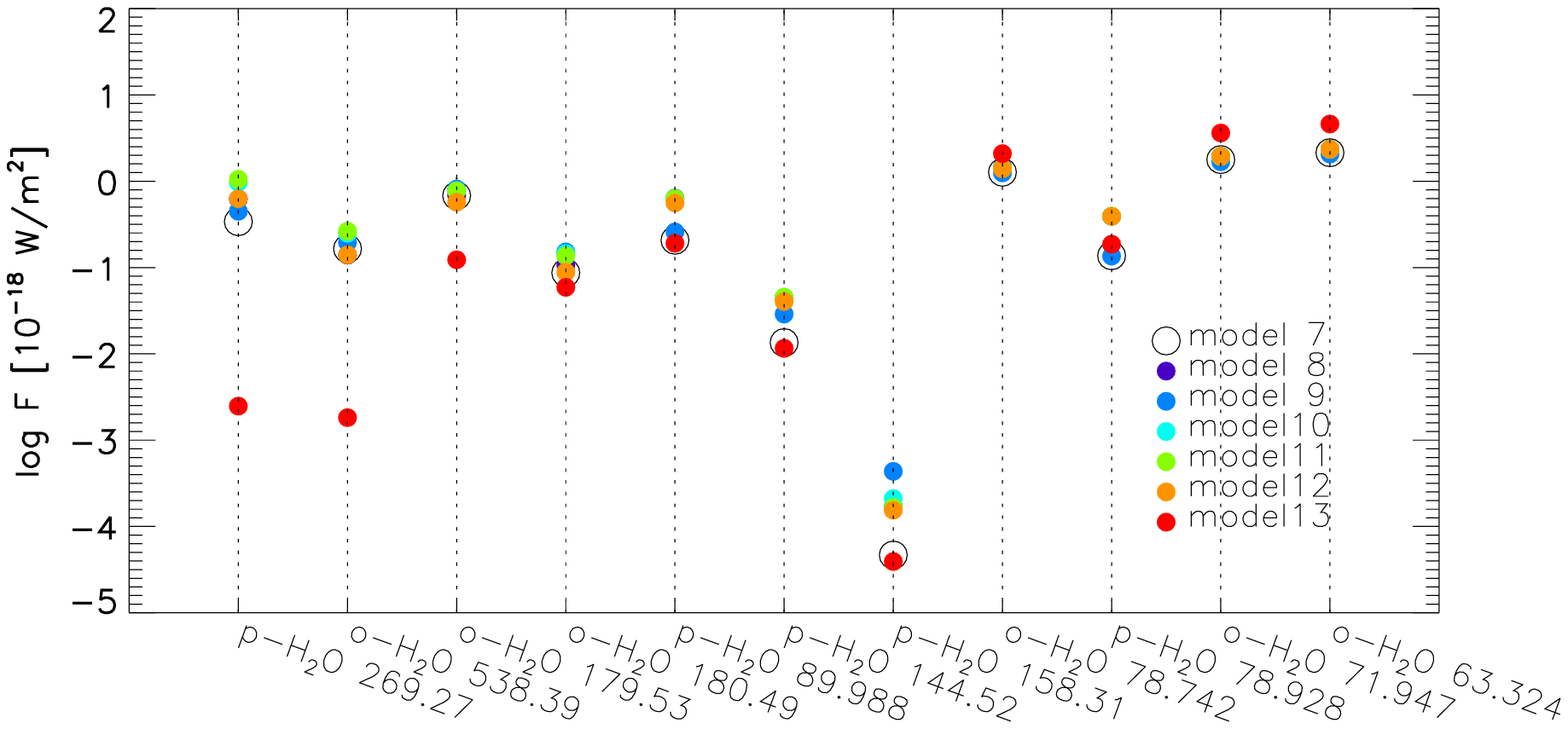}}
\vspace*{-9mm}
\caption{Sensitivity of the water lines in the model series without grain surface chemistry (top) and with grain surface chemistry (bottom).} 
\label{fig:waterlines}
\end{figure}

A comparison of water line fluxes for all models of the series is presented in Fig.~\ref{fig:waterlines}. In most cases, the fluxes change by less than a factor 3 with respect to the standard model of the series (model~1 and model~7). The largest difference in water line fluxes are seen for the model with $T_{\rm gas}\!=\!T_{\rm dust}$ (model 5). Due to the lower gas temperatures, the line formation region for most water lines becomes smaller, except for the HIFI line at $538.29~\mu$m, which originates in both cases between 30 and 200~AU. All lines with a low excitation temperature (both HIFI lines and the PACS $179.53$ and $180.49~\mu$m lines) are less affected. Increasing the C/O ratio to $1.86$ (model 4), affects the HIFI water lines leading to a factor $30\!-\!50$ lower fluxes. 

If grain surface reactions are added (model~1 versus model~7), the fundamental water lines get weaker by a factor two. The higher excitation lines do not change at all since they all originate in much warmer gas where grain surface formation of water plays no role. Including the two channels for water photodesorption (H$_2$O and OH$+$H) does not change the line fluxes of the HIFI water lines by more than 35\%. The PACS lines change typically by 10\%. Changing the adsorption energy to the higher value of $5600$~K, changes the higher excitation water line fluxes by less than 20\%, while it changes the HIFI lines by up to a factor $1.6$.

If the inner disk has a higher gas-to-dust mass ratio of 100 (model~14), the PACS high-excitation water lines change by a factor of a few to 10. While this brings the $89.988~\mu$m line flux into agreement with the observed one, it does not violate any upper limits within $3\sigma$.

\begin{figure}[!htbp]
\vspace*{-5mm}
\hspace*{-0.7cm}
\includegraphics[width=10cm]{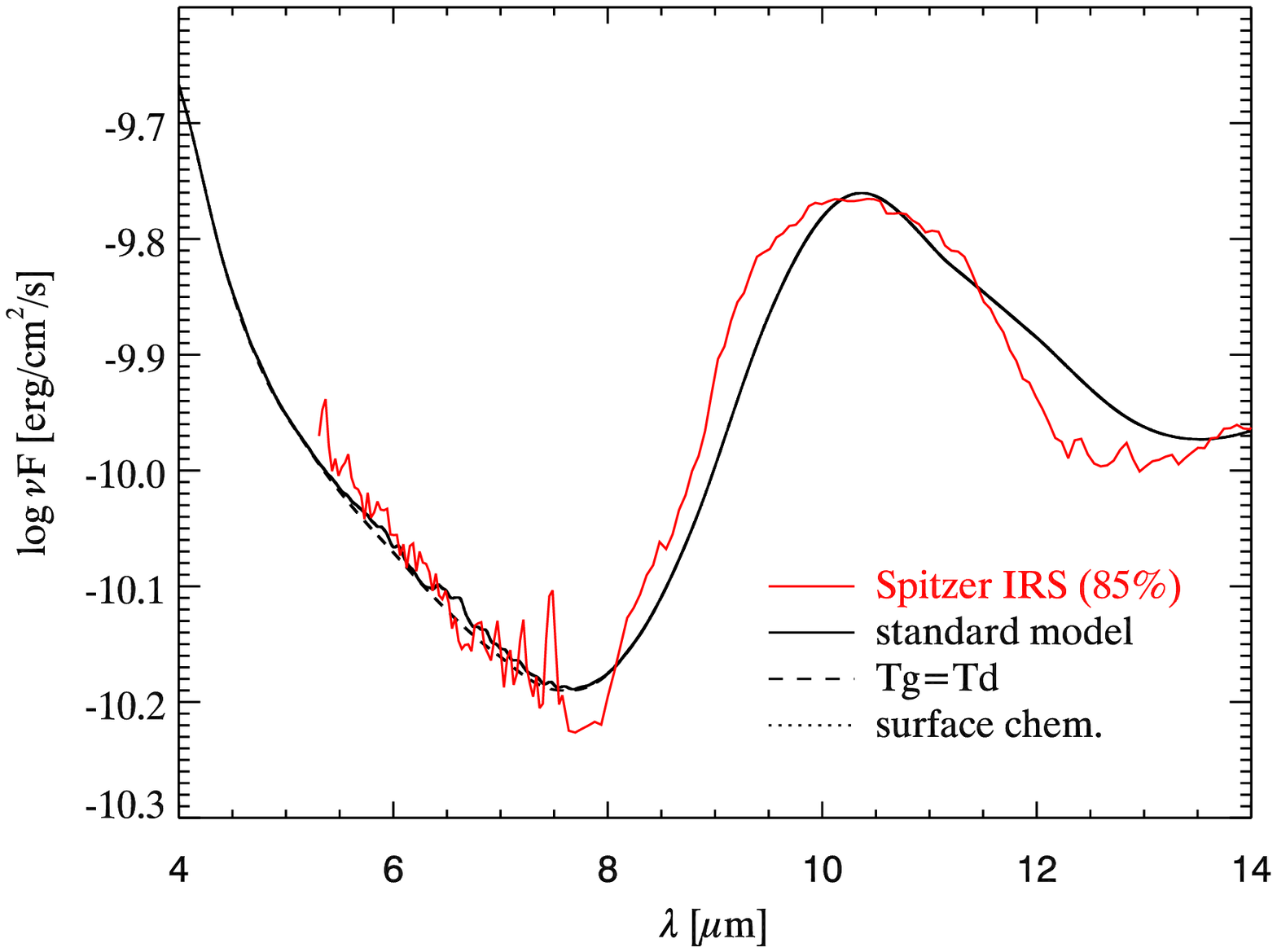}
\vspace*{-8mm}
\caption{TW Hya low-resolution IRS spectrum (red) scaled by a factor 0.85. Overplotted in black are the TW Hya models (`standard' model - solid line; model with $T_{\rm gas}\!=\!T_{\rm dust}$ - dashed line; model with water formation on dust surfaces - dotted line, indistinguishable from solid line) convolved to give a resolution of 100.}
\label{plot:IRS}
\end{figure}


If the photodesorption process is omitted (model~13), the low energy water lines such as the $538.29~\mu$m line become a factor 100 weaker, but also much broader ($\sim\!3$~km/s) because they now originate in the inner disk. The `standard' model predicts a FWHM of 1.0~km/s for the $538.29~\mu$m line and a FWHM of 1.14~km/s for the $269.27~\mu$m line. The observed FWHM of the two lines are 0.96~km/s and 1.17~km/s respectively \citep{Hogerheijde2011}. Our model~7 can fit the observed HIFI water lines without an assumption of icy grain settling; the lines are optically thick and also very sensitive to the details of the radiative transfer (e.g. number of collision partners, radiative transfer method, see Sect.~\ref{stateq-water}). The o/p water vapor mass ratio in the model is 0.3, a factor $\sim\!2$ lower than the one in the model by \citet{Hogerheijde2011}.

\citet{Szulagyi2012} published a Spitzer low resolution short wavelength spectrum of TW Hya (Fig.~\ref{plot:IRS}). Degraded to the appropriate  resolution of 100, our `standard' model does not show any signatures of water between 5 and 8~$\mu$m that stick out above the Spitzer noise level. However, this is partially due to the poor resolution of the SL Spitzer mode. 
Spitzer spectra in the $20-40$~$\mu$m range have a resolution of 600 and there water lines are clearly visible next to a few OH lines. Table~\ref{tab:mid-IR-Spitzerlines} lists a selection of high-excitation line fluxes in the near- and mid-IR wavelength range calculated from simplified vertical escape probability. Note that a more detailed comparison with Spitzer observations needs to take into account that many water lines overlap at this resolution, allowing only blends to be measured instead of individual line fluxes.

\begin{table}[htb]
\caption{Selected CO, water and H$_2$ lines with their observed fluxes --- n.a.\ denotes not available --- \citep{Salyk2007, Najita2010, Zhang2013} in the near- and mid-IR wavelength range from the `standard model' (model 1) and the model with 100 times higher gas mass inside 4~AU (model~14).}
\begin{tabular}{l|lll|ll}
\hline
\hline
Species      & $\lambda$ & line                                    & obs.   & \multicolumn{2}{c}{model line flux}  \\
                     & [$\mu$m]     & ident.                                & $F_{\rm line}$ &  \multicolumn{2}{c}{[$10^{-18}$~W/m$^2$]} \\[2mm]
                     &                       &                                           &                 & model 1         & model~14 \\
                \hline
CO               & 4.700           & v$=$1-0\,P4                             & 7.2 &    3.33(-1)     & 50.8  \\
CO               & 4.759           & v$=$2-1\,P4                             & n.a. &   7.66(-2)      & 25.1  \\
CO               & 5.041           & v$=$1-0\,P36                           & n.a. &   2.80(-2)      & 35.7 \\
o-H$_2$     & 17.036         &  S(1)                                   & 12.1 &     3.43(-1)      & 4.71 \\
o-H$_2$O & 17.36            & 11$_{29}$-10$_{110}$ & n.a. &    3.21(-1)       & 6.91 \\
o-H$_2$O & 23.86            & 9$_{82}$-8$_{71}$       & 466.9$^1$ &     2.19             & 20.1 \\
o-H$_2$O & 30.87            & 8$_{54}$-7$_{43}$       & 188.3$^1$ &     5.00             & 23.7 \\
\hline \\
\end{tabular}
\tablefoottext{1}{At the low resolution of Spitzer, the observed fluxes are for a blend of multiple water lines, while the model flux is for the individual line.}
\label{tab:mid-IR-Spitzerlines}
\end{table}

\subsubsection{The statistical equilibrium and line transfer of water}
\label{stateq-water}

\begin{figure*}[!htbp]
\vspace*{-0mm}
\begin{center}
\includegraphics[width=14cm]{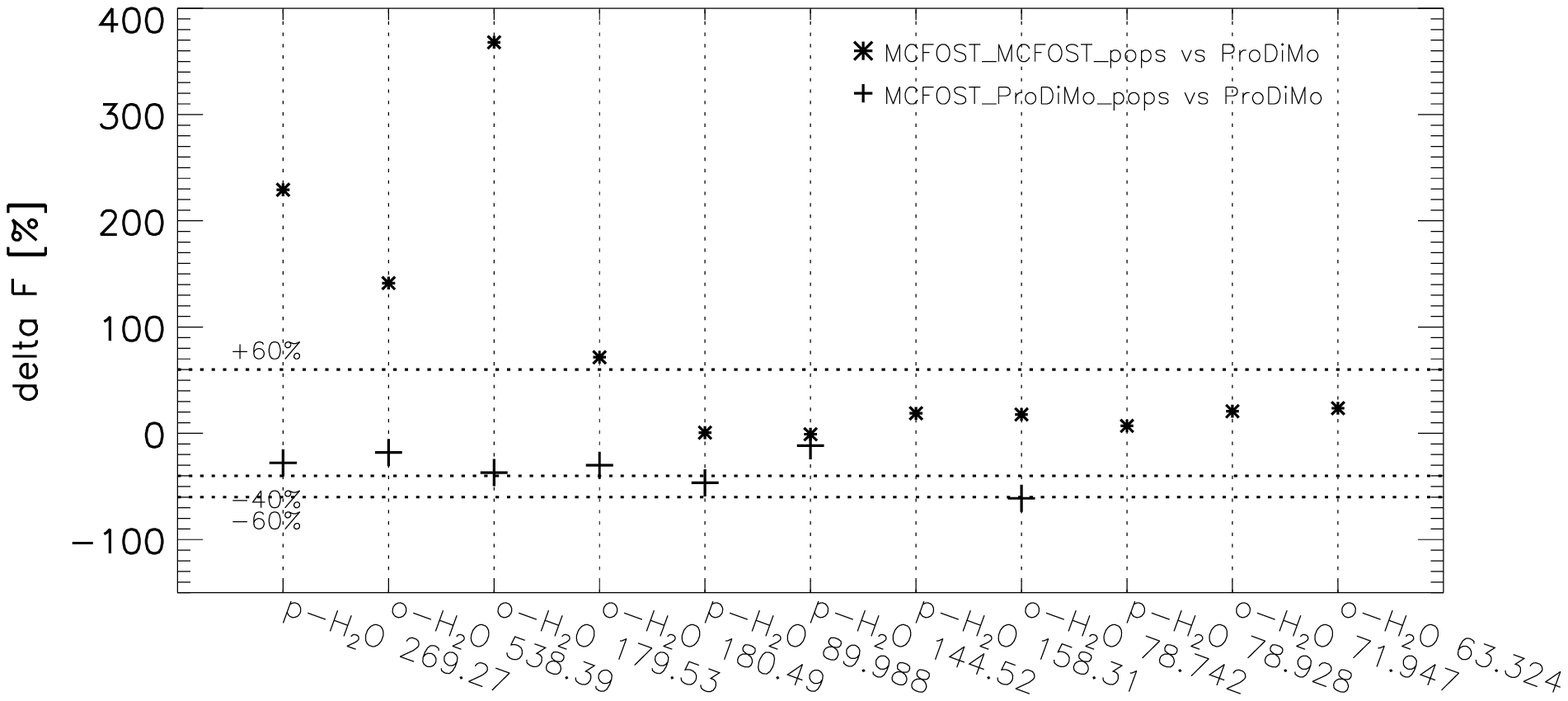}
\end{center}
\vspace*{-10mm}
\caption{The percentage change in line fluxes for a series of water lines. The plus signs are a comparison between the fluxes from the `standard' model and the fluxes computed with MCFOST using the level populations of the `standard' model (large water molecule, 70x70). The asterisks are a comparison between the `standard' model with the new LAMDA file and the small water molecule and the MCFOST run. Note that MCFOST contains only H$_2$ collisions.}
\label{fig:water-comparison}
\end{figure*}

We tested the formal water line radiative transfer of ProDiMo against the 3D Monte Carlo radiative transfer code MCFOST \citep{Pinte2006,Pinte2009} for a series of our water lines. The line radiative transfer is implemented such that it uses the level populations from ProDiMo as starting point. To remove gridding uncertainties and possible interpolation uncertainties, we used the same grid (100x70) and line fluxes calculated with an inclination angle of $0^{\rm o}$. For the three lines originating in the outer disk, 538.29, 269.27 and 179.53~$\mu$m, the agreement of the line flux predictions is within 40\% (Fig.~\ref{fig:water-comparison}). Some of the higher excitation lines deviate by up to 60\%.

\begin{figure*}[!htbp]
\vspace*{-0mm}
 \begin{center}
\includegraphics[width=14cm]{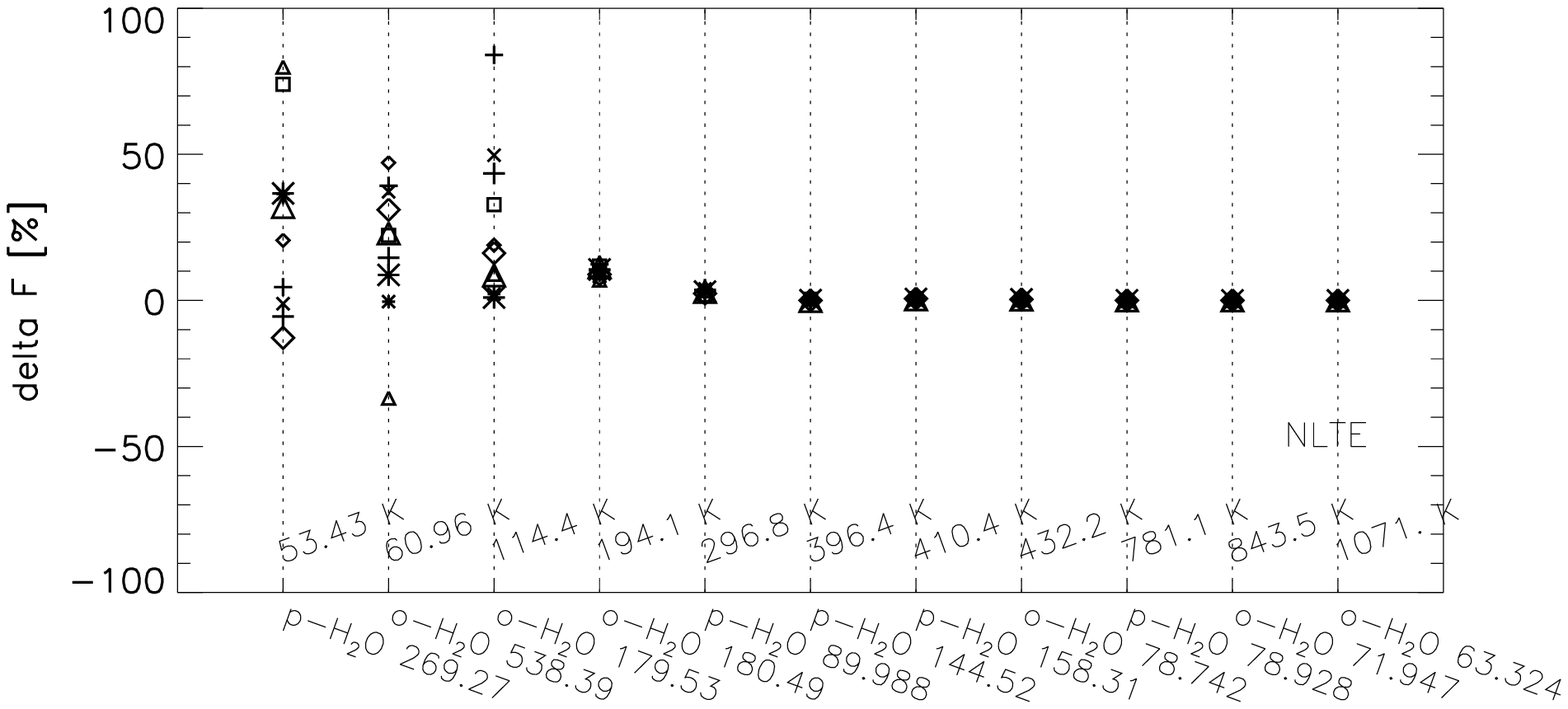}
\end{center}
\vspace*{-10mm}
\caption{The percentage change in line fluxes for a series of water lines due to varying collision cross sections randomly within a factor 10. The symbols correspond to 10 different sets of collision cross sections. The labeling of the x-axis shows the excitation temperature of the upper level (top) and the line identification with wavelength (bottom).}
\label{fig:sensitivity-water}
\end{figure*}

\begin{figure*}[!htbp]
\includegraphics[width=9cm]{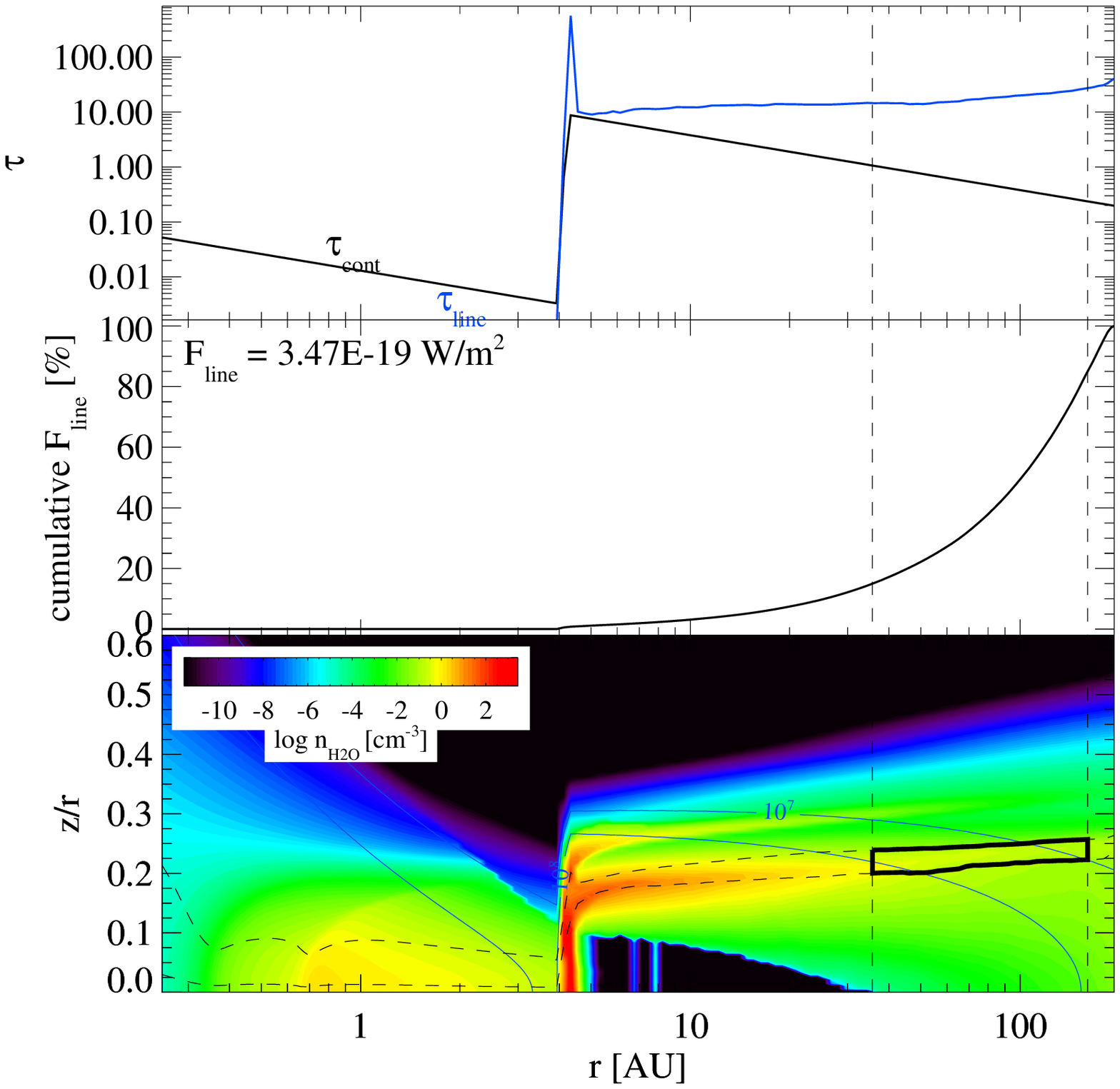}
\includegraphics[width=9cm]{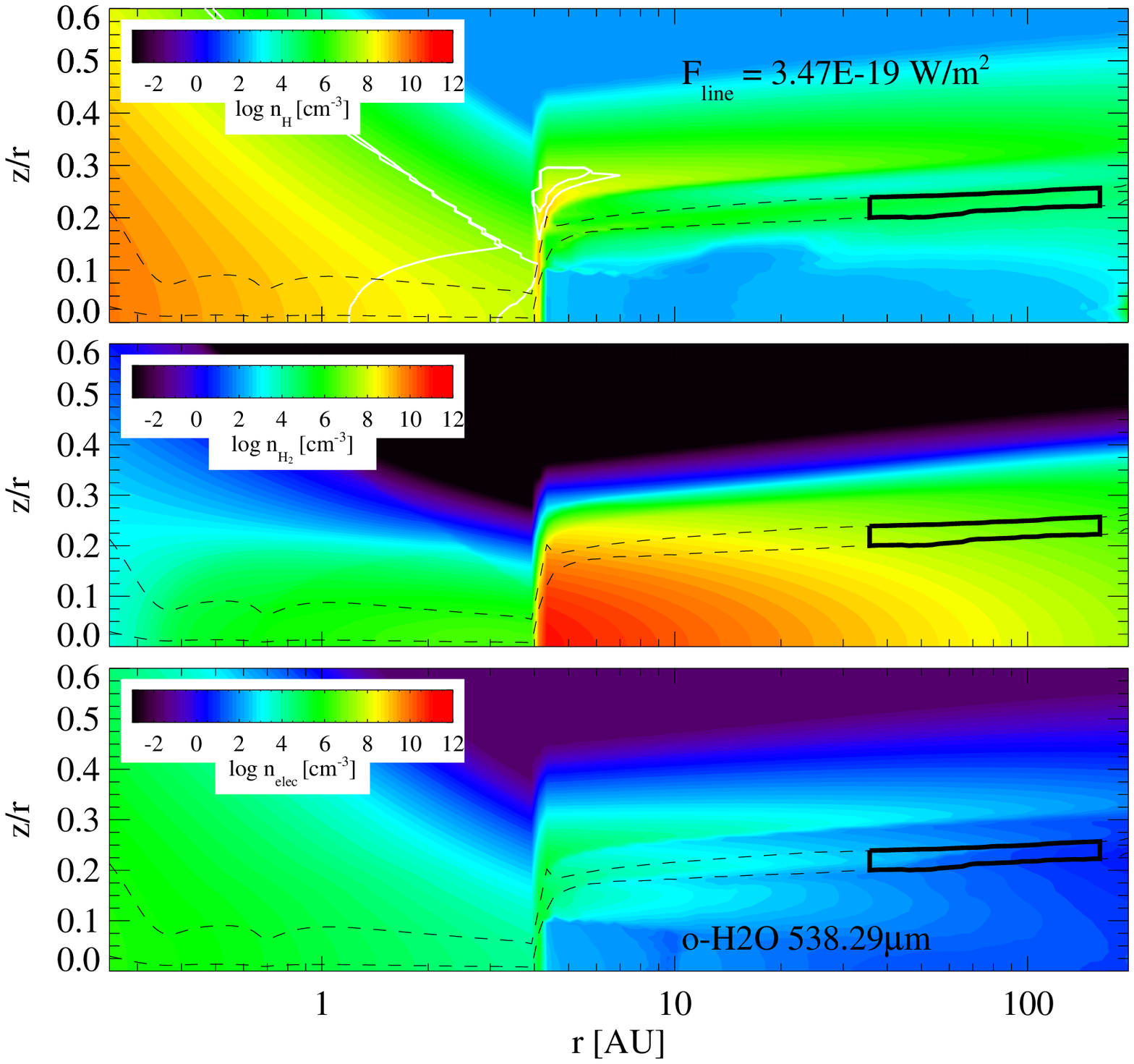}
\vspace*{-2mm}
\caption{Left: Optical depth, cumulative line fluxes and zone where the water 538.29~$\mu$m line originates in our `standard' model. The optical depth is obtained from vertical escape probability which is a good approximation in the case of the TW Hya disk which has a low inclination of $7^{\rm o}$. The zone of line formation outlined by black dashed contours that denote where 15 to 85\% of the flux build up vertically and radially; the black solid line encloses thus 50\% of the total line emission. The blue contours show total hydrogen number densities of $10^7$ and $10^8$~cm$^{-3}$. Right: density of the collision partners H, H$_2$ and e$^-$ with the line forming region of the 538.29~$\mu$m line superimposed.}
\label{fig:H2Ocumulative_lineflux}
\end{figure*}

The HIFI water lines are slightly sub-thermal in our disk model; line fluxes are a factor $\sim\!3$ lower than LTE. At the same time, the lines are optically thick (Fig.~\ref{fig:H2Ocumulative_lineflux}) and the dust provides a strong continuum flux at $300\!-\!500~\mu$m; continuum optical depth across the fundamental water line (538.29 and 269.27~$\mu$m) emission region are between 0.1 and a few (Fig.~\ref{fig:H2Ocumulative_lineflux}). In the absence of this continuum, the lines would be optically thick, but effectively thin; this means that water line photons are resonantly scattered until they reach the disk surface and escape. Resonant scattering of water line photons has been studied for example by \citet{Poelman2007}. Whether the water line photons can resonantly escape depends critically on the details of the local dust opacity, i.e.\ what is the probability that a line photon escapes after a certain number of scatter event without encountering a dust grain. In our model, the excitation temperature of the upper level is very close to the radiative temperature, indicating that far-IR pumping is efficient. With increasing excitation energy, the excitation temperature deviates more from the radiative temperature. In all cases, the gas temperature is higher than the excitation temperature.

A comparison between excitation temperatures from the $1\!+\!1$D escape probability and from the Monte Carlo method shows clear differences due to the limited amount of directions used in escape probability. Even though absolute differences may not be large, the low excitation lines react very sensitively to small changes in level populations. With the caveat of using a different number of collision partners in ProDiMo and MCFOST, Fig.~\ref{fig:water-comparison} shows that the low excitation line fluxes can vary by a factor $2\!-\!4$ between the two codes. This is partially due to differences in collision cross sections and partially due to the radiative transfer method for calculating level populations. 

Fig.~\ref{fig:H2Ocumulative_lineflux} shows the densities of the main collision partners H, H$_2$ and electrons in the `standard' model. In the main water line forming region, atomic hydrogen densities are only about an order of magnitude smaller than H$_2$. Collisions between water and atomic hydrogen are included based on a scaling of the respective H$_2$ collision cross sections. However, this approach could be wrong by an order of magnitude given the generally higher reactivity of hydrogen. To assess the relevance of the collision cross sections on determining level populations for water, we carry out a sensitivity analysis where we vary randomly the cross sections $C_{ij}$ within an order of magnitude
\begin{equation}
\log C_{ij} = \log C_{ij} + (2 r -1) \,\,\,
\end{equation}
with $r$ being a random number between 0 and 1. We ran ten `standard' models with randomly varied cross sections and compare the resulting line fluxes in Fig.~\ref{fig:sensitivity-water} to those of model~1. The only three lines affected are those with low excitation energies (the two HIFI lines and the PACS $179.53~\mu$m line). They change within a factor two, while all other high excitation lines stay within 10\%. A possible explanation is that the high excitation lines originate close to the inner rim of the outer disk, where the dust temperature is higher ($70\!-\!110$~K) and the level populations could be dominated by IR pumping \citep{Daniel2012}. A possibility to quantify this radiative pumping is to compare the excitation temperature of a lines upper level $T_{\rm ex}$ with the temperature of the radiation field at that wavelength $T_{\rm rad}$ and the local gas temperature $T_{\rm g}$. The 63.3~$\mu$m line is one of the water lines that forms at the inner rim of the outer disk. Its ratio $T_{\rm ex}/T_{\rm rad}$ across the line forming region is $\sim\!1$, while the ratio $T_{\rm ex}/T_{\rm g}$ is close to 0.2.

\begin{table*}[htb]
\caption{Ortho/para water line ratios of lines that are close in excitation energy in the model series. 'std' stands for the 'standard model' and SS for self-shielding.}
\begin{tabular}{l|lllll|llll}
       \hline
       \hline
  line ratios & \multicolumn{5}{c|}{without grain surface reactions} &  \multicolumn{4}{c}{with grain surface reactions} \\
                     & std & with  & 
                     $T_g=T_d$ & without & $\delta=0.01$ & std & without water &100xGG &  photodes. \\
                     &  & X-rays &
                      & SS & ($<4$~AU) &  & adsorption & & yields \\
       \hline
 538.29/269.27 & 0.49 & 0.48 & 
 0.45 & 0.50 & 0.25 & 0.49 & 0.44 & 0.27 & 0.25\\
 78.74/78.93 & 9.3 & 10.0 & 
 63.6 & 9.2 & 1.95 & 9.3 & 9.2 & 3.6 & 3.6 \\
\hline
\end{tabular}
\label{tab:ortho-para-water}
\end{table*}

\subsubsection{The water o/p line ratios}

The observed o/p line flux ratio for the fundamental lines is 0.27 \citep{Hogerheijde2011}. Table~\ref{tab:ortho-para-water} shows the line o/p ratio from our model series for two sets of ortho and para water lines being very close in excitation energy: the ratio of the fundamental 538.29 and 269.27~$\mu$m lines and the ratio of the high excitation 78.74 and 78.93~$\mu$m lines.

If the lines are optically thick, the o/p line ratio does not necessarily reflect the chemical o/p water ratio. In our model, the fundamental lines are optically thick as shown in Fig.~\ref{fig:H2Ocumulative_lineflux}. The lines originate in layers, where the total hydrogen number density is close to the critical densities for these lines at low temperature ($n_{\rm cr}\!=\!2\!\times\!10^7$ and $2\!\times\!10^8$~cm$^{-3}$ respectively for the 538.29 and the 269.27~$\mu$m line at $T\!\sim\!50$~K). The level populations are slightly sub thermal leading to line fluxes that are a factor $3.6$ (538.29~$\mu$m) and $4.3$ (269.27~$\mu$m) smaller than the LTE values.

Experiments by \citet{Abouaf-Marguin2009} show that nuclear spin conversion of water captured in a xenon matrix is fast ($\sim\!100$~min) at very low temperatures of a few Kelvin. Theoretical calculations of \citet{Buntkowsky2008} show that this happens in an ice matrix even on timescales of milliseconds. Recent experiments \citep{Hama2011} suggest that the o/p ratio of thermally desorbed water --- originally deposited at 8~K --- is always close to three, which is the high temperature limit. It remains unclear however, whether photodesorption affects the o/p ratio of the liberated water ice. In addition, our `standard' model shows that most of the water vapour is re-formed in the gas phase and should thus rather reflect the LTE o/p ratio corresponding to $T_{\rm gas}$ (Fig.~\ref{fig:ortho-para-water}).

\subsubsection{Lines from inner disk and inner rim of the outer disk}
\label{subsect:innerrimlines}

The lines originating inside $r\! \lesssim \!6$~AU are the CH$^+$ lines, the CO high J rotational lines and the OH hyperfine structure lines. Since they form in warmer disk regions, grain surface chemistry plays no significant role. The C/O ratio changes the high J CO rotational lines, increasing fluxes by up to a factor $\sim\!10$ (see Fig.~\ref{fig:finestructurelines}). $T_{\rm gas}\!=\!T_{\rm dust}$ is important for the OH lines (less formation of OH at low temperatures) and the high J CO rotational lines (lower excitation temperatures).

The molecular abundances inside 4~AU change substantially in model~14 with a gas-to-dust mass ratio of 100 in the inner disk. This increases the line fluxes from high excitation lines such as the CO high J rotational lines, CO ro-vibrational lines and water high excitation lines. Table~\ref{tab:mid-IR-Spitzerlines} lists selected line fluxes in addition to the ones listed in Table~\ref{tab:PACSmodfluxes1} and \ref{tab:PACSmodfluxes2}. Fluxes change by one to two orders of magnitude, thus bringing them closer to the observed values \citep[our PACS data and][]{Najita2010, Salyk2007}.

\section{Comparison with observations}


\begin{figure*}[!htbp]
\vspace*{-2mm}
\begin{center}
\includegraphics[width=14cm]{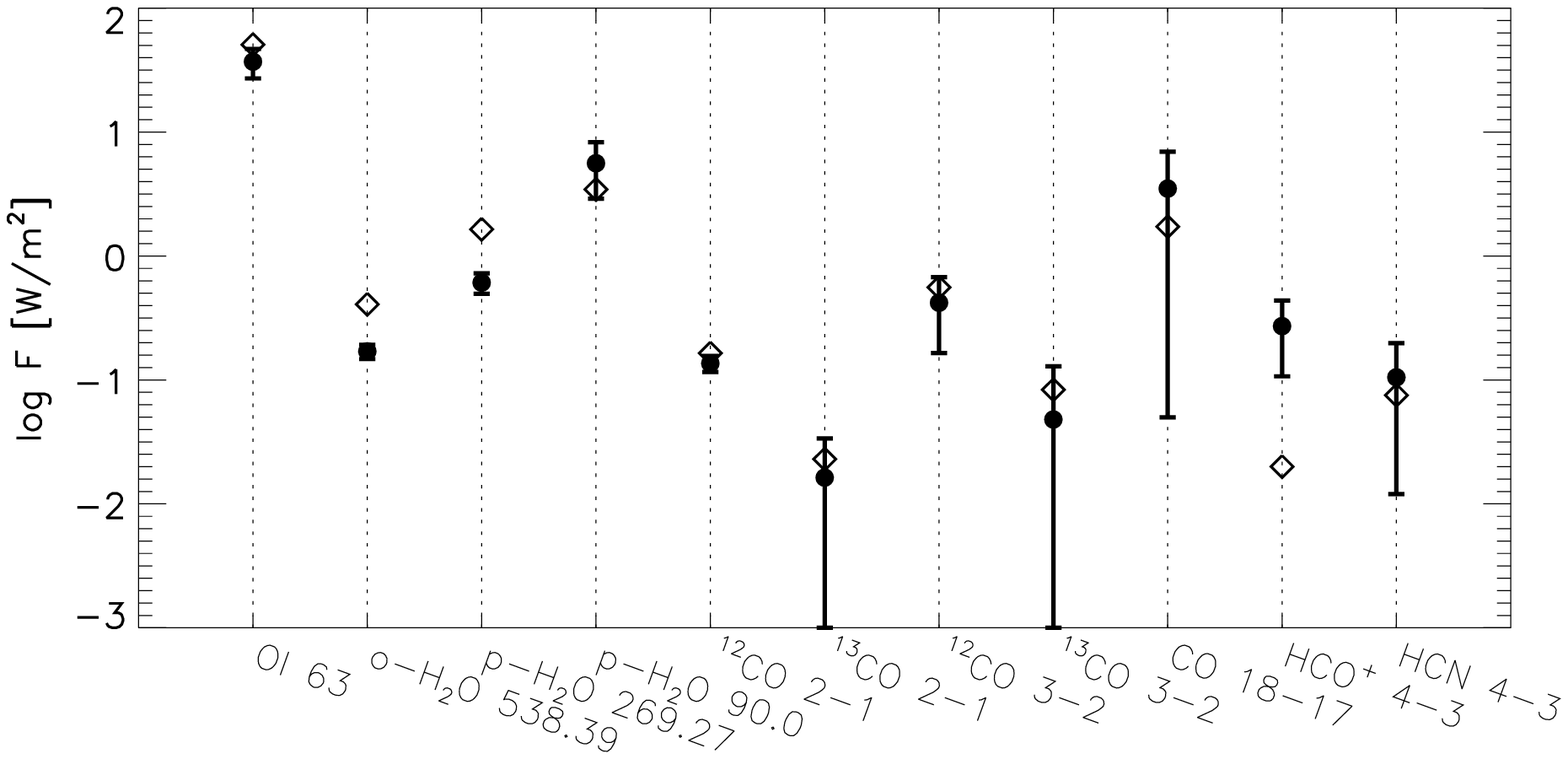}
\end{center}
\vspace*{-3mm}
\caption{Comparison between selected observed line fluxes and the standard model with two zones (inside 4~AU a gas-to-dust mass ratio of 0.01 and outside 1.0). HCO$^+$ fluxes are from the model with a 1000 times lower metallicity. Observed fluxes are denoted by filled circles with $3\sigma$ error bars and model fluxes by diamonds.}
\label{plot:fluxes-obs}
\end{figure*}

Fig.~\ref{plot:fluxes-obs} shows a comparison between our observed line fluxes and the two-zone standard model (inside 4~AU a gas-to-dust mass ratio of 0.01 and outside 1.0)
except for the HCO$^+$ line which is taken from the extremely low (1/1000) metal abundance model (model 3). The agreement is for most lines within a factor two, also for the low-excitation water lines. Given the results from the sensitivity analysis in Sect.~\ref{sect:emergentfluxes}, pushing the models any further is of limited value.

The model flux for HCO$^+$\,J=4--3 is still too weak. The line originates in a very thin layer and is optically thin based on a simple vertical escape probability estimate. Scaling the gas mass up by a factor 10 in the `standard' model, changes the HCO$^+$ line flux only by 20\%. 

Not shown in Fig.~\ref{plot:fluxes-obs} is the reasonable agreement of the CO ro-vib lines in the model with a more canonical gas-to-dust mass ratio of 100 in the inner disk (see Table~\ref{tab:mid-IR-Spitzerlines}) with the observed CO ro-vib fluxes of \citet{Salyk2007} ($2\!-\!12\!\times\!10^{-18}$~W/m$^{-2}$). The FWHM of $\sim\!13$~km/s found in the modeled lines is also consistent with the observed data. In that very same model, the p-H$_2$O line flux at 89.99~$\mu$m reaches $3.54\!\times\!10^{-18}$~W/m$^2$, while all other water lines originating in the inner disk stay below their $3\sigma$ upper limits from the Herschel/PACS observations.

 Other groups found a much higher gas-to-dust mass ratio for the outer disk of TW~Hya \citep{Gorti2011, Bergin2013}. However, these works have not included the strong constraint from the $^{12}$CO/$^{13}$CO line ratio that \citet{Thi2010a} found decisive in setting the gas mass of the outer disk (assuming canonical C and O abundances and equilibrium chemistry). Sect.~\ref{subsect:CoverO} and \ref{subsect:waterlines} and \ref{subsect:innerrimlines} show that increasing the C/O ratio has opposing effects on the fluxes of the fundamental water lines and high J CO lines, while leaving the CO submm lines almost unchanged (see also Table~\ref{tab:PACSmodfluxes1}). The stability of the CO submm isotopic line ratio (see Sect.~\ref{subsect:submmlines}) makes it difficult to reconcile the observations with a higher gas mass in the outer disk.

 If the gas-to-dust ratio is indeed lower than the ISM in the outer disk, the possibility that this ratio increases towards the inner disk is interesting in the context of planet formation models and disk evolution. If gas evolution is related to the orbital timescale, it should be faster in the inner disk. Finding a primordial gas-to-dust mass ratio inside 4~AU and a lower gas-to-dust mass ratio in the outer disk  would exclude the possibility that gas and dust migrate together from the outer disk to the inner disk. Evolution of both parts of the disk are either de-coupled or gas and dust do not evolve co-spatially. An interesting conclusion is that the depletion in surface density in the gas at 4~AU is only a factor $\sim\!30$, while the dust is depleted by a factor $\sim\!3000$ with respect to the outer disk. It could thus mean that dust is much more evolved inside 4~AU than the gas.


\section{Conclusion}

We report here the first detection of the far-IR p-H$_2$O line at $89.99~\mu$m in the disk around TW Hya from recent observations with the Herschel/PACS instrument in the framework of the Open Time Key Program GASPS. In addition, the CO\,J=18--17 line at $144.78~\mu$m is detected. Many previous studies aimed at deriving disk geometry, structure, extent and dust content for TW Hya by `fitting' dust SEDs, images and gas emission lines simultaneously leading to sometimes different disk structures/masses. Before distinguishing between these models, a careful study of the robustness of a set of used line diagnostics is needed.

This paper systematically investigated the robustness of the disk chemistry and line emission from uncertainties in underlying physical/chemical processes and numerical methods besides the usual disk structure/extent/geometry. We adopt a fixed disk structure and a limited set of IR and submm lines from O, C$^+$, OH, CO, CH$^+$, H$_2$O, HCO$^+$ and HCN. 

The water lines from the disk studied here are sensitive to those often `hidden' processes and model ingredients. The gas temperature ($T_{\rm g}\!=\!T_{\rm d}$), ice formation, surface chemistry, photo-desorption yields, metal abundances can affect water line fluxes by a factor of a few. In addition, details of the radiative transfer method and uncertain hydrogen collision cross sections can affect the low excitation water lines by a similar factor. Experimental and/or theoretical collision rates for H$_2$O with atomic hydrogen are needed for ro-vibrational levels in order to diminish some of the uncertainties.


The stability of the CO, HCN and HCO$^+$ lines --- originating from the cold gas in the outer disk --- studied here against uncertainties in the intrinsic physical and chemical parameters such as grain surface chemistry, ice formation, and desorption yields, confirms the robust diagnostic power of these lines also for protoplanetary disk research. 

The TW Hya observations listed above --- especially the new PACS high excitation line detections --- are consistent within a factor two with a disk model in which gas and dust are well mixed and which has a gas-to-dust mass ratio of $\sim\!1$ in the outer disk (beyond 4~AU) and a higher gas-to-dust mass ratio of $\sim\!100$ in the optically thin inner disk.

\begin{acknowledgement}
GM is supported by Ramon y Cajal grant RYC-2011-07920. IK acknowledges
funding by an NWO MEERVOUD grant. WFT, PW and IK acknowledge funding from
the EU FP7-2011 under Grant Agreement nr.\ 284405. CP acknowledges funding from the European Commission's
7$^\mathrm{th}$ Framework Program (contract PERG06-GA-2009-256513) and from
Agence Nationale pour la Recherche (ANR) of France under contract
ANR-2010-JCJC-0504-01. Part of the computations presented in this paper were performed at the Service
Commun de Calcul Intensif de l'Observatoire de Grenoble (SCCI) on the
super-computer funded by Agence Nationale pour la Recherche under
contracts ANR-07-BLAN-0221, ANR-2010-JCJC-0504-01 and ANR-2010-JCJC-0501-01.
\end{acknowledgement}

\bibliography{reference}

\begin{thebibliography}{98}
\expandafter\ifx\csname natexlab\endcsname\relax\def\natexlab#1{#1}\fi

\bibitem[{Abouaf-Marguin {et~al.}(2009)Abouaf-Marguin, Vasserot, Pardanaud, \&
  Michaut}]{Abouaf-Marguin2009}
Abouaf-Marguin, L., Vasserot, A.-M., Pardanaud, C., \& Michaut, X. 2009,
  Chemical Physics Letters, 480, 82

\bibitem[{{Abrahamsson} {et~al.}(2007){Abrahamsson}, {Krems}, \&
  {Dalgarno}}]{Abrahamsson2007}
{Abrahamsson}, E., {Krems}, R.~V., \& {Dalgarno}, A. 2007, \apj, 654, 1171

\bibitem[{{Adamson} {et~al.}(1994){Adamson}, {Farhat}, {Morter}, {Glass},
  {Curl}, \& {Phillips}}]{Adamson1994}
{Adamson}, J., {Farhat}, S., {Morter}, C., {et~al.} 1994, J. Chem. Phys., 98,
  5665

\bibitem[{{Aikawa} \& {Herbst}(2001)}]{Aikawa2001}
{Aikawa}, Y. \& {Herbst}, E. 2001, \aap, 371, 1107

\bibitem[{{Aikawa} {et~al.}(1996){Aikawa}, {Miyama}, {Nakano}, \&
  {Umebayashi}}]{Aikawa1996}
{Aikawa}, Y., {Miyama}, S.~M., {Nakano}, T., \& {Umebayashi}, T. 1996, \apj,
  467, 684

\bibitem[{{Anders} \& {Grevesse}(1989)}]{Anders1989}
{Anders}, E. \& {Grevesse}, N. 1989, \gca, 53, 197

\bibitem[{{Andersson} \& {van Dishoeck}(2008)}]{Andersson2008}
{Andersson}, S. \& {van Dishoeck}, E.~F. 2008, \aap, 491, 907

\bibitem[{{Arasa} {et~al.}(2010){Arasa}, {Andersson}, {Cuppen}, {van Dishoeck},
  \& {Kroes}}]{Arasa2010}
{Arasa}, C., {Andersson}, S., {Cuppen}, H., {van Dishoeck}, E., \& {Kroes}, G.
  2010, Journal of Chemical Physics, 132, 184510

\bibitem[{{Aresu} {et~al.}(2011){Aresu}, {Kamp}, {Meijerink}, {Woitke}, {Thi},
  \& {Spaans}}]{Aresu2011}
{Aresu}, G., {Kamp}, I., {Meijerink}, R., {et~al.} 2011, \aap, 526, A163

\bibitem[{{Aresu} {et~al.}(2012){Aresu}, {Meijerink}, {Kamp}, {Spaans}, {Thi},
  \& {Woitke}}]{Aresu2012}
{Aresu}, G., {Meijerink}, R., {Kamp}, I., {et~al.} 2012, \aap, 547, A69

\bibitem[{{Balakrishnan} {et~al.}(2002){Balakrishnan}, {Yan}, \&
  {Dalgarno}}]{Balakrishnan2002}
{Balakrishnan}, N., {Yan}, M., \& {Dalgarno}, A. 2002, \apj, 568, 443

\bibitem[{{Barber} {et~al.}(2006){Barber}, {Tennyson}, {Harris}, \&
  {Tolchenov}}]{Barber2006}
{Barber}, R.~J., {Tennyson}, J., {Harris}, G.~J., \& {Tolchenov}, R.~N. 2006,
  \mnras, 368, 1087

\bibitem[{{Barrado Y Navascu{\'e}s}(2006)}]{Barrado2006}
{Barrado Y Navascu{\'e}s}, D. 2006, \aap, 459, 511

\bibitem[{{Bell} {et~al.}(1998){Bell}, {Berrington}, \& {Thomas}}]{Bell1998}
{Bell}, K.~L., {Berrington}, K.~A., \& {Thomas}, M.~R.~J. 1998, \mnras, 293,
  L83

\bibitem[{{Bergin} {et~al.}(2003){Bergin}, {Calvet}, {D'Alessio}, \&
  {Herczeg}}]{Bergin2003}
{Bergin}, E., {Calvet}, N., {D'Alessio}, P., \& {Herczeg}, G.~J. 2003, \apjl,
  591, L159

\bibitem[{{Bergin} {et~al.}(2013){Bergin}, {Cleeves}, {Gorti}, {Zhang},
  {Blake}, {Green}, {Andrews}, {Evans}, {Henning}, {{\"O}berg}, {Pontoppidan},
  {Qi}, {Salyk}, \& {van Dishoeck}}]{Bergin2013}
{Bergin}, E.~A., {Cleeves}, L.~I., {Gorti}, U., {et~al.} 2013, \nat, 493, 644

\bibitem[{{Bethell} \& {Bergin}(2009)}]{Bethell2009}
{Bethell}, T. \& {Bergin}, E. 2009, Science, 326, 1675

\bibitem[{{Bockel{\'e}e-Morvan}(2010)}]{Bockelee-Morvan2010}
{Bockel{\'e}e-Morvan}, D. 2010, in EAS Publications Series, Vol.~41, EAS
  Publications Series, ed. {T.~Montmerle, D.~Ehrenreich, \& A.-M.~Lagrange},
  313--324

\bibitem[{Buntkowsky {et~al.}(2008)Buntkowsky, Limbach, Walaszek, Adamczyk, Xu,
  Breitzke, Schweitzer, Gutmann, W\"{a}chtler, Amadeu, Tietze, \&
  Chaudret}]{Buntkowsky2008}
Buntkowsky, G., Limbach, H.-H., Walaszek, B., {et~al.} 2008, Z. Phys. Chem.,
  222, 1049

\bibitem[{{Carr} \& {Najita}(2008)}]{Carr2008}
{Carr}, J.~S. \& {Najita}, J.~R. 2008, Science, 319, 1504

\bibitem[{{Ceccarelli} {et~al.}(2005){Ceccarelli}, {Dominik}, {Caux},
  {Lefloch}, \& {Caselli}}]{Ceccarelli2005}
{Ceccarelli}, C., {Dominik}, C., {Caux}, E., {Lefloch}, B., \& {Caselli}, P.
  2005, \apjl, 631, L81

\bibitem[{{Cecchi-Pestellini} {et~al.}(2002){Cecchi-Pestellini}, {Bodo},
  {Balakrishnan}, \& {Dalgarno}}]{Cecchi-Pestellini2002}
{Cecchi-Pestellini}, C., {Bodo}, E., {Balakrishnan}, N., \& {Dalgarno}, A.
  2002, \apj, 571, 1015

\bibitem[{{Chaparro Molano} \& {Kamp}(2012)}]{Chaparro2012a}
{Chaparro Molano}, G. \& {Kamp}, I. 2012, \aap, 537, A138

\bibitem[{{Chapillon} {et~al.}(2010){Chapillon}, {Parise}, {Guilloteau},
  {Dutrey}, \& {Wakelam}}]{Chapillon2010}
{Chapillon}, E., {Parise}, B., {Guilloteau}, S., {Dutrey}, A., \& {Wakelam}, V.
  2010, \aap, 520, A61

\bibitem[{{Ciesla} \& {Cuzzi}(2006)}]{Ciesla2006}
{Ciesla}, F.~J. \& {Cuzzi}, J.~N. 2006, \icarus, 181, 178

\bibitem[{{Daniel} {et~al.}(2011){Daniel}, {Dubernet}, \&
  {Grosjean}}]{Daniel2011}
{Daniel}, F., {Dubernet}, M.-L., \& {Grosjean}, A. 2011, \aap, 536, A76

\bibitem[{{Daniel} {et~al.}(2010){Daniel}, {Dubernet}, {Pacaud}, \&
  {Grosjean}}]{Daniel2010}
{Daniel}, F., {Dubernet}, M.-L., {Pacaud}, F., \& {Grosjean}, A. 2010, \aap,
  517, A13

\bibitem[{Daniel {et~al.}(2012)Daniel, Goicoechea, Cernicharo, Dubernet, \&
  Faure}]{Daniel2012}
Daniel, F., Goicoechea, J., Cernicharo, J., Dubernet, M.-L., \& Faure, A. 2012,
  \aap, submitted

\bibitem[{{Dent} {et~al.}(2013){Dent}, {Thi}, {Kamp}, {Williams}, {Menard},
  {Andrews}, {Ardila}, {Aresu}, {Augereau}, {Barrado y Navascues}, {Brittain},
  {Carmona}, {Ciardi}, {Danchi}, {Donaldson}, {Duchene}, {Eiroa}, {Fedele},
  {Grady}, {de Gregorio-Molsalvo}, {Howard}, {Hu{\'e}lamo}, {Krivov},
  {Lebreton}, {Liseau}, {Martin-Zaidi}, {Mathews}, {Meeus},
  {Mendigut{\'{\i}}a}, {Montesinos}, {Morales-Calderon}, {Mora}, {Nomura},
  {Pantin}, {Pascucci}, {Phillips}, {Pinte}, {Podio}, {Ramsay}, {Riaz},
  {Riviere-Marichalar}, {Roberge}, {Sandell}, {Solano}, {Tilling}, {Torrelles},
  {Vandenbusche}, {Vicente}, {White}, \& {Woitke}}]{Dent2013}
{Dent}, W.~R.~F., {Thi}, W.~F., {Kamp}, I., {et~al.} 2013, \pasp, 125, 477

\bibitem[{{Dominik} {et~al.}(2005){Dominik}, {Ceccarelli}, {Hollenbach}, \&
  {Kaufman}}]{Dominik2005}
{Dominik}, C., {Ceccarelli}, C., {Hollenbach}, D., \& {Kaufman}, M. 2005,
  \apjl, 635, L85

\bibitem[{{Dubernet} {et~al.}(2006){Dubernet}, {Daniel}, {Grosjean}, {Faure},
  {Valiron}, {Wernli}, {Wiesenfeld}, {Rist}, {Noga}, \&
  {Tennyson}}]{Dubernet2006}
{Dubernet}, M.-L., {Daniel}, F., {Grosjean}, A., {et~al.} 2006, \aap, 460, 323

\bibitem[{{Dubernet} {et~al.}(2009){Dubernet}, {Daniel}, {Grosjean}, \&
  {Lin}}]{Dubernet2009}
{Dubernet}, M.-L., {Daniel}, F., {Grosjean}, A., \& {Lin}, C.~Y. 2009, \aap,
  497, 911

\bibitem[{{Faure} \& {Josselin}(2008)}]{Faure2008}
{Faure}, A. \& {Josselin}, E. 2008, \aap, 492, 257

\bibitem[{{Flower}(1999)}]{Flower1999}
{Flower}, D.~R. 1999, \mnras, 305, 651

\bibitem[{{Fogel} {et~al.}(2011){Fogel}, {Bethell}, {Bergin}, {Calvet}, \&
  {Semenov}}]{Fogel2011}
{Fogel}, J.~K.~J., {Bethell}, T.~J., {Bergin}, E.~A., {Calvet}, N., \&
  {Semenov}, D. 2011, \apj, 726, 29

\bibitem[{{Fraser} {et~al.}(2001){Fraser}, {Collings}, {McCoustra}, \&
  {Williams}}]{Fraser2001}
{Fraser}, H.~J., {Collings}, M.~P., {McCoustra}, M.~R.~S., \& {Williams}, D.~A.
  2001, \mnras, 327, 1165

\bibitem[{{Glassgold} {et~al.}(2009){Glassgold}, {Meijerink}, \&
  {Najita}}]{Glassgold2009}
{Glassgold}, A.~E., {Meijerink}, R., \& {Najita}, J.~R. 2009, \apj, 701, 142

\bibitem[{{Gorti} {et~al.}(2011){Gorti}, {Hollenbach}, {Najita}, \&
  {Pascucci}}]{Gorti2011}
{Gorti}, U., {Hollenbach}, D., {Najita}, J., \& {Pascucci}, I. 2011, \apj, 735,
  90

\bibitem[{{Green}(1980)}]{Green1980}
{Green}, S. 1980, \apjs, 42, 103

\bibitem[{{G{\"u}del} {et~al.}(2007){G{\"u}del}, {Briggs}, {Arzner}, {Audard},
  {Bouvier}, {Feigelson}, {Franciosini}, {Glauser}, {Grosso}, {Micela},
  {Monin}, {Montmerle}, {Padgett}, {Palla}, {Pillitteri}, {Rebull}, {Scelsi},
  {Silva}, {Skinner}, {Stelzer}, \& {Telleschi}}]{Guedel2007}
{G{\"u}del}, M., {Briggs}, K.~R., {Arzner}, K., {et~al.} 2007, \aap, 468, 353

\bibitem[{{Hama} {et~al.}(2011){Hama}, {Watanabe}, {Kouchi}, \&
  {Yokoyama}}]{Hama2011}
{Hama}, T., {Watanabe}, N., {Kouchi}, A., \& {Yokoyama}, M. 2011, \apjl, 738,
  L15

\bibitem[{{Hammami} {et~al.}(2009){Hammami}, {Owono Owono}, \&
  {St{\"a}uber}}]{Hammami2009}
{Hammami}, K., {Owono Owono}, L.~C., \& {St{\"a}uber}, P. 2009, \aap, 507, 1083

\bibitem[{{Hogerheijde} {et~al.}(2011){Hogerheijde}, {Bergin}, {Brinch},
  {Cleeves}, {Fogel}, {Blake}, {Dominik}, {Lis}, {Melnick}, {Neufeld},
  {Pani{\'c}}, {Pearson}, {Kristensen}, {Y{\i}ld{\i}z}, \& {van
  Dishoeck}}]{Hogerheijde2011}
{Hogerheijde}, M.~R., {Bergin}, E.~A., {Brinch}, C., {et~al.} 2011, Science,
  334, 338

\bibitem[{{Hollenbach} {et~al.}(2009){Hollenbach}, {Kaufman}, {Bergin}, \&
  {Melnick}}]{Hollenbach2009}
{Hollenbach}, D., {Kaufman}, M.~J., {Bergin}, E.~A., \& {Melnick}, G.~J. 2009,
  \apj, 690, 1497

\bibitem[{{Jaquet} {et~al.}(1992){Jaquet}, {Staemmler}, {Smith}, \&
  {Flower}}]{Jaquet1992}
{Jaquet}, R., {Staemmler}, V., {Smith}, M.~D., \& {Flower}, D.~R. 1992, Journal
  of Physics B Atomic Molecular Physics, 25, 285

\bibitem[{{Jenkins}(2009)}]{Jenkins2009}
{Jenkins}, E.~B. 2009, \apj, 700, 1299

\bibitem[{{Kamp} {et~al.}(2010){Kamp}, {Tilling}, {Woitke}, {Thi}, \&
  {Hogerheijde}}]{Kamp2010}
{Kamp}, I., {Tilling}, I., {Woitke}, P., {Thi}, W., \& {Hogerheijde}, M. 2010,
  \aap, 510, A260000+

\bibitem[{{Kastner} {et~al.}(2002){Kastner}, {Huenemoerder}, {Schulz},
  {Canizares}, \& {Weintraub}}]{Kastner2002}
{Kastner}, J.~H., {Huenemoerder}, D.~P., {Schulz}, N.~S., {Canizares}, C.~R.,
  \& {Weintraub}, D.~A. 2002, \apj, 567, 434

\bibitem[{{Kastner} {et~al.}(1997){Kastner}, {Zuckerman}, {Weintraub}, \&
  {Forveille}}]{Kastner1997}
{Kastner}, J.~H., {Zuckerman}, B., {Weintraub}, D.~A., \& {Forveille}, T. 1997,
  Science, 277, 67

\bibitem[{{Klippenstein} {et~al.}(2010){Klippenstein}, {Georgievskii}, \&
  {McCall}}]{KGM10}
{Klippenstein}, S.~J., {Georgievskii}, Y., \& {McCall}, B.~J. 2010,
  J.Phys.Chem.A, 114, 278

\bibitem[{{Korolov} {et~al.}(2009){Korolov}, {Plasil}, {Kotrik}, {Dohnal}, \&
  {Glosik}}]{K09}
{Korolov}, I., {Plasil}, R., {Kotrik}, T., {Dohnal}, P., \& {Glosik}, J. 2009,
  International Journal of Mass Spectrometry, 280, 144

\bibitem[{{Lim} {et~al.}(1999){Lim}, {Rabad{\'a}n}, \& {Tennyson}}]{Lim1999}
{Lim}, A.~J., {Rabad{\'a}n}, I., \& {Tennyson}, J. 1999, \mnras, 306, 473

\bibitem[{{Mamajek}(2005)}]{Mamajek2005}
{Mamajek}, E.~E. 2005, \apj, 634, 1385

\bibitem[{{McCall}(2006)}]{McCall2006}
{McCall}, B.~J. 2006, Royal Society of London Philosophical Transactions Series
  A, 364, 2953

\bibitem[{{Meeus} {et~al.}(2010){Meeus}, {Pinte}, {Woitke}, {Montesinos},
  {Mendigut{\'{\i}}a}, {Riviere-Marichalar}, {Eiroa}, {Mathews},
  {Vandenbussche}, {Howard}, {Roberge}, {Sandell}, {Duch{\^e}ne}, {M{\'e}nard},
  {Grady}, {Dent}, {Kamp}, {Augereau}, {Thi}, {Tilling}, {Alacid}, {Andrews},
  {Ardila}, {Aresu}, {Barrado}, {Brittain}, {Ciardi}, {Danchi}, {Fedele}, {de
  Gregorio-Monsalvo}, {Heras}, {Huelamo}, {Krivov}, {Lebreton}, {Liseau},
  {Martin-Zaidi}, {Mora}, {Morales-Calderon}, {Nomura}, {Pantin}, {Pascucci},
  {Phillips}, {Podio}, {Poelman}, {Ramsay}, {Riaz}, {Rice}, {Solano}, {Walker},
  {White}, {Williams}, \& {Wright}}]{Meeus2010}
{Meeus}, G., {Pinte}, C., {Woitke}, P., {et~al.} 2010, \aap, 518, L124

\bibitem[{{Meijerink} {et~al.}(2012){Meijerink}, {Aresu}, {Kamp}, {Spaans},
  {Thi}, \& {Woitke}}]{Meijerink2012}
{Meijerink}, R., {Aresu}, G., {Kamp}, I., {et~al.} 2012, \aap, 547, A68

\bibitem[{{Najita} {et~al.}(2011){Najita}, {{\'A}d{\'a}mkovics}, \&
  {Glassgold}}]{Najita2011}
{Najita}, J.~R., {{\'A}d{\'a}mkovics}, M., \& {Glassgold}, A.~E. 2011, \apj,
  743, 147

\bibitem[{{Najita} {et~al.}(2010){Najita}, {Carr}, {Strom}, {Watson},
  {Pascucci}, {Hollenbach}, {Gorti}, \& {Keller}}]{Najita2010}
{Najita}, J.~R., {Carr}, J.~S., {Strom}, S.~E., {et~al.} 2010, \apj, 712, 274

\bibitem[{{{\"O}berg} {et~al.}(2009){{\"O}berg}, {Linnartz}, {Visser}, \& {van
  Dishoeck}}]{Oeberg2009}
{{\"O}berg}, K.~I., {Linnartz}, H., {Visser}, R., \& {van Dishoeck}, E.~F.
  2009, \apj, 693, 1209

\bibitem[{{Offer} {et~al.}(1994){Offer}, {van Hemert}, \& {van
  Dishoeck}}]{Offer1994}
{Offer}, A., {van Hemert}, M., \& {van Dishoeck}, E. 1994, J. Che. Phys., 100,
  362

\bibitem[{{Ott}(2010)}]{Ott2010}
{Ott}, S. 2010, in Astronomical Society of the Pacific Conference Series, Vol.
  434, Astronomical Data Analysis Software and Systems XIX, ed. {Y.~Mizumoto,
  K.-I.~Morita, \& M.~Ohishi}, 139--+

\bibitem[{{Pascucci} {et~al.}(2012){Pascucci}, {Gorti}, \&
  {Hollenbach}}]{Pascucci2012}
{Pascucci}, I., {Gorti}, U., \& {Hollenbach}, D. 2012, \apjl, 751, L42

\bibitem[{{Pascucci} {et~al.}(2011){Pascucci}, {Sterzik}, {Alexander},
  {Alencar}, {Gorti}, {Hollenbach}, {Owen}, {Ercolano}, \&
  {Edwards}}]{Pascucci2011}
{Pascucci}, I., {Sterzik}, M., {Alexander}, R.~D., {et~al.} 2011, \apj, 736, 13

\bibitem[{{Pinte} {et~al.}(2009){Pinte}, {Harries}, {Min}, {Watson},
  {Dullemond}, {Woitke}, {M{\'e}nard}, \& {Dur{\'a}n-Rojas}}]{Pinte2009}
{Pinte}, C., {Harries}, T.~J., {Min}, M., {et~al.} 2009, \aap, 498, 967

\bibitem[{{Pinte} {et~al.}(2006){Pinte}, {M{\'e}nard}, {Duch{\^e}ne}, \&
  {Bastien}}]{Pinte2006}
{Pinte}, C., {M{\'e}nard}, F., {Duch{\^e}ne}, G., \& {Bastien}, P. 2006, \aap,
  459, 797

\bibitem[{{Poelman} {et~al.}(2007){Poelman}, {Spaans}, \&
  {Tielens}}]{Poelman2007}
{Poelman}, D.~R., {Spaans}, M., \& {Tielens}, A.~G.~G.~M. 2007, \aap, 464, 1023

\bibitem[{{Poglitsch} {et~al.}(2010){Poglitsch}, {Waelkens}, {Geis},
  {Feuchtgruber}, {Vandenbussche}, {Rodriguez}, {Krause}, {Renotte}, {van
  Hoof}, {Saraceno}, {Cepa}, {Kerschbaum}, {Agn{\`e}se}, {Ali}, {Altieri},
  {Andreani}, {Augueres}, {Balog}, {Barl}, {Bauer}, {Belbachir}, {Benedettini},
  {Billot}, {Boulade}, {Bischof}, {Blommaert}, {Callut}, {Cara}, {Cerulli},
  {Cesarsky}, {Contursi}, {Creten}, {De Meester}, {Doublier}, {Doumayrou},
  {Duband}, {Exter}, {Genzel}, {Gillis}, {Gr{\"o}zinger}, {Henning},
  {Herreros}, {Huygen}, {Inguscio}, {Jakob}, {Jamar}, {Jean}, {de Jong},
  {Katterloher}, {Kiss}, {Klaas}, {Lemke}, {Lutz}, {Madden}, {Marquet},
  {Martignac}, {Mazy}, {Merken}, {Montfort}, {Morbidelli}, {M{\"u}ller},
  {Nielbock}, {Okumura}, {Orfei}, {Ottensamer}, {Pezzuto}, {Popesso},
  {Putzeys}, {Regibo}, {Reveret}, {Royer}, {Sauvage}, {Schreiber}, {Stegmaier},
  {Schmitt}, {Schubert}, {Sturm}, {Thiel}, {Tofani}, {Vavrek}, {Wetzstein},
  {Wieprecht}, \& {Wiezorrek}}]{Poglitsch2010}
{Poglitsch}, A., {Waelkens}, C., {Geis}, N., {et~al.} 2010, \aap, 518, L2+

\bibitem[{{Pontoppidan} {et~al.}(2010{\natexlab{a}}){Pontoppidan}, {Salyk},
  {Blake}, \& {K{\"a}ufl}}]{Pontoppidan2010a}
{Pontoppidan}, K.~M., {Salyk}, C., {Blake}, G.~A., \& {K{\"a}ufl}, H.~U.
  2010{\natexlab{a}}, \apjl, 722, L173

\bibitem[{{Pontoppidan} {et~al.}(2010{\natexlab{b}}){Pontoppidan}, {Salyk},
  {Blake}, {Meijerink}, {Carr}, \& {Najita}}]{Pontoppidan2010}
{Pontoppidan}, K.~M., {Salyk}, C., {Blake}, G.~A., {et~al.} 2010{\natexlab{b}},
  \apj, 720, 887

\bibitem[{{Qi} {et~al.}(2003){Qi}, {Kessler}, {Koerner}, {Sargent}, \&
  {Blake}}]{Qi2003}
{Qi}, C., {Kessler}, J.~E., {Koerner}, D.~W., {Sargent}, A.~I., \& {Blake},
  G.~A. 2003, \apj, 597, 986

\bibitem[{{Qi} {et~al.}(2006){Qi}, {Wilner}, {Calvet}, {Bourke}, {Blake},
  {Hogerheijde}, {Ho}, \& {Bergin}}]{Qi2006}
{Qi}, C., {Wilner}, D.~J., {Calvet}, N., {et~al.} 2006, \apjl, 636, L157

\bibitem[{{Riviere-Marichalar} {et~al.}(2012){Riviere-Marichalar},
  {M{\'e}nard}, {Thi}, {Kamp}, {Montesinos}, {Meeus}, {Woitke}, {Howard},
  {Sandell}, {Podio}, {Dent}, {Mendigut{\'{\i}}a}, {Pinte}, {White}, \&
  {Barrado}}]{Riviere-Marichalar2012}
{Riviere-Marichalar}, P., {M{\'e}nard}, F., {Thi}, W.~F., {et~al.} 2012, \aap,
  538, L3

\bibitem[{{Salyk} {et~al.}(2007){Salyk}, {Blake}, {Boogert}, \&
  {Brown}}]{Salyk2007}
{Salyk}, C., {Blake}, G.~A., {Boogert}, A.~C.~A., \& {Brown}, J.~M. 2007,
  \apjl, 655, L105

\bibitem[{{Salyk} {et~al.}(2008){Salyk}, {Pontoppidan}, {Blake}, {Lahuis}, {van
  Dishoeck}, \& {Evans}}]{Salyk2008}
{Salyk}, C., {Pontoppidan}, K.~M., {Blake}, G.~A., {et~al.} 2008, \apjl, 676,
  L49

\bibitem[{{Sch{\"o}ier} {et~al.}(2005){Sch{\"o}ier}, {van der Tak}, {van
  Dishoeck}, \& {Black}}]{Schoier2005}
{Sch{\"o}ier}, F.~L., {van der Tak}, F.~F.~S., {van Dishoeck}, E.~F., \&
  {Black}, J.~H. 2005, \aap, 432, 369

\bibitem[{{Semenov} \& {Wiebe}(2011)}]{Semenov2011}
{Semenov}, D. \& {Wiebe}, D. 2011, \apjs, 196, 25

\bibitem[{{Szul{\'a}gyi} {et~al.}(2012){Szul{\'a}gyi}, {Pascucci},
  {{\'A}brah{\'a}m}, {Apai}, {Bouwman}, \& {Mo{\'o}r}}]{Szulagyi2012}
{Szul{\'a}gyi}, J., {Pascucci}, I., {{\'A}brah{\'a}m}, P., {et~al.} 2012, ArXiv
  e-prints

\bibitem[{{Tennyson} {et~al.}(2001){Tennyson}, {Zobov}, {Williamson},
  {Polyansky}, \& {Bernath}}]{Tennyson2001}
{Tennyson}, J., {Zobov}, N.~F., {Williamson}, R., {Polyansky}, O.~L., \&
  {Bernath}, P.~F. 2001, Journal of Physical and Chemical Reference Data, 30,
  735

\bibitem[{{Thi} {et~al.}(2010{\natexlab{a}}){Thi}, {Mathews}, {M{\'e}nard},
  {Woitke}, {Meeus}, {Riviere-Marichalar}, {Pinte}, {Howard}, {Roberge},
  {Sandell}, {Pascucci}, {Riaz}, {Grady}, {Dent}, {Kamp}, {Duch{\^e}ne},
  {Augereau}, {Pantin}, {Vandenbussche}, {Tilling}, {Williams}, {Eiroa},
  {Barrado}, {Alacid}, {Andrews}, {Ardila}, {Aresu}, {Brittain}, {Ciardi},
  {Danchi}, {Fedele}, {de Gregorio-Monsalvo}, {Heras}, {Huelamo}, {Krivov},
  {Lebreton}, {Liseau}, {Martin-Zaidi}, {Mendigut{\'{\i}}a}, {Montesinos},
  {Mora}, {Morales-Calderon}, {Nomura}, {Phillips}, {Podio}, {Poelman},
  {Ramsay}, {Rice}, {Solano}, {Walker}, {White}, \& {Wright}}]{Thi2010a}
{Thi}, W.-F., {Mathews}, G., {M{\'e}nard}, F., {et~al.} 2010{\natexlab{a}},
  \aap, 518, L125+

\bibitem[{{Thi} {et~al.}(2011{\natexlab{a}}){Thi}, {M{\'e}nard}, {Meeus},
  {Martin-Za{\"i}di}, {Woitke}, {Tatulli}, {Benisty}, {Kamp}, {Pascucci},
  {Pinte}, {Grady}, {Brittain}, {White}, {Howard}, {Sandell}, \&
  {Eiroa}}]{Thi2011b}
{Thi}, W.-F., {M{\'e}nard}, F., {Meeus}, G., {et~al.} 2011{\natexlab{a}}, \aap,
  530, L2

\bibitem[{{Thi} {et~al.}(2004){Thi}, {van Zadelhoff}, \& {van
  Dishoeck}}]{Thi2004}
{Thi}, W.-F., {van Zadelhoff}, G.-J., \& {van Dishoeck}, E.~F. 2004, \aap, 425,
  955

\bibitem[{{Thi} {et~al.}(2010{\natexlab{b}}){Thi}, {Woitke}, \&
  {Kamp}}]{Thi2010b}
{Thi}, W.-F., {Woitke}, P., \& {Kamp}, I. 2010{\natexlab{b}}, \mnras, 407, 232

\bibitem[{{Thi} {et~al.}(2011{\natexlab{b}}){Thi}, {Woitke}, \&
  {Kamp}}]{Thi2011a}
{Thi}, W.-F., {Woitke}, P., \& {Kamp}, I. 2011{\natexlab{b}}, \mnras, 412, 711

\bibitem[{{Tilling} {et~al.}(2012){Tilling}, {Woitke}, {Meeus}, {Mora},
  {Montesinos}, {Riviere-Marichalar}, {Eiroa}, {Thi}, {Isella}, {Roberge},
  {Martin-Zaidi}, {Kamp}, {Pinte}, {Sandell}, {Vacca}, {M{\'e}nard},
  {Mendigut{\'{\i}}a}, {Duch{\^e}ne}, {Dent}, {Aresu}, {Meijerink}, \&
  {Spaans}}]{Tilling2012}
{Tilling}, I., {Woitke}, P., {Meeus}, G., {et~al.} 2012, \aap, 538, A20

\bibitem[{{Turpin} {et~al.}(2010){Turpin}, {Stoecklin}, \&
  {Voronin}}]{Turpin2010}
{Turpin}, F., {Stoecklin}, T., \& {Voronin}, A. 2010, \aap, 511, A28

\bibitem[{{Vacca} \& {Sandell}(2011)}]{Vacca2011}
{Vacca}, W.~D. \& {Sandell}, G. 2011, \apj, 732, 8

\bibitem[{{van Dishoeck} {et~al.}(2011){van Dishoeck}, {Kristensen}, {Benz},
  {Bergin}, {Caselli}, {Cernicharo}, {Herpin}, {Hogerheijde}, {Johnstone},
  {Liseau}, {Nisini}, {Shipman}, {Tafalla}, {van der Tak}, {Wyrowski},
  {Aikawa}, {Bachiller}, {Baudry}, {Benedettini}, {Bjerkeli}, {Blake},
  {Bontemps}, {Braine}, {Brinch}, {Bruderer}, {Chavarr{\'{\i}}a}, {Codella},
  {Daniel}, {de Graauw}, {Deul}, {di Giorgio}, {Dominik}, {Doty}, {Dubernet},
  {Encrenaz}, {Feuchtgruber}, {Fich}, {Frieswijk}, {Fuente}, {Giannini},
  {Goicoechea}, {Helmich}, {Herczeg}, {Jacq}, {J{\o}rgensen}, {Karska},
  {Kaufman}, {Keto}, {Larsson}, {Lefloch}, {Lis}, {Marseille}, {McCoey},
  {Melnick}, {Neufeld}, {Olberg}, {Pagani}, {Pani{\'c}}, {Parise}, {Pearson},
  {Plume}, {Risacher}, {Salter}, {Santiago-Garc{\'{\i}}a}, {Saraceno},
  {St{\"a}uber}, {van Kempen}, {Visser}, {Viti}, {Walmsley}, {Wampfler}, \&
  {Y{\i}ld{\i}z}}]{vanDishoeck2011}
{van Dishoeck}, E.~F., {Kristensen}, L.~E., {Benz}, A.~O., {et~al.} 2011,
  \pasp, 123, 138

\bibitem[{{Van Zadelhoff} {et~al.}(2003){Van Zadelhoff}, {Aikawa},
  {Hogerheijde}, \& {van Dishoeck}}]{vanZadelhoff2003}
{Van Zadelhoff}, G.-J., {Aikawa}, Y., {Hogerheijde}, M.~R., \& {van Dishoeck},
  E.~F. 2003, \aap, 397, 789

\bibitem[{{Westley} {et~al.}(1995){Westley}, {Baragiola}, {Johnson}, \&
  {Baratta}}]{Westley1995}
{Westley}, M.~S., {Baragiola}, R.~A., {Johnson}, R.~E., \& {Baratta}, G.~A.
  1995, \nat, 373, 405

\bibitem[{{Wilner} {et~al.}(2000){Wilner}, {Ho}, {Kastner}, \&
  {Rodr{\'{\i}}guez}}]{Wilner2000}
{Wilner}, D.~J., {Ho}, P.~T.~P., {Kastner}, J.~H., \& {Rodr{\'{\i}}guez}, L.~F.
  2000, \apjl, 534, L101

\bibitem[{{Wilson} \& {Bell}(2002)}]{Wilson2002}
{Wilson}, N.~J. \& {Bell}, K.~L. 2002, \mnras, 337, 1027

\bibitem[{{Wilson}(1999)}]{Wilson1999}
{Wilson}, T.~L. 1999, Reports on Progress in Physics, 62, 143

\bibitem[{{Woitke} {et~al.}(2009{\natexlab{a}}){Woitke}, {Kamp}, \&
  {Thi}}]{Woitke2009a}
{Woitke}, P., {Kamp}, I., \& {Thi}, W.-F. 2009{\natexlab{a}}, \aap, 501, 383

\bibitem[{{Woitke} {et~al.}(2011){Woitke}, {Riaz}, {Duch{\^e}ne}, {Pascucci},
  {Lyo}, {Dent}, {Phillips}, {Thi}, {M{\'e}nard}, {Herczeg}, {Bergin}, {Brown},
  {Mora}, {Kamp}, {Aresu}, {Brittain}, {de Gregorio-Monsalvo}, \&
  {Sandell}}]{Woitke2011}
{Woitke}, P., {Riaz}, B., {Duch{\^e}ne}, G., {et~al.} 2011, \aap, 534, A44

\bibitem[{{Woitke} {et~al.}(2009{\natexlab{b}}){Woitke}, {Thi}, {Kamp}, \&
  {Hogerheijde}}]{Woitke2009b}
{Woitke}, P., {Thi}, W.-F., {Kamp}, I., \& {Hogerheijde}, M.~R.
  2009{\natexlab{b}}, \aap, 501, L5

\bibitem[{{Woodall} {et~al.}(2007){Woodall}, {Ag{\'u}ndez}, {Markwick-Kemper},
  \& {Millar}}]{UMIST2007}
{Woodall}, J., {Ag{\'u}ndez}, M., {Markwick-Kemper}, A.~J., \& {Millar}, T.~J.
  2007, \aap, 466, 1197

\bibitem[{{Yang} {et~al.}(2010){Yang}, {Stancil}, {Balakrishnan}, \&
  {Forrey}}]{Yang2010}
{Yang}, B., {Stancil}, P.~C., {Balakrishnan}, N., \& {Forrey}, R.~C. 2010,
  \apj, 718, 1062

\bibitem[{{Zhang} {et~al.}(2013){Zhang}, {Pontoppidan}, {Salyk}, \&
  {Blake}}]{Zhang2013}
{Zhang}, K., {Pontoppidan}, K.~M., {Salyk}, C., \& {Blake}, G.~A. 2013, \apj,
  766, 82

\end{thebibliography}

\appendix

\section{Atomic and molecular data}
\label{app:moldata}

The level populations required for the line radiative transfer are calculated from statistical equilibrium and escape probability \citep[see][for details]{Woitke2009a}. Table~\ref{Tab:atom-mole-data} summarizes the number of levels, lines and collision partners employed for each atom, ion and molecule and also provides the respective references for the collision cross sections. The data itself was taken from the LAMDA database \citep{Schoier2005} for all of them except water, which is discussed in more detail in the following paragraph.

Computed collision rates with p-H$_2$ and o-H$_2$ between rotational levels of water are taken from \citet{Daniel2011,Daniel2010} and \citet{Dubernet2009,Dubernet2006}. Collision rates with He from \citet{Green1980} are taken into account. The ro-vibrational level energies and transition frequencies for the 441 lowest ortho-water levels and 413 lowest para-water levels ($E_{\mathrm{up}}\!<\!5000~{\rm cm}^{-1}$) are drawn from the work of \citet{Tennyson2001} and \citet{Barber2006}. The collision rates with H$_2$ and electrons for the ro-vibrational levels and covering temperatures between 200 and 500~K were computed by \citet{Faure2008}. We extrapolated those rates down to 10~K assuming the rate at 200~K for the collisions with electrons.  In the absence of any other data, the collision rates with atomic hydrogen are scaled by a factor $\sqrt{2}$ from the collision rates with H$_2$.

\begin{table}[h]
\caption{Non-LTE model atoms, ions and molecules. The second column provides the number of levels and lines included.}
\begin{tabular}{cccc}
\hline
\hline
Species & levels/lines & Coll. partners & References \\
\hline
O\,I &  3/3 & p-H$_2$, o-H$_2$, H, H$^+$, e$^{-}$ & 1,1,2,3,4\\
C\,II & 8/14 & p-H$_2$, o-H$_2$, H, e$^{-}$ & 3,3,3,5 \\
CH$^+$ & 16/15 & H$_2$, e$^{-}$ & 6,7\\
HCO$^+$ & 31/30 & H$_2$ & 8\\
CO & 41/40 & p-H$_2$, o-H$_2$, H, He, & 9,9,10,11\\
OH & 24/95 & p-H$_2$, o-H$_2$ & 12,12\\
o-H$_2$O & 411/4248 &  He, H, H$_2$, e$^-$  & 13,14,14,14\\
p-H$_2$O & 413/3942 &  He, H, H$_2$, e$^-$  & 13,14,14,14\\
HCN & 30/29 & H$_2$ & 3\\
\hline
\end{tabular}
\label{Tab:atom-mole-data}
\\ \tablefoottext{1}{\citet{Jaquet1992};}
\tablefoottext{2}{\citet{Abrahamsson2007};}
\tablefoottext{3}{\citet{Schoier2005};}
\tablefoottext{4}{\citet{Bell1998};}
\tablefoottext{5}{\citet{Wilson2002};}
\tablefoottext{6}{\citet{Hammami2009, Turpin2010}, rates for He scaled by 1.39;}
\tablefoottext{7}{\citet{Lim1999};}
\tablefoottext{8}{\citet{Flower1999};}
\tablefoottext{9}{\citet{Yang2010};}
\tablefoottext{10}{\citet{Balakrishnan2002};}
\tablefoottext{11}{\citet{Cecchi-Pestellini2002};}
\tablefoottext{12}{\citet{Offer1994};}
\tablefoottext{13}{\citet{Green1980};}
\tablefoottext{14}{\citet{Faure2008}}
\end{table}

\section{Metal abundances}
\label{app:metalabus}

HCO$^+$ is a molecule that is known to be very sensitive to the ionization in the disk surface, especially the electron abundance \citep{Qi2003}. It is formed in the outer disk surface through ion-molecule reactions such as H$_3^+$\,$+$\,CO, CO$^+$\,$+$\,H$_2$ and predominately destroyed by electronic recombination. This is essentially PDR chemistry and can be well understood using the standard PDR parameter $\chi/n_{\langle H \rangle}$. The HCO$^+$ abundance peaks in a thin surface layer at a $\chi/n_{\langle H \rangle}$ of $10^{-4}$, where the total hydrogen number densities are of the order of $\sim\!10^7$~cm$^{-3}$ (Fig.~\ref{fig:standard-basic}). The thickness of this HCO$^+$ layer increases if ionizing radiation can penetrate deeper, e.g.\ in the case of X-rays and/or if self-shielding is switched off. \citet{vanZadelhoff2003} showed the differences in the thickness of the layer due to different radiative transfer methods and stellar input spectra. However, the effect on the HCO$^+$ line fluxes was of the order of 30\% or less.

The HCO$^+$~4--3 line flux as measured by \citet{Thi2004} is $0.272\!\times\!10^{-18}$~W/m$^2$, which is roughly two orders or magnitude above our predicted flux from the standard model (low ISM abundances from Table~\ref{Tab:elementabus}). Hence, in Table~\ref{HCO+chemistry} we carefully revisited several recombination rates from the recent literature and the overall metal abundance.

\begin{table}[h]
\caption{New rate coefficients for HCO$^+$ chemistry using the standard UMIST rate parametrization.}
\begin{center}
\begin{tabular}{llr}
\hline
\hline
\hspace*{-1mm}reaction \hspace*{-2mm}& rate coefficient $R_{ij}$\hspace*{-5mm} & ref. \\
                \hline
\hspace*{-1mm}H$_3^++$CO$\rightarrow$HCO$^++$H$_2$\hspace*{-2mm} &  $1.36\,10^{-9}  \left(\frac{T}{300~{\rm K}}\right)^{-0.142} \exp{\left(\frac{3.41}{T}\right)}$ \hspace*{-5mm}& 1 \\
\hspace*{-1mm}H$_3^++$O$\rightarrow$OH$^++$H$_2$\hspace*{-2mm}  & $1.14\,10^{-9} \left(\frac{T}{300~{\rm K}}\right)^{-0.156} \exp{\left(\frac{-1.41}{T}\right)}$ \hspace*{-5mm}& 1 \\
\hspace*{-1mm}H$_3^++$CO$\rightarrow$HOC$^++$H$_2$\hspace*{-2mm} & $8.49\,10^{-10} \left(\frac{T}{300~{\rm K}}\right)^{0.0661} \exp{\left(\frac{-5.21}{T}\right)}$ \hspace*{-5mm}& 1 \\
\hspace*{-1mm}HCO$^++$e$^-\rightarrow$CO$+$H\hspace*{-2mm} &  $2.0\,10^{-7}  \left(\frac{T}{300~{\rm K}}\right)^{-1.30}$  \hspace*{-5mm} & 2 \\
\hspace*{-1mm}DCO$^++$e$^-\rightarrow$CO$+$D\hspace*{-2mm}  & $1.70\,10^{-7}   \left(\frac{T}{300~{\rm K}}\right)^{1.10}$ \hspace*{-5mm} & 2 \\ 
\hline
\end{tabular}
\end{center}
{(1) \citet{KGM10}; (2) \citet{K09}}
\label{HCO+chemistry}
\end{table}

The low metal abundances from Table~\ref{Tab:elementabus} are still high enough (electron donors are mostly the metals) to provide electron abundances of $10^{-5}$ below $\chi/n_{\langle H \rangle}$ of $10^{-4}$. This makes electronic recombination very efficient and destroys HCO$^+$ in those layers. Changing to the new rates presented in Table~\ref{HCO+chemistry} does not affect the thickness of the HCO$^+$ layer significantly. For all the other tests carried out in this paper, we thus stick to the standard UMIST rates.

The H$_3^+$ recombination rate is still disputed \citep[e.g.][]{McCall2006}. We use in this study the UMIST rates for the reactions H$_3^+$\,$+$\,e$^- \rightarrow$~H$_2$\,$+$\,H ($k_1$) or $\rightarrow 3\,$H ($k_2$)
\begin{eqnarray}
k_1 & = & 4.36\times 10^{-8} \left(\frac{T}{300~{\rm K}}\right)^{-0.52} ~~{\rm cm^3~s^{-1}}\\ \nonumber
k_2 & = & 2.34 \times 10^{-8} \left(\frac{T}{300~{\rm K}}\right)^{-0.52}~~{\rm cm^3~s^{-1}}~~, \nonumber 
\end{eqnarray}
where $T$ is the gas temperature in K. Lowering the H$_3^+$ recombination rate by a factor 10 has no significant impact on the HCO$^+$ abundance.


\section{Water formation on grain surfaces}
\label{app:watersurf}

The model used here for the formation of water on grain surfaces is based on the one described by \citet{Hollenbach2009}. The network captures only the Eley-Rideal mechanism of prompt reaction with a gas phase H atom. In addition, water desorption can occur either intact or in the form of OH\,$+$\,H.
%

\citet{Fogel2011} report that they included this surface formation channel of water into their chemical network. The reactions 
\begin{eqnarray}
{\rm H\#} + {\rm O} \rightarrow {\rm OH\#} \\ \label{Eq:chem1}
{\rm H\#} + {\rm OH} \rightarrow {\rm H_2O\#} \label{Eq:chem2}
\end{eqnarray}
which are reported as their Eq.(16) and (17) most likely play a very minor role. Unless the environment is highly molecular (all H in the form of H$_2$), any H adsorbed to the surface (H\#) will have a much larger chance to encounter another H atom impinging from the gas phase before it encounters an O or OH impinging from the gas phase. These reactions are also not included in the reaction network suggested by \citet{Hollenbach2009}. We here strictly follow the latter reaction network and include in two steps the formation and desorption of O and OH ice and then the surface reaction of O and OH ice with impinging H-atoms (Eley-Rideal mechanism). The formation of water ice and its thermal desorption, desorption through cosmic rays and photodesorption were already included before \citep[see][]{Woitke2009a}.

For a direct reaction of a surface species $i$ with an impinging atom/molecule $j$ from the gas phase, the reaction rate can be written as
\begin{equation}
R_{ij}  =  n_j v_{\rm th,j} \pi \langle a^2 \rangle n_{\rm d} \left( \frac{N_{i\#}}{N_{\rm surf}} \right)~~~{\rm cm^{-3}~s^{-1}} 
\end{equation}
where the last term in brackets denotes the covering fraction for a single grain. $a$ is here the radius of the dust grain, $n_{\rm d}$ the density of dust grains, $N_{i\#}$ the total number of ice particles of species $i$ and $N_{\rm surf}$ the total number of surface sites on a single grain. To rewrite this in terms of volume densities (as used in {\sc ProDiMo}), the latter two can be calculated from the number density of ice species $i$, $n_{i\#}$ and the number of surface sites per surface area on a grain $n_{\rm surf}$ [cm$^{-2}$]
\begin{equation}
R_{ij}   =  n_j v_{\rm th,j} \pi \langle a^2 \rangle n_{\rm d} \left( \frac{n_{i\#}}{n_{\rm surf} 4 \pi \langle a^2 \rangle n_{\rm d} N_{\rm Lay}} \right)~~~{\rm cm^{-3}~s^{-1}} 
\end{equation}
where $N_{Lay}$ is the number of active layers on a grain surface. The thermal velocity $v_{\rm th,j}$ of a species $j$ is calculated as
\begin{equation}
v_{\rm th,j} = \sqrt{\frac{kT}{2\pi m_j}}
\end{equation}
where $m_j$ is the mass of the species $j$. We need to take into account that the number of surface ice layers taking part in the chemistry is limited. The total density of surface species is calculated as
\begin{equation}
n_{\rm tot}^{\rm ice} = \sum_i n_{i\#}
\end{equation}
and the number density of active surface sites $n_{\rm act}$ is defined as
\begin{equation}
n_{\rm act} = 4 \pi \langle a^2 \rangle n_{\rm surf} n_{\rm d} N_{Lay}
\end{equation}
The total reaction rate is then given by
\begin{equation}
R_{ij} = \begin{cases} k_{ij} \cdot n_j n_{i\#}, & {\rm if}~n_{\rm tot}^{\rm ice} < n_{\rm act} \\ k_{ij} \cdot n_j n_{act} \frac{n_{i\#}}{n_{\rm tot}^{\rm ice}}, & {\rm if}~n_{\rm tot}^{\rm ice} \geq n_{\rm act} \end{cases}
\label{Eq:GG1}
\end{equation}
where the rate coefficient $k_{ij}$ is given as
\begin{equation}
k_{ij} = \frac{v_{\rm th,j} \pi \langle a^2 \rangle}{n_{\rm act}}~~~{\rm cm^{3}~s^{-1}} 
\end{equation}
The reaction rates for the two surface reactions
\begin{eqnarray}
\nonumber {\rm O\#} + {\rm H} \rightarrow {\rm OH\#} \\
{\rm OH\#} + {\rm H} \rightarrow {\rm H_2O\#} 
\label{Eq:watersurf}
\end{eqnarray}
are calculated from Eq.(\ref{Eq:GG1}). In fact by assuming the simple geometric cross section, we most likely maximize the reaction rate. 

Another change to the water chemistry, if surface reactions are included, is that we allow for two reaction channels for water photodesorption, (a) H$_2$O\#\,$+$\,$h\nu$ $\rightarrow$ H$_2$O and (b) H$_2$O\# $+$ $h\nu$ $\rightarrow$ OH\,$+$\,H. The yields for the two reactions are taken from \citet{Hollenbach2009}, $10^{-3}$ and $2\!\times\!10^{-3}$ respectively. The adsorption energy is assumed to be 4800~K. Thermal as well as cosmic ray desorption are still assumed to desorb the water molecule as a whole.

\section{Line fluxes from model series}
\label{app:linefluxes}

Table~\ref{tab:PACSmodfluxes1} and \ref{tab:PACSmodfluxes2} list the modeled emission line fluxes for the entire series of 14 models described in Table~\ref{Tab:modelseries}. The following two subsections discuss a number of key diagnostics in addition to water: the fine structure lines and a series of submm lines.

\subsection{Fine structure lines}

\begin{figure}[bth]
\includegraphics[width=9.5cm]{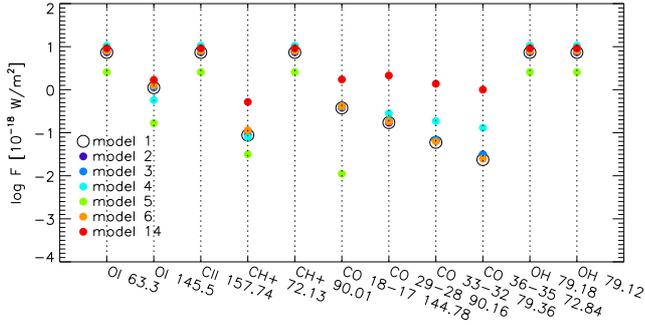}
\vspace*{-9mm}
\caption{Sensitivity of the fine structure lines, CH$^+$ lines, CO high J rotational lines and OH hyperfine structure lines in the model series without grain surface chemistry.} 
\label{fig:finestructurelines}
\end{figure}

Grain surface chemistry does not affect the fine structure line fluxes. Fig.~\ref{fig:finestructurelines} illustrates that the model with $T_{\rm gas}\!=\!T_{\rm dust}$ has the largest impact on the fine structure lines of oxygen and carbon; the [O\,{\sc i}]\,63~$\mu$m line is a factor $6.3$ weaker than the standard model. The change in line flux caused by introducing O ice or switching off the self-shielding is below 20\%. Taking into account X-rays at the level observed for TW Hya, $L_{\rm X}\!=\!1.3\!\times\!10^{30}$~erg/s, increases the [O\,{\sc i}] line fluxes by only 5\%. 

For a high C/O ratio of $1.86$, the [O\,{\sc i}]\,63 and 145\,$\mu$m fine structure line fluxes are lower by a factor $\sim\!1.6$ and 1.9 respectively (compared to the `standard' model). The [OI]\,63\,$\mu$m line changes from being optically thick in the standard model to being thin in the model with lower oxygen abundance. This explains why both lines do not change by the same factor.

\subsection{Submm lines of CO, HCO$^+$ and HCN}
\label{subsect:submmlines}

The observed $^{12}$CO/$^{13}$CO line ratio is $\sim\!10$, smaller than the expected optically thin value of $69$ \citep[isotopologue ratio][]{Wilson1999}. In our `standard' model, the line ratio is $\sim\!4$. If the gas mass were higher, the $^{13}$CO line would become optically thick as well, thus moving the line ratio to even smaller values \citep[see][]{Thi2010a}. Since both lines are highly optically thick at disk gas masses above $M_{\rm gas}\!=\!3\!\times\!10^{-3}$~M$_\odot$, different photodissociation rates for $^{13}$CO will not change the main conclusion. 

\begin{figure}[bth]
\includegraphics[width=9.5cm]{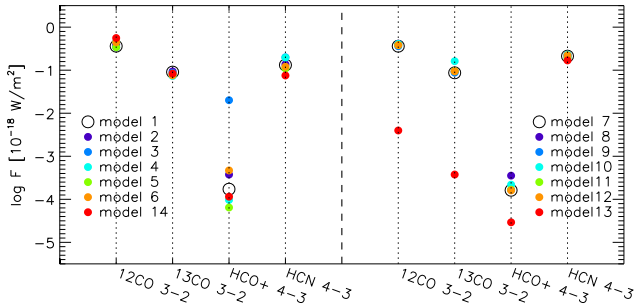}
\vspace*{-9mm}
\caption{Sensitivity of the submm lines from the outer disk in the model series without grain surface chemistry (left) and with grain surface chemistry (right).} 
\label{fig:submmlines}
\end{figure}

The HCN fluxes from the model do not change by more than 50\% throughout the series of models presented in this work. Even when changing the C/O ratio to $1.86$, the HCN rotational line fluxes increase only by a factor 1.5. 

Also, the CO submm line fluxes are very stable (within $0.3$~dex). The only process impacting those lines is the lack of photodesorption. In that case, all oxygen remains in the form of water ice on the grains, thus preventing the formation of CO in the gas phase.

HCO$^+$ is the most sensitive line in our study. Its flux reacts to the presence of X-rays because the H$_3^+$ abundances increase. The line forms in the upper disk, where the assumption $T_{\rm gas}\!=\!T_{\rm dust}$ is not valid; assuming the lower dust temperature for the gas produces a lower HCO$^+$ line flux. Since the line originates in a layer with fairly high density ($n_{\langle H \rangle}\!\sim\! 10^7$~cm$^{-3}$), the line is close to LTE. The strongest impact occurs for a changing metallicity. If the metal abundance is lowered with respect to the standard model by two/three orders of magnitude (S, Na, Mg, Si, Fe, and PAH abundance), the model HCO$^+$ line flux becomes $5.66\!\times\!10^{-21}$/$1.88\!\times\!10^{-20}$~W/m$^2$. The thickness of the HCO$^+$ layer grows because of the lower electron abundance, thus building up higher column densities. Table~\ref{tab:metallicity} shows that the [Fe\,{\sc ii}]\,25.99~$\mu$m line is anticorrelated with the HCO$^+$ line. When the metallicity changes by 3~dex, the HCO$^+$ line becomes a factor 116 stronger, while the iron fine structure line decreases by a factor 1000. On the other hand, HCN line fluxes are not affected and the water line fluxes and [O\,{\sc i}] line flux do not change by more than 3\%. The TW Hya upper limit reported from Spitzer/IRS spectra for the [Fe\,{\sc ii}]\,25.99~$\mu$m line (Carr private communication) is $\sim\!10^{-17}$~W/m$^2$, but the presence of additional spectral features makes this estimate rather difficult. At this level, the Spitzer upper limit on [Fe\,{\sc ii}] is less stringent in fixing the metal/electron abundance than the HCO$^+$ line. \citet{Semenov2011} modeled a disk with strong  turbulent diffusion (a diffusion coefficient of $\sim\!10^{18}$~cm$^2$/s in the outer disk regions) and show that the mixing can enhance the HCO$^+$ abundance in the thin surface layers by up to two orders of magnitude. 

\begin{table}[h]
\caption{Line fluxes [$10^{-18}$\,W/m$^2$] as a function of metallicity. The notation (-2) is an abbreviation for $\times\!10^{-2}$.}
\begin{tabular}{l|ccc}
\hline
\hline
metallicity    & HCO$^+$~4--3 & HCN~4--3 & [Fe\,{\sc ii}]\,25.99\,$\mu$m \\
\hline
low-metal           & 1.61(-4)                & 0.134 & 1.50 \\
low-metal/100   & 5.66(-3)                & 0.186 & 1.47(-2) \\
low-metal/1000 & 1.88(-2)                & 0.185 & 1.54(-3) \\
\end{tabular}
\label{tab:metallicity}
\end{table}

This model series confirms once more the robust diagnostic of the standard submm lines (CO, HCN) for deriving the physics/geometry of the outer regions of protoplanetary disks. This is partly due to the relatively simple chemistry, but also due to simpler line radiative transfer (LTE). Similar studies are required to identify robust tracers for the inner disk structure such as the CO ro-vibrational lines. 

\onecolumn
\begin{landscape}

\begin{table*}
\caption{Model series without grain surface reactions: Line fluxes are in $10^{-18}$~W/m$^2$. For non-detections, $3\sigma$ upper limits are listed.}
\begin{tabular}{lll|lllllll}
\hline
\hline
Species & wavelength & observed & standard & with X-rays & low-metal & C/O$=\!1.86$ & 
 $T_g\!=\!T_d$ & without & $\delta\!=\!0.01$ \\
  & [$\mu$m] & line flux &  &  & $\times\!10^{-3}$ &  & 
    & self-shielding & (inside 4~AU)\\
\hline
O & 63.18 & 3.7(1) $\pm$ 3.3  &5.88(1) & 6.14(1) & 5.77(1) & 3.69(1) & 
 9.38  & 6.91(1) & 5.08(1) \\
  & 145.52 & $<3.5$ &1.11  & 1.19  & 1.12  & 5.71(-1) &
    1.67(-1) & 1.25  & 1.68  \\
C\,{\sc ii} & 157.74 & $< 3.6$ &7.35  & 7.43  & 7.76  & 1.05(1) & 
2.56  & 7.98  & 9.18 \\
o-H2O & 538.29 &  1.7(-1)\tablefootmark{a} $\pm$ 7.5(-3)&3.27(-1) & 3.37(-1) & 2.49(-1) & 6.97(-3) & 
1.90(-1) & 3.90(-1) & 4.08(-1) \\
o-H2O & 180.49 & $<8.3$ &8.77(-2) & 1.11(-1) & 8.92(-2) & 7.81(-2) & 
3.17(-2) & 1.14(-1) & 5.18(-1) \\
o-H2O & 179.52 & $<7.9$ &9.06(-1) & 1.07  & 6.04(-1) & 1.95(-1) & 
4.15(-1) & 1.21  & 1.53  \\
o-H2O  & 78.74 & $<6.0$ &1.27  & 1.40  & 1.44  & 2.46  & 
5.55(-2) & 1.53  & 5.94  \\
o-H2O & 71.95 &   $<6.4$ &1.78  & 1.87  & 1.94  & 3.94  & 
8.04(-3) & 2.03  & 6.75  \\
o-H2O & 63.32 &  $<7.2$ &2.14  & 2.23  & 2.37  & 4.86  & 
1.84(-3) & 2.44  & 8.03  \\
p-H2O & 269.27 & 6.1(-1)\tablefootmark{a} $\pm$ 3.8(-2) &6.68(-1) & 7.07(-1) & 9.94(-1) & 2.38(-2) & 
4.18(-1) & 7.82(-1) & 1.64  \\
p-H2O & 158.31 & $<5.2$ &4.67(-5) & 5.02(-5) & 2.21(-4) & 1.42(-4) & 
5.55(-6) & 6.81(-5) & 3.68(-3) \\
p-H2O & 144.52 & $<6.6$ &1.35(-2) & 1.49(-2) & 4.79(-2) & 4.23(-2) & 
1.77(-3) & 1.94(-2) & 4.65(-1) \\
p-H2O & 89.988\tablefootmark{b} & 5.6 $\pm$ 0.9 &2.08(-1) & 2.43(-1) & 5.95(-1) & 6.98(-1) & 
3.49(-2) & 2.86(-1) & 3.45  \\
p-H2O & 78.93 &  $<6.4$ &1.37(-1) & 1.40(-1) & 4.08(-1) & 5.98(-1) & 
8.73(-4) & 1.67(-1) & 3.04 \\
CH$^+$ J=5-4 & 72.14 & $<7.1$ &5.19(-2) & 5.31(-2) & 5.58(-2) & 2.09(-1) & 
2.76(-3) & 5.36(-2) & 1.59(-2) \\
CH$^+$ J=4-3 & 90.02\tablefootmark{c} &  &4.82(-2) & 4.94(-2) & 5.08(-2) & 1.63(-1) & 
4.87(-3) & 5.02(-2) & 6.58(-3) \\
CH$^+$ J=3-2 & 119.85 &  &2.62(-2) & 2.63(-2) & 2.70(-2) & 7.17(-2) & 
5.91(-3) & 2.77(-2) & 2.06(-3) \\
CH$^+$ J=2-1 & 179.59\tablefootmark{c} & $<7.8$ &1.08(-2) & 1.06(-2) & 1.11(-2) & 2.32(-2) & 
4.62(-3) & 1.22(-2) & 1.09(-3) \\
$^{12}$CO J=3-2 & 866.96 & 4.2(-1) $\pm$ 8.5(-2)\tablefootmark{d} &3.61(-1) & 3.91(-1) & 3.41(-1) & 4.60(-1) & 
3.41(-1) & 4.46(-1) & 5.59(-1) \\
$^{13}$CO J=3-2 & 906.85 & 4.8(-2) $\pm$ 2.7(-2)\tablefootmark{d} &9.03(-2) & 9.37(-2) & 8.69(-2) & 7.32(-2) & 
8.19(-2) & 7.61(-2) & 8.34(-2) \\
CO J=18-17 & 144.78 & 3.5 $\pm$ 1.2 &3.77(-1) & 3.80(-1) & 3.95(-1) & 4.03(-1) & 
1.11(-2) & 4.08(-1) & 1.73  \\
CO J=29-28 & 90.16 & $<3.4$ &1.72(-1) & 1.71(-1) & 1.84(-1) & 2.83(-1) & 
9.98(-8) & 1.82(-1) & 2.14 \\
CO J=33-32 & 79.36 & $<8.3$ &5.97(-2) & 6.66(-2) & 6.98(-2) & 1.87(-1) & 
6.13(-9) & 6.30(-2) & 1.37 \\
CO J=36-35 & 72.84 & $<4.2$ &2.38(-2) & 3.13(-2) & 3.10(-2) & 1.30(-1) & 
1.83(-9) & 2.49(-2) & 1.00 \\
HCO$^+$ J=4-3 & 840.38 & 2.72(-1) $\pm$ 5.5(-2)\tablefootmark{d} &1.72(-4) & 3.76(-4) & 2.00(-2) & 9.67(-5) & 
6.45(-5) & 4.69(-4) & 1.15(-4) \\
HCN J=4-3 & 845.66 & 1.05(-1) $\pm$ 3.1(-2)\tablefootmark{d} &1.32(-1) & 1.37(-1) & 1.99(-1) & 1.96(-1) & 
1.10(-1) & 1.21(-1) & 7.53(-2) \\
OH   & 79.12 & $<7.8$ &1.66  & 1.75  & 1.82  & 1.63  & 
1.44(-1) & 2.02  & 6.63  \\
OH   & 79.18 & $<7.8$ &1.72  & 1.82  & 1.89  & 1.70  & 
1.50(-1) & 2.10  & 6.68  \\
\hline
\end{tabular}
\\ \tablefoottext{a}{\citet{Hogerheijde2011}, Herschel/HIFI;}
\tablefoottext{b}{this line is a blend with CH$^+$ at 90.02~$\mu$m;}
\tablefoottext{c}{these lines are blended with water;}
\tablefoottext{d}{\citet{Thi2004}, JCMT;}
\label{tab:PACSmodfluxes1}
\end{table*}

\begin{table*}
\caption{Model series with grain surface reactions: Line fluxes are in $10^{-18}$~W/m$^2$. For non-detections, $3\sigma$ upper limits are listed.}
\begin{tabular}{lll|lllllll}
\hline
\hline
Species & wavelength & observed & standard & with X-rays & $100\times$GG & without H$_2$O & photodesorption & $N_{\rm lay}\!=\!10$ & no water \\
 & [$\mu$m] & line flux &  &  &  & adsorption & yields &  & photodesoprtion \\
\hline
O & 63.18 & 3.7(1) $\pm$ 3.3  &5.93(1) & 6.78(1) & 6.01(1) & 5.93(1) & 6.04(1) & 6.06(1) & 2.94(1) \\
  & 145.52 & $<3.5$ &1.11  & 1.25  & 1.16  & 1.12  & 1.17  & 1.16  & 8.31(-1) \\
C\,{\sc ii} & 157.74 & $< 3.6$ &7.52  & 7.94  & 7.70  & 7.52  & 7.70  & 7.85  & 1.57(1) \\
o-H2O & 538.29 &  1.7(-1)\tablefootmark{a} $\pm$ 7.5(-3) &1.65(-1) & 1.39(-1) & 2.49(-1) & 1.96(-1) & 2.64(-1) & 1.40(-1) & 1.82(-3) \\
o-H2O & 180.49 & $<8.3$ &8.69(-2) & 1.07(-1) & 1.45(-1) & 1.51(-1) & 1.36(-1) & 8.86(-2) & 5.89(-2) \\
o-H2O & 179.52 & $<7.9$ &6.80(-1) & 6.89(-1) & 7.79(-1) & 8.11(-1) & 7.73(-1) & 5.72(-1) & 1.23(-1) \\
o-H2O  & 78.74 & $<6.0$ &1.27  & 1.46  & 1.41  & 1.25  & 1.42  & 1.39  & 2.09  \\
o-H2O & 71.95 &   $<6.4$ &1.78  & 1.95  & 1.94  & 1.68  & 1.94  & 1.94  & 3.62  \\
o-H2O & 63.32 &  $<7.2$ &2.14  & 2.35  & 2.34  & 2.06  & 2.34  & 2.34  & 4.60  \\
p-H2O & 269.27 & 6.1(-1)\tablefootmark{a} $\pm$ 3.8(-2)&3.40(-1) & 6.21(-1) & 9.70(-1) & 4.48(-1) & 1.05  & 6.28(-1) & 2.48(-3) \\
p-H2O & 158.31 & $<5.2$ &4.67(-5) & 1.68(-4) & 2.11(-4) & 4.35(-4) & 1.66(-4) & 1.54(-4) & 3.93(-5) \\
p-H2O & 144.52 & $<6.6$ &1.35(-2) & 4.41(-2) & 4.47(-2) & 2.89(-2) & 4.55(-2) & 4.04(-2) & 1.16(-2) \\
p-H2O & 89.988\tablefootmark{b} & 5.6 $\pm$ 0.9 &2.07(-1) & 6.36(-1) & 6.30(-1) & 2.59(-1) & 6.29(-1) & 5.61(-1) & 1.91(-1) \\
p-H2O & 78.93 &  $<6.4$ &1.37(-1) & 3.92(-1) & 3.92(-1) & 1.36(-1) & 3.93(-1) & 3.92(-1) & 1.86(-1) \\
CH$^+$ J=5-4 & 72.14 & $<7.1$ &5.19(-2) & 5.61(-2) & 5.59(-2) & 5.19(-2) & 5.59(-2) & 5.59(-2) & 3.92(-1) \\
CH$^+$ J=4-3 & 90.02\tablefootmark{c} &  &4.82(-2) & 5.04(-2) & 5.07(-2) & 4.82(-2) & 5.07(-2) & 5.07(-2) & 5.65(-1) \\
CH$^+$ J=3-2 & 119.85 &  &2.62(-2) & 2.65(-2) & 2.69(-2) & 2.62(-2) & 2.69(-2) & 2.69(-2) & 4.82(-1) \\
CH$^+$ J=2-1 & 179.59\tablefootmark{c} & $<7.8$ &1.08(-2) & 1.07(-2) & 1.10(-2) & 1.08(-2) & 1.10(-2) & 1.10(-2) & 2.24(-1) \\
$^{12}$CO J=3-2 & 866.96 & 4.2(-1) $\pm$ 8.5(-2)\tablefootmark{d} &3.61(-1) & 4.11(-1) & 4.08(-1) & 3.61(-1) & 3.83(-1) & 3.83(-1) & 3.97(-3) \\
$^{13}$CO J=3-2 & 906.85 & 4.8(-2) $\pm$ 2.7(-2)\tablefootmark{d} &8.68(-2) & 9.31(-2) & 1.62(-1) & 8.77(-2) & 9.48(-2) & 9.42(-2) & 3.77(-4) \\
CO J=18-17 & 144.78 & 3.5 $\pm$ 1.2 &3.77(-1) & 3.93(-1) & 4.08(-1) & 3.77(-1) & 4.08(-1) & 4.08(-1) & 4.81(-1) \\
CO J=29-28 & 90.16 & $<3.4$ &1.72(-1) & 1.72(-1) & 1.79(-1) & 1.72(-1) & 1.79(-1) & 1.79(-1) & 2.95(-1) \\
CO J=33-32 & 79.36 & $<8.3$ &5.97(-2) & 6.97(-2) & 6.78(-2) & 5.97(-2) & 6.78(-2) & 6.78(-2) & 1.62(-1) \\
CO J=36-35 & 72.84 & $<4.2$ &2.38(-2) & 3.35(-2) & 3.01(-2) & 2.38(-2) & 3.01(-2) & 3.01(-2) & 9.56(-2) \\
HCO$^+$ J=4-3 & 840.38 & 2.72(-1) $\pm$ 5.5(-2)\tablefootmark{d} &1.62(-4) & 3.55(-4) & 2.19(-4) & 1.64(-4) & 1.67(-4) & 1.61(-4) & 2.92(-5) \\
HCN J=4-3 & 845.66 & 1.05(-1) $\pm$ 3.1(-2)\tablefootmark{d} &2.12(-1) & 2.23(-1) & 2.31(-1) & 2.02(-1) & 1.94(-1) & 2.22(-1) & 1.68(-1) \\
OH   & 79.12 & $<7.8$ &1.66  & 1.84  & 1.85  & 1.69  & 1.84  & 1.81  & 2.31  \\
OH   & 79.18 & $<7.8$ &1.72  & 1.91  & 1.91  & 1.75  & 1.91  & 1.87  & 2.38  \\
\hline
\end{tabular}
\\ \tablefoottext{a}{\citet{Hogerheijde2011}, Herschel/HIFI;}
\tablefoottext{b}{this line is a blend with CH$^+$ at 90.02~$\mu$m;}
\tablefoottext{c}{these lines are blended with water;}
\tablefoottext{d}{\citet{Thi2004}, JCMT;}
\label{tab:PACSmodfluxes2}
\end{table*}
\end{landscape}
\twocolumn

\clearpage

\end{document}